\documentclass[12pt]{article}

\usepackage[cp1251]{inputenc}
\usepackage[russian]{babel}

\usepackage{russcorr}

\usepackage{amsmath, amssymb, amsthm, amsfonts}
\usepackage{fullpage}
\usepackage{graphicx}
\usepackage{longtable}

\newtheorem{theorem}{Теорема}[subsection]
\newtheorem{lemma}[theorem]{Лемма}
\newtheorem{corollary}[theorem]{Следствие}
\newtheorem{proposition}[theorem]{Предложение}
\newtheorem{conjecture}{Гипотеза}
\theoremstyle{definition}

\newtheorem{problem}[conjecture]{Проблема}

\newcommand{\N}{\mathbb{N}}
\newcommand{\Z}{\mathbb{Z}}
\newcommand{\B}{\mathbb{B}}
\newcommand{\R}{\mathbb{R}}
\renewcommand{\epsilon}{\varepsilon}
\newcommand{\Per}{\mathcal{P}}
\newcommand{\EP}{\mathcal{EP}}
\newcommand{\GAP}{\mathcal{GAP}}
\newcommand{\AP}{\mathcal{AP}}
\newcommand{\PAP}{\mathcal{PAP}}
\newcommand{\EAP}{\mathcal{EAP}}
\newcommand{\Rec}{\mathcal{R}}
\newcommand{\ER}{\mathcal{ER}}

\newcommand{\QP}{\mathcal{QP}}
\newcommand{\MQP}{\mathcal{MQP}}

\newcommand{\M}{\mathcal{M}}

\newcommand{\pr}{\mathop{\mathrm{pr}}} % prefix function for x \in \EAP
\newcommand{\p}{\mathop{\mathrm{p}}}   % subword complexity

\newcommand{\Fac}{\mathop{\mathrm{Fac}}}
\renewcommand{\r}{\mathop{\mathrm R}} % regulator
\newcommand{\rd}{\mathop{\mathrm r}} % regulator - distance
\newcommand{\s}{\mathop{\mathrm s}}   % splitting
\renewcommand{\b}{\mathop{\mathrm b}} % alphabet of blocks when split
\newcommand{\am}{\mathop{\mathrm{AM}}}   % aperiodicity measure
\newcommand{\F}{\mathcal{F}}

\newcommand{\thue}{\mathbf t} % Thue-Morse sequence
\newcommand{\fib}{\mathbf f} % Fibonacci sequence
\newcommand{\kolak}{\mathbf k} % Kolakoski sequence

\newcommand{\T}[1]{\mathop{\mathrm{T}\langle \N, <, #1\rangle}} % theory of order
\newcommand{\Tp}[1]{\mathop{\mathrm{T'}\langle \N, <, #1\rangle}} % theory of order modified
\newcommand{\MT}[1]{\mathop{\mathrm{MT}\langle \N, <, #1\rangle}} % monadic theory of order

\newcommand{\tocontents}[1]{\addcontentsline{toc}{section}{#1}}

\date{\today}

\begin{document}

\noindent УДК 510.5, 510.6, 519.101, 517.938

\bigskip

\begin{center}
{\LARGE Последовательности, близкие к периодическим%
\footnote{Работа частично поддержана грантами НШ-5351.2006.1 (Совет
поддержки научных школ) и 05-01-02803, 06-01-00122 (РФФИ).}}
\end{center}

\begin{center}
\large Ан.~А.~Мучник\footnote{Институт Новых Технологий. Андрей
Альбертович Мучник (24.02.1958 --- 18.03.2007),
см.~\cite{MuchnikObituary07}.}, %
Ю.~Л.~Притыкин\footnote{Московский Государственный Университет
им.~М.~В.~Ломоносова, \texttt{yura@mccme.ru}.%,\texttt{http://www.mccme.ru/\~{}yura/}
}, %
А.~Л.~Семёнов\footnote{Вычислительный Центр им.~А.~А.~Дородницына
РАН, \texttt{alsemenov@umail.ru}.}
%\large Ан.~А.~Мучник\footnote{ }, %
%Ю.~Л.~Притыкин\footnote{ }, %
%А.~Л.~Семёнов\footnote{ }
\end{center}

\bigskip

\centerline\today

\medskip

\begin{abstract}
В работе даётся обзор понятий и результатов, связанных со ставшими
классическими и новыми обобщениями понятия периодической
последовательности. Обсуждаются вопросы, относящиеся к почти
периодичности в таких областях, как комбинаторика слов,
символическая динамика, выразимость в логических теориях,
алгоритмическая вычислимость, колмогоровская сложность, теория
чисел.
\end{abstract}

\tableofcontents

%%%%%%%%%%%%%%%%%%%%%%%%%%%%%%%%%%%%%%%%%%%%%%%%%%%%%%%%%%%%%%%%%%
%%%%%%%%%%%%%%%%%%%%%%%%%%%%%%%%%%%%%%%%%%%%%%%%%%%%%%%%%%%%%%%%%%
%%%%%%%%%%%%%%%%%%%%%%%%%%%%%%%%%%%%%%%%%%%%%%%%%%%%%%%%%%%%%%%%%%

\section{Введение}
\label{Introduction}

Символическая динамика мотивируется очень широким классом
ситуаций~--- любое наблюдение за физическим процессом и измерение
каких-то его параметров в процессе наблюдения можно трактовать как
дискретизацию непрерывной динамической системы.

Идея символической динамики восходит к 19 веку, к работам Пуанкаре,
Адамара. Многие авторы упоминают в этой связи именно Адамара и его
работу 1898 года~\cite{Hadam1898}, например, Биркгофф говорит о
``символах, по существу введённых Адамаром'' (\cite{Birk35}, взято
из~\cite{HasselKatok02}). Продолжением работы Адамара является
работа Морса 1921 года~\cite{Morse21}. Однако как самостоятельный
предмет символическая динамика впервые изучалась в серии из двух
работ Морса и Хедлунда 1938~--- 1940 годов
\cite{MorseHedl38,MorseHedl40}.

Идея заключается в сопоставлении траектории в непрерывной
динамической системе последовательности букв конечного алфавита.
Понимая траекторию как движение точки в некотором пространстве, мы,
во-первых, делим пространство на конечное количество областей и,
во-вторых, выбираем дискретное множество моментов времени (например,
через каждую фиксированную единицу времени), после чего смотрим, в
каких областях оказывается точка в эти моменты. Траектории точки
сопоставляется бесконечная последовательность букв конечного
алфавита, где буквы соответствуют областям пространства. Естественно
возникает противоположная задача~--- насколько хорошо такая
символическая последовательность описывает исходную непрерывную
ситуацию.

Скажем теперь то же самое более формально. (Дающиеся в настоящем
введении определения, нужные в дальнейшем, будут систематически
повторены в разделах~\ref{Preliminaries} и~\ref{Classes}.)
\emph{Топологическая динамическая
система}\label{def-topological-dynamical-system}~--- это
топологическое пространство $V$ с заданным на нём непрерывным
отображением $f\colon V \to V$. Пусть $V, f$~--- топологическая
динамическая система, $A_1$, \dots, $A_k$~--- попарно
непересекающиеся открытые подмножества $V$, и $x_0$~--- точка,
орбита которой $\{f^n(x_0) : n \in \N\}$ лежит в $\bigcup_{i = 1}^k
A_k$. Определим последовательность $x\colon \N \to \{1,\dots,k\}$
условием $f^n(x_0) \in A_{x(n)}$. Таким образом, (некоторым) точкам
пространства $V$ мы сопоставили последовательности букв алфавита $A
= \{1,\dots,k\}$. При этом применению отображения $f$ соответствует
операция \emph{левого сдвига}\label{def-left-shift} $L$:
$L(a_0a_1a_2a_3\dots) = a_1a_2a_3\dots$

Ясно, что периодические последовательности описывают только самые
простые ситуации. Однако оказывается, что в довольно широком классе
ситуаций возникающая символическая последовательность будет обладать
свойствами, близкими к свойствам периодической последовательности.
Это наблюдение можно формализовывать разными способами, примеры
точных результатов такого типа~---
теоремы~\ref{dynamics-almost-periodicity},
\ref{dynamics-metrics-almost-periodicity},
\ref{dynamics-torus-effectively-almost-periodicity} ниже.

Так мы приходим к желанию определить класс последовательностей, в
каком-нибудь смысле близких к периодическим. Пусть $d(f,g)$~---
некоторая мера близости функций $f$ и $g$, измеряемая
неотрицательным действительным числом. Понятие почти периодичности
для функций в общем виде можно сформулировать так: функция $f\colon
\R \to \R$ \emph{почти периодическая относительно
$d$}\label{def-almost-periodic-function}, если для любого $\epsilon
> 0$ множество сдвигов $T$, для которых $d(f(x), f(x + T)) <
\epsilon$, \emph{относительно плотно}, то есть имеет ограниченные
пропуски (или, другими словами, элементы этого множества есть на
любом отрезке длины $C$ для достаточно большой константы $C$). В
качестве меры близости функций $d(g, h)$ можно брать норму разности
$\|f - g\|$ для различных норм $\|\cdot\|$: равномерной нормы
$\|f\|_\infty = \sup_x |f(x)|$ (получаем \emph{почти периодические
по Бору
функции}\label{def-almost-periodic-function-Bohr}~\cite{Bohr25}),
нормы Безиковича $\|f\|_{B,p} = \limsup_{x \to \infty} \left(
\frac1{2x} \int_{-x}^x |f(x)|^p\,dx \right)^{1/p}$ (получаем
\emph{почти периодические по Безиковичу
функции}\label{def-almost-periodic-function-Besicovitch}~\cite{Besic32}),
многих других.

По аналогии, для последовательностей в достаточно большой общности
можно дать такое определение. Пусть $d(x, y)$~--- какая-то мера
близости последовательностей, измеряемая неотрицательным
действительным числом, причём $d(x,x) = 0$ для любого~$x$. Будем
говорить, что множество $M \subseteq \N$ натуральных чисел
относительно плотно, если расстояние между соседними элементами
этого множества ограничено. Другими словами, $M$ \emph{относительно
плотно}\label{def-relatively-dense-set}, если для некоторой
константы $l$ и для всех $i$ имеем $M \cap \{i, i + 1, i + 2, \dots,
i + l - 1\} \ne \varnothing$. Последовательность $x$ \emph{близка к
периодической относительно $d$}\label{def-close-to-periodic}, если
для любого $\epsilon > 0$ существует такое относительно плотное
множество $M$, что для любого $m \in M$ имеем $d(L^m(x),x) <
\epsilon$. Тогда ясно, что если $x$ периодична с периодом $p$, то
$d(L^m(x),x) = 0$ для всех $m \in M = \{pn : n \in \N\}$, то есть
$x$ является близкой к периодической по нашему определению.

Одной из естественных мер близости на множестве последовательностей
является \emph{канторова метрика}\label{def-Cantor-distance}
$d_C(x,y) = 2^{-n}$, где $n = \min\{k : x(k) \ne y(k)\}$. Используя
такую метрику в определении выше, мы получаем определение почти
периодической последовательности. По-другому это определение можно
сформулировать так: последовательность $x$ почти периодическая, если
для каждого её подслова $u$ найдётся такое натуральное $l$, что на
каждом отрезке длины~$l$ последовательности $x$ найдётся вхождение
слова~$u$. (Далее мы будем рассматривать функцию, сопоставляющую
длине слова $u$ максимальное подходящее для всех слов этой длины
число $l$~--- она называется регулятор почти периодичности.) Для
широкого класса динамических систем символическая
последовательность, определённая выше, будет почти периодической.
Вот примеры результатов такого типа.

\begin{theorem}[\cite{MuchSemUsh03}]\label{dynamics-almost-periodicity}
Если пространство $V$ компактно и орбита каждой точки плотна в $V$,
то последовательность $x$ почти периодическая.
\end{theorem}

Более того, несложно показать, что каждая почти периодическая
последовательность может быть получена таким образом, как описано в
теореме~\ref{dynamics-almost-periodicity}. Для этого надо взять
пространство всех последовательностей с топологией, порождённой
метрикой $d_C$, для каждой буквы $a$ в качестве области $A_a$ взять
множество последовательностей, начинающихся с $a$, и в качестве
отображения $f$ взять отображение $L$ левого сдвига, а в качестве
$V$ взять замыкание $x$ относительно действия~$f$, после чего
замкнуть получившееся множество относительно метрики~$d_C$.
Подробнее см., например,~\cite{Lothaire02}.

Следующий факт о траекториях в метрических пространствах~--- частный
случай теоремы~\ref{dynamics-almost-periodicity}.

\begin{theorem}[\cite{MuchSemUsh03}]
\label{dynamics-metrics-almost-periodicity} Если $V$~--- компактное
метрическое пространство, и $f$~--- изометрия, то последовательность
$x$ почти периодическая.
\end{theorem}

Теоремы~\ref{dynamics-almost-periodicity} и
\ref{dynamics-metrics-almost-periodicity} в том или ином виде уже
можно считать фольклором (доказательства можно найти, например,
в~\cite{MuchSemUsh03}). Видимо, это уже не совсем так для результата
теоремы~\ref{dynamics-torus-effectively-almost-periodicity}.

На протяжении всего настоящего обзора нас будут интересовать
эффективные в алгоритмическом смысле результаты и понятия.
Эффективно почти периодической последовательностью мы называем почти
периодическую последовательность, которая вычислима, и для которой
эффективно по $n$ можно находить такое $l$, что все подслова длины
$n$ последовательности входят в каждое подслово длины~$l$.

Пусть $T^s$~--- $s$-мерный тор $[0, 1)^s$ с естественной метрикой,
индуцированной метрикой в $\R^s$. Множество в $\R^s$ называется
\emph{алгебраическим}\label{def-algebraic-set}, если является
решением системы (строгих или нестрогих) полиномиальных неравенств с
целыми коэффициентами. Множество в $\R^s$ называется
\emph{полуалгебраическим}\label{def-semialgebraic-set}, если
является объединением алгебраических множеств. Множество в $T^s$
полуалгебраическое, если является пересечением полуалгебраического
множества в $\R^s$ с тором $T^s$.

\begin{theorem}[\cite{MuchSemUsh03}]
\label{dynamics-torus-effectively-almost-periodicity} Пусть $V$~---
$s$-мерный тор, $x_0$~--- алгебраическая точка в $V$, $f$~--- сдвиг
на вектор с алгебраическими координатами, все $A_i$~---
полуалгебраические. Тогда последовательность $x$ эффективно почти
периодическая.
\end{theorem}

Вернёмся к определению близости к периодичности относительно меры
близости. Из других естественных функций $d$, которые можно взять в
этом определении, рассмотрим дискретный аналог \emph{расстояния
Безиковича}\label{def-Besicovitch-distance}: $d_B(x, y) = \liminf
\frac1n \#\{i: 0 \leqslant i \leqslant n - 1, x(i) \ne y(i)\}$.
Такая функция уже не будет метрикой, потому что возможно $d_B(x, y)
= 0$ при $x \ne y$. Последовательности, близкие к периодическим
относительно $d_B$, будем называть \emph{почти периодическими по
Безиковичу}\label{def-almost-periodic-Besicovich}. Это определение,
введённое в~\cite{MorseHedl38} (под другим названием~--- там такие
последовательности назывались почти периодическими, а те, что мы в
этой статье называем почти периодическими, назывались
рекуррентными), является дискретным вариантом определения почти
периодических по Безиковичу функций (см.~выше). Мы обращаемся к
этому определению в несколько другом контексте, когда говорим о мере
апериодичности последовательности (см.
раздел~\ref{Aperiodicity-measure}).

Топологическая динамическая система, удовлетворяющая свойству из
условия теоремы~\ref{dynamics-almost-periodicity}, называется
\emph{минимальной}\label{def-minimal-dynamical-system}. Один из
синонимов почти периодической последовательности~--- минимальная
последовательность. Почти периодические последовательности
действительно обладают следующим свойством минимальности. Обозначим
через $\Fac(x)$ множество всех конечных подслов
последовательности~$x$. Почти периодические последовательности
обладают минимальным множеством подслов среди всех
последовательностей, в следующем смысле.

\begin{proposition}\label{minimality-property}
\begin{enumerate}
\item Для любых бесконечных последовательностей $x$ и $y$ если $x$
почти периодическая и $\Fac(y) \subseteq \Fac(x)$, то $\Fac(y) =
\Fac(x)$.
\item Для любой последовательности $y$ существует почти
периодическая последовательность $x$, такая что $\Fac(x) \subseteq
\Fac(y)$.
\end{enumerate}
\end{proposition}

Доказательство предложения~\ref{minimality-property} можно найти,
например, в~\cite{Lothaire97}. О динамических системах можно
прочитать, например, в~\cite{KatokHass99-rus}.

Предложение~\ref{minimality-property}~--- пример комбинаторного
свойства почти периодических последовательностей. Комбинаторные
результаты появляются уже в~\cite{MorseHedl38,MorseHedl40}. Кроме
того, конечные и бесконечные слова и до этого изучались с
комбинаторной точки зрения. Зарождением соответствующей области~---
комбинаторики слов~--- принято считать работы Туэ
\cite{Thue1906,Thue12}, в которых он, в частности, изучает свойства
последовательности, которую мы обсуждаем в разделе~\ref{Thue}, и
доказывает её бескубность (подробнее о работах Туэ
см.~в~\cite{Berst92}). Важно отметить, что Туэ не имел в виду
каких-то конкретных применений своих результатов и считал
рассматриваемые им вопросы представляющими самостоятельный интерес
(см.~\cite{AlShall99}). Сейчас комбинаторика слов~--- активная
область, и в настоящем обзоре много примеров соответствующих
результатов.

Ещё один класс результатов, обсуждаемых в настоящем обзоре~---
логические. В~\cite{Buchi62} была доказана разрешимость монадической
теории следования на натуральных числах, при помощи сопоставления
формул теории конечным автоматам на бесконечных последовательностях.
После этого возникает естественный вопрос о разрешимости этой
теории, расширенной какой-то функцией. Конечно, ответ на этот вопрос
положительный, если функция выразима в исходной теории, но запас
таких функций невелик~--- это периодические функции, возможно, с
предпериодом (т.~е. опять периодичность является простейшим
случаем). Оказывается, что для функций с конечным количеством
значений, то есть последовательностей в конечном алфавите, можно
получить критерий разрешимости, если эта последовательность обладает
свойствами, близкими к свойствам периодических последовательностей,
и для этого полезно изучить, как ведут себя конечные автоматы,
запущенные на таких последовательностях. Так
в~\cite{Sem79-rus,Sem84-diss} появляется определение обобщённо почти
периодических последовательностей. При этом, как показывает
теорема~\ref{GAP-first-order-quantum-elimination}, обобщённо почти
периодические последовательности возникают не просто как класс
последовательностей, для которых оказалось удобно доказать какие-то
результаты, а как ответ на естественный вопрос. Мы обсуждаем
соответствующие результаты в разделах~\ref{Automata_Mappings},
\ref{Logic}, кроме того, эти вопросы являются одной из основных тем
во всём обзоре в целом. О другого рода связях символической динамики
и теории автоматов, которые мы в настоящем обзоре не затрагиваем,
можно прочитать в~\cite{BealPerr97}.

Помимо всего перечисленного, в этом обзоре затрагиваются и другие
вопросы, связанные с теорией чисел, алгеброй, теорией алгоритмов.

\subsection{Определения и обозначения}
\label{Preliminaries}

Введём некоторые основные понятия и обозначения.

\emph{Алфавитом}\label{def-alphabet} мы называем произвольное
конечное множество. Обычно алфавит мы обозначаем буквами $A$, $B$ и
т.~п. Его элементы мы называем \emph{буквами}\label{def-letter} или
\emph{символами}\label{def-symbol}. Стандартный бинарный алфавит из
двух символов $\{0, 1\}$ обозначим $\B$, множество натуральных чисел
$\{0, 1, 2,\dots\}$ через $\N$. По умолчанию мы рассматриваем только
конечные алфавиты, и при рассмотрении бесконечных алфавитов
оговариваем это особо. Для конечного множества $X$ через $\#X$ или
$|X|$ обозначим количество элементов в нём.

Будем рассматривать \emph{последовательности}\label{def-sequence}
над $A$~--- отображения $x\colon\N\to A$. Множество всех таких
последовательностей, наделённое метрикой $d_C$, образует стандартное
канторовское компактное метрическое пространство, которое мы
обозначаем~$A^\N$. Таким образом, $\lim_{n\to\infty}u_n = x$, если
$\forall i\,\exists n\,\forall m > n\ u_m(i) = x(i)$ (это
определение годится и для конечных слов~$u_n$). Иногда
последовательности называют также \emph{бесконечными
словами}\label{def-infinite-word}.

Конечная последовательность букв в алфавите~$A$ называется
\emph{словом}\label{def-word} над или в $A$. Через $A^*$ обозначим
множество всех конечных слов над $A$, включая пустое
слово~$\Lambda$. Через $|u|$ будем обозначать длину слова $u$, через
$|u|_a$~--- количество букв $a$ в слове $u$. Если $i\leqslant j$
натуральные, через $[i,j]$ обозначим отрезок натурального ряда с
концами в $i$ и $j$, то есть множество $\{i, i+1, i+2,\dots,j\}$.
Отрезок $[0,n]$ будем обозначать просто $[n]$. Через $x[i,j]$
обозначим отрезок последовательности $x$~--- слово $x(i)x(i+1)\dots
x(j)$. Говорят, что $[i,j]$~--- \emph{вхождение в последовательность
$x$ слова}\label{def-occurrence} $u\in A^*$, если $x[i,j] = u$.
Слово $u \ne \Lambda$ называется \emph{фактором}\label{def-factor}
или \emph{подсловом}\label{def-subword} последовательности $x$, если
$u$ входит в $x$. Множество всевозможных факторов последовательности
$x$ будем обозначать $\Fac(x)$, множество всевозможных факторов
длины $m$~--- $\Fac_m(x)$. Факторы вида $x[0,i]$ называются
\emph{префиксами}\label{def-prefix} $x$, последовательности вида
$x(i)x(i+1)x(i+2)\dots$~--- \emph{суффиксами}\label{def-suffix} $x$
и обозначаются $x[i,\infty)$. Мы представляем себе
последовательность расположенной горизонтально и идущей слева
направо до бесконечности, поэтому будем использовать выражения
``левее'' и ``правее'' для меньших и б\'ольших индексов
соответственно.

\emph{Подсловной сложностью}\label{def-subword-complexity}
последовательности $x$ называется такая функция $\p_x\colon \N \to
\N$, что $\p_x(n)$ равно количеству слов длины $n$, входящих в
последовательность~$x$.

Пусть $A$, $B$~--- конечные алфавиты. Отображение $\phi\colon A^*\to
B^*$ называется \emph{морфизмом}\label{def-morphism}, если для любых
$u,v\in A^*$ выполнено $\phi(uv) = \phi(u)\phi(v)$. Ясно, что
морфизм полностью определяется своими значениями на однобуквенных
словах. Морфизм \emph{нестирающий}\label{def-non-erasing-morphism},
если $|\phi(a)| \geqslant 1$ для всех $a \in A$. Морфизм называется
\emph{$k$-равномерным}\label{def-uniform-morphism}, если $|\phi(a)|
= k$ для всех $a \in A$. 1-равномерный морфизм называется
\emph{кодированием}\label{def-coding}. Если $x$~---
последовательность букв алфавита $A$, по определению положим
  $$
 \phi(x) = \phi(x(0))\phi(x(1))\phi(x(2))\dots
  $$
Подробнее о морфизмах, а также о строящихся с помощью них
последовательностях~--- автоматных и морфических~---
см.~разделы~\ref{Automatic} и~\ref{Morphic}.

\subsection{Основные классы последовательностей. Регуляторы}
\label{Classes}

Последовательность $x$ \emph{периодическая}\label{def-periodic},
если для некоторого $T \in \N$, называемого
\emph{периодом}\label{def-period}, имеем $x(i) = x(i + T)$ для
любого $i \in \N$. В соответствии с общепринятым соглашением,
периодом мы называем также и слово $x[0, T - 1]$. Будем называть
последовательность \emph{заключительно
периодической}\label{def-eventually-periodic}, если то же выполнено
для всех $i$, начиная с некоторого $K$. Тогда $x[0, K - 1]$
называется \emph{предпериодом}\label{def-preperiod}. Предпериодом
будем называть также и число $K$. Последовательность, не являющаяся
заключительно периодической, называется
\emph{апериодической}\label{def-aperiodic}. Множество всех
периодических последовательностей обозначим $\Per$, множество
заключительно периодических обозначим $\EP$. Рассмотрим некоторые
расширения этих классов.

Последовательность $x$ называется \emph{почти
периодической}\label{def-almost-periodic}, если для каждого её
фактора $u$ найдётся такое натуральное $l$, что на каждом отрезке
длины~$l$ последовательности $x$ найдётся вхождение слова~$u$. Тем
самым, любое слово, входящее в почти периодическую
последовательность, входит в неё бесконечное количество раз. Через
$\AP$ (almost periodic) будем обозначать класс всех таких
последовательностей. Ясно, что для проверки почти периодичности
достаточно только убедиться в повторяемости с ограниченными
интервалами всех префиксов, а не всех факторов. Почти периодические
последовательности называют также \emph{равномерно
рекуррентными}\label{def-uniformly-recurrent-sequence} и
\emph{минимальными}\label{def-minimal-sequence}.

Будем называть последовательность $x$ \emph{заключительно почти
периодической}\label{def-eventually-almost-periodic}, если некоторый
её суффикс почти периодичен. Класс всех таких последовательностей
обозначим~$\EAP$ (eventually almost periodic).

Последовательность $x$ называется \emph{обобщённо почти
периодической}\label{def-generalized-almost-periodic}, если для
каждого её фактора $u$, входящего в неё бесконечное число раз,
найдётся такое натуральное $l$, что на каждом отрезке длины~$l$
последовательности $x$ найдётся вхождение слова~$u$. Класс всех
таких последовательностей обозначим через~$\GAP$ (generalized almost
periodic).

Если $x\in\EAP$, то минимальное такое $n$, что $x[n,\infty)\in\AP$,
будем называть \emph{минимальным
префиксом}\label{def-minimal-prefix} и обозначать~$\pr(x)$. Заметим,
что для любого $m\geqslant\pr(x)$ имеем $x[m,\infty)\in\AP$.

И наконец, назовём последовательность $x$
\emph{рекуррентной}\label{def-recurrent}, если каждое слово, которое
в неё входит, обязательно входит бесконечное количество раз. Ясно,
что если последовательность рекуррентная и обобщённо почти
периодическая, то она почти периодическая. Класс рекуррентных
последовательностей будем обозначать $\Rec$. Последовательность
\emph{заключительно рекуррентная}\label{def-eventually-recurrent},
если некоторый её суффикс рекуррентен. Класс таких
последовательностей обозначим $\ER$.

\emph{Регулятором почти
периодичности}\label{def-almost-periodicity-regulator}
последовательности $x \in \GAP$ назовём функцию $\r_x\colon \N \to
\N$, которая на числе $n$ равна минимальному такому $l$, что каждое
слово длины $n$, которое входит в $x$ бесконечное количество раз,
встретится на любом отрезке длины $l$ последовательности $x$, а
также любое слово длины $n$, которое входит в $x$ конечное
количество раз, не входит в $x[l,\infty)$ (второе важно только для
обобщённо почти периодических последовательностей, не являющихся
почти периодическими). Часто вместо регулятора нам будет достаточно
рассматривать только какую-то верхнюю оценку на него, то есть
функцию $f$, такую что $f(n) \geqslant \r_x(n)$ для всех~$n$~--- в
этом случае мы пишем $f \geqslant \r_x$. Если говорить более точно,
регулятор почти периодичности объединяет в себе две функции: одна
следит за расстояниями между вхождениями слов, входящих бесконечное
количество раз, а другая следит за тем префиксом, которым
ограничивается вхождение слов, входящих конечное количество раз.
Иногда стоит эти функции разделять явно, хотя на протяжении
настоящей статьи мы этого делать не будем.

Будем говорить, что последовательность $x$ \emph{эффективно
обобщённо почти
периодическая}\label{def-effective-generalized-almost-periodic},
если $x$ вычислима, обобщённо почти периодична, и некоторая оценка
сверху на регулятор почти периодичности последовательности $x$ также
вычислима. Мы получим эквивалентное определение, если заменим
требование вычислимости некоторой оценки сверху на регулятор
требованием вычислимости регулятора в точности.

\begin{proposition}\label{regulator-bound-vs-precise}
Пусть последовательность $x \in \GAP$ вычислима, и вычислима
некоторая функция $f \geqslant \r_x$. Тогда функция $\r_x$
вычислима.
\end{proposition}
\begin{proof}
Слова длины $n$, входящие в $x$ бесконечно много раз~--- это слова
длины $n$, входящие в $x[f(n),2f(n)]$. Все остальные слова
длины~$n$, которые входят в $x[0,f(n) + n]$, входят в $x$ конечное
количество раз.

Пусть $l_1$ равно минимальному такому $l$, что все слова длины $n$,
входящие в $x$ конечное количество раз, не входят в $x[l,\infty]$.
Поскольку $l_1 \leqslant f(n)$, такое $l_1$ можно найти перебором.

Аналогично предыдущему, можно найти множество $K$ слов длины $f(n)$,
хотя бы один раз входящих в~$x$. Теперь достаточно найти минимальное
такое $l_2$, что каждое подслово длины $n$, встречающееся в $x$
бесконечно много раз, встречается в каждом подслове длины $l_2$
каждого из слов множества $K$. Тогда выполнено, что любое подслово
$x$ длины $n$, встречающееся в $x$ бесконечно много раз, входит в
любое подслово $x$ длины~$l_2$.

Таким образом, $\r_x(n) = \max(l_1,l_2)$.
\end{proof}

Назовём последовательность \emph{эффективно почти
периодической}\label{def-effective-almost-periodic}, если она почти
периодична и эффективно обобщённо почти периодична. Определение
эффективной почти периодичности можно упростить.

\begin{proposition}\label{effective-AP-equivalent-definition}
Почти периодическая последовательность $x$ эффективно почти
периодична тогда и только тогда, когда $x$ вычислима и множество
подслов $\Fac(x)$ разрешимо.
\end{proposition}
\begin{proof}
$\Rightarrow$. Пусть $x \in \AP$ вычислима, и мы умеем вычислять
некоторую оценку $f \geqslant \r_x$. Тогда чтобы найти все слова
длины $n$ в $x$, достаточно взять произвольное слово длины $f(n)$ в
$x$: множество его подслов длины $n$ в точности является множеством
подслов длины $n$ всей последовательности $x$, в соответствии с
определением почти периодичности.

$\Leftarrow$. Пусть $x \in \AP$ вычислима и множество $\Fac(x)$
разрешимо. Чтобы найти значение $\r_x(n)$, будем перебирать все
натуральные числа по очереди, начиная, скажем, с $n$. Проверяя число
$m$, мы смотрим, верно ли, что каждое из слов множества $\Fac_m(x)$
(подслова $x$ длины $m$) содержит все слова из $\Fac_n(x)$ (подслова
$x$ длины $n$). Если это так, то $\r_x(n) \leqslant m$, а иначе
$\r_x(n) > m$. Поскольку $x \in \AP$, до бесконечности мы проверять
не можем и когда-нибудь найдём подходящее~$m$.
\end{proof}

Несложно видеть, что $\Per\subset\AP\subset\EAP\subset\GAP$.
Оказывается, все эти включения строгие. Например, известная
последовательность Туэ~--- Морса $\thue = 0110100110010110\dots$
(см.~раздел~\ref{Thue})~--- пример последовательности из $\AP$, но
не из $\Per$ (более того, класс $\AP$ континуален, тогда как $\Per$
счётный, доказательство см.~в~\cite{Jac70} или~\cite{MuchSemUsh03}).
Неравенство $\AP \subsetneq \EAP$ очевидно (можно взять $10000\ldots
\in \EAP \setminus \AP$). Неравенство $\EAP \subsetneq \GAP$ было
доказано в~\cite{Prit06b-rus} (см.~также в~\cite{Prit06c-rus}).
Кроме того, можно заметить следующие включения
$\Per\subset\EP\subset\EAP$, все из которых тоже, очевидно, строгие.

Введённые классы изображены на рис.~\ref{classes-of-sequences}
\footnote{По техническим причинам калиграфические буквы на этом
рисунке заменены на обычные.}.

\begin{figure}[h]
\centering
\includegraphics{survey_almper_pictures.4}
\caption{Классы последовательностей.} \label{classes-of-sequences}
\end{figure}

Назовём последовательность \emph{точно почти
периодической}\label{def-precisely-almost-periodic}, если каждое
слово, которое в неё входит, входит в некоторой арифметической
прогрессии. Более формально, $x$~--- точно почти периодическая, если
для любого входящего в неё слова $u$ найдутся такие $a, d \in \N$,
что $x[a + id, a + id + |u| - 1] = u$ для всех $i \geqslant 0$. При
этом $u$ может входить и где угодно ещё в последовательности~--- мы
не требуем, чтобы все вхождения слова $u$ образовывали
арифметическую прогрессию. Ясно, что любая точно почти периодическая
последовательность почти периодична. Класс точно почти периодических
последовательностей обозначим $\PAP$. На
с.~\pageref{Rumyantsev-method} описывается метод построения
последовательностей (использованный в доказательстве
теоремы~\ref{Rumyantsev-Ushakov-complex_almost_periodic}), с помощью
которого можно строить точно почти периодические последовательности.
Но этим методом получаются не все точно почти периодические
последовательности.

Пусть $\phi\colon A^*\to B^*$~--- морфизм, $x$ обобщённо почти
периодическая последовательность над алфавитом $A$.
В~\cite{MuchSemUsh03} было показано, что если последовательность
$\phi(x)$ бесконечна, то она обобщённо почти периодична. Ясно тогда,
что для $x$~--- почти периодической $\phi(x)$ будет почти
периодической (если бесконечна). Действительно, достаточно показать,
что для любого слова $u = x[i,j]$ слово $\phi(u) = \phi(x(i))\dots
\phi(x(j))$ встречается в $\phi(x)$ бесконечно много раз. Но это
следует из определения $\phi(x)$ и из того, что $x$ почти
периодична, и значит, слово $u$ встречается в ней бесконечно много
раз. Очевидно, для $x$~--- заключительно почти периодической
$\phi(x)$ будет снова заключительно почти периодической (если она
бесконечна). Несложно видеть, что классы эффективно обобщённо почти
периодических и эффективно почти периодических последовательностей
также замкнуты относительно действий морфизмов.

\subsection{Первый пример: последовательность Туэ~--- Морса}
\label{Thue}

\emph{Последовательность Туэ~--- Морса}\label{def-Thue-Morse}~---
одна из самых популярных последовательностей в комбинаторике слов,
встречается в самых разных сюжетах и служит классическим примером
или контрпримером ко многим определениям и утверждениям. Хорошим
обзором по свойствам этой последовательности является
статья~\cite{AlShall99} (которую мы во многом используем в изложении
этого раздела).

У этой последовательности много различных эквивалентных между собой
определений. Дадим некоторые из них.

Проще всего эту последовательность получить так. Запишем символ: 0.
Далее инвертируем его (то есть заменим на 1) и припишем к
имеющемуся: 01. Теперь имеющийся блок опять инвертируем (то есть
заменим в нём 0 на 1 и 1 на 0) и припишем к имеющемуся блоку: 0110.
Будем продолжать так дальше до бесконечности. Получим
последовательность конечных слов:
$$
\begin{tabular}{l}
$u_0 = 0$\\
$u_1 = 01$\\
$u_2 = 0110$\\
$u_3 = 01101001$\\
$u_4 = 0110100110010110$\\
$u_5 = 01101001100101101001011001101001$\\
\dots
\end{tabular}
$$

По построению каждое следующее конечное слово начинается с
предыдущего. Поэтому все эти слова являются началом некоторого
одного (предельного) бесконечного слова, которое и является
последовательностью Туэ~--- Морса:
  $$
\thue = 0 1 1 0 1 0 0 1 1 0 0 1 0 1 1 0 1 0 0 1 0 1 1 0 0 1 1 0 1 0
0 1\dots
  $$

Можно дать и такое определение. Назовём последовательность $\thue =
t_0t_1t_2t_3t_4\dots$ последовательностью Туэ~--- Морса, если
выполнено $t_0 = 0$ и $t_{2n} = t_n$, $t_{2n + 1} = \bar t_n$ для
всех~$n$, где \ $\bar{}$\ \ означает инвертирование~--- замену всех
0 на 1 и всех 1 на 0.

Дадим и ещё одно определение. Пусть $s_2(n)$ обозначает сумму цифр
натурального числа $n$, записанного в двоичной системе счисления.
Определим двоичную последовательность $\thue =
t_0t_1t_2t_3t_4\dots$, так что $t_n = s_2(n) \pmod 2$. Обобщением
такого определения является понятие автоматной последовательности
(см.~раздел~\ref{Automatic}).

Четвёртое определение, которое мы дадим, такое. Пусть $x$~---
переменная. Тогда бесконечное произведение $\prod_{i \geqslant 0} (1
- x^{2^i}) = (1 - x)(1 - x^2)(1 - x^4)\dots = \sum_{i \geqslant 0}
f_i x$ определяет последовательность $f_i$. Определим двоичную
последовательность $\thue = t_0t_1t_2t_3t_4\dots$, так что
$(-1)^{t_i} = f_i$.

Можно легко доказать, что четыре вышеперечисленных определения
эквивалентны и задают одну и ту же последовательность. Есть и другие
способы определения. Например, последовательность Туэ~--- Морса~---
простейший пример морфической последовательности~--- неподвижной
точки морфизма $0 \to 01$, $1 \to 10$ (см.~разделы~\ref{Automatic},
\ref{Morphic}).

Одним из самых ранних случаев появления этой последовательности
можно считать работы Туэ: в \cite{Thue1906} и \cite{Thue12} он
строит эту последовательность как пример бескубной бесконечной
последовательности. Здесь \emph{кубами}\label{def-cube} мы называем
слова вида $uuu$, где $u$~--- некоторое непустое слово, а
\emph{бескубной}\label{def-cube-free}~--- последовательность, в
которую не входят кубы. Более того, можно доказать, что в $\thue$ не
входят слова вида $auaua$, где $a$~--- некоторый символ, а $u$~---
некоторое слово. Как следствие, можно построить
\emph{бесквадратную}\label{def-square-free} (то есть не содержащую
\emph{квадратов}\label{def-square}~--- слов вида $uu$ для
непустого~$u$) последовательность в алфавите из трёх символов.

Впоследствии последовательность Туэ~--- Морса много раз
переоткрывали, отсюда и много других имён, которые она носит.
В~\cite{Morse21} Морс доказывает про эту последовательность, что она
почти периодическая, но не периодическая. В~\cite{Euwe29}
голландский шахматист Макс Эйве переоткрывает последовательность
Туэ~--- Морса и использует её для доказательства существования
бесконечной шахматной партии. (Впоследствии для решения этой же
задачи её опять используют Морс и Хедлунд
\cite{Morse38,MorseHedl44}, не зная о работе Эйве.)

Следующую задачу, в которой также неявно появляется
последовательность Туэ~--- Морса, можно отнести к теории чисел.
Впервые её сформулировал Пруэ \cite{Prouhet1851} и позднее, более
чем через 50 лет, Тарри и Эскотт. Теперь эта задача обычно
называется ``задача Пруэ~--- Тарри~--- Эскотта'' или ``задача о
мультистепенях''.

Задача заключается в следующем. Можно ли разделить множество чисел
$\{0, 1, 2, \dots, 2^{N - 1}\}$ на два множества $I$ и $J$, так
чтобы было выполнено равенство
  $$
\sum_{i \in I} i^k = \sum_{j \in J} j^k
  $$
для $k = 0, 1, 2, \dots, t$?

Пруэ доказал, что это возможно при $N = t + 1$. Более точно, по
существу Пруэ удалось доказать следующее свойство последовательности
$\thue$. Определим
  $$
I = \{i \in \{0, 1, 2, \dots, 2^{N - 1}\} : t_i = 0\},
  $$
  $$
J = \{j \in \{0, 1, 2, \dots, 2^{N - 1}\} : t_j = 1\}.
  $$
Тогда для всех $0 \leqslant k \leqslant N - 1$ выполнено
  $$
\sum_{i \in I} i^k = \sum_{j \in J} j^k.
  $$

Видимо, статья Пруэ \cite{Prouhet1851}~--- это самое раннее (но
неявное) использование последовательности Туэ~--- Морса.

Следующий теоретико-числовой вопрос о последовательности Туэ~---
Морса был задан Гельфондом в 1968 году. Четвёртое определение из
данных нами выше задаёт последовательность $f_i = (-1)^{t_i}$.
Несложно показать, что $|\sum_{i = 0}^N f_i| \leqslant 1$ для любого
$N$. Имея в виду третье определение на основе суммы цифр в двоичном
представлении, это значит, что натуральные числа практически в
равной степени разделены на те, у которых сумма цифр чётна, и те, у
которых сумма цифр нечётна. Гельфонд спросил, верно ли, что сумма
цифр у простых чисел в двоичной системе счисления может быть
асимптотически в равной степени чётной и нечётной~\cite{Gelf68}.
Недавно был получен положительный ответ на этот вопрос:
в~\cite{MaudRiv08} показано, что $|\sum_{p \leqslant x} f_p| =
o(\pi(x))$ при $x \to \infty$, где сумма берётся по простым числам
$p$, а $\pi(x)$ равно количеству простых чисел, не
превосходящих~$x$. См.~также~\cite{Tao08} о~трудностях проверки
простоты числа по его представлению в какой-то системе счисления.

Последовательность Туэ~--- Морса встречается также в решении
знаменитой проблемы Бернсайда: верно ли, что каждая
конечно-порождённая группа, удовлетворяющая соотношению $x^n = 1$,
конечна? Ответ положительный (и хорошо известный) для $n = 2$.
Однако для больших $n$ ответ отрицательный: как показали Новиков и
Адян \cite{AdNov68I,AdNov68II,AdNov68III}, бесконечная группа
$\Gamma(m,n)$ с $m$ порождающими и с соотношением $x^n = 1$
существует для всех $m > 1$ и для всех нечётных $n \geqslant 4381$.
Изучению свойств бесконечных групп $\Gamma(m,n)$ на основе
усовершенствованных методов из \cite{AdNov68I,AdNov68II,AdNov68III}
посвящена монография Адяна \cite{Adyan75}, где, в частности, оценка
4381 улучшена до 655.

Один из самых первых шагов в решении проблемы Бернсайда состоит в
построении бескубной последовательности в алфавите из двух символов,
в качестве которой Адян и Новиков используют последовательность
Туэ~--- Морса. Говоря более точно, авторы ссылаются на работу
Аршона, в которой построена эта последовательность.

Советский математик Соломон Ефимович Аршон родился в 1892 году и был
репрессирован в конце 1930-х. Предположительное время ареста конец
1938~--- начало 1939 года. До этого Аршон был главным редактором
издательств ``Редакция технико-теоретической литературы при Академии
Наук СССР'' и/или издательства ОНТИ. Он был снят с работы в июне
1938 года (см.~\cite{Arshon-arrest,Vilenkin91}).

В работе~\cite{Arshon37b} (которая фактически копирует более раннюю
работу \cite{Arshon37a}), Аршон строит серию последовательностей, по
одной для каждого натурального~$n$, удовлетворяющую специальному
свойству ассиметричности. Для $n = 2$ последовательность Аршона
совпадает с последовательностью Туэ~--- Морса. Таким образом, этот
результат~--- обобщение результатов Туэ о последовательностях,
избегающих слов вида $auaua$, где $a$~--- буква, $u$~--- слово.

Докажем в заключение, что последовательность Туэ~--- Морса почти
периодична. Это следует из
теоремы~\ref{block-product-almost-periodic}, но мы приведём здесь
рассуждение явно. Заметим, что достаточно доказать, что для каждого
$n$ найдётся такое $l_n$, что на каждом отрезке длины $l_n$ в
$\thue$ найдётся вхождение слова $u_n$ (в обозначениях первого
определения из начала раздела). Действительно, все остальные
подслова содержатся в одном из $u_n$, поскольку $\thue = \lim_{n \to
\infty} u_n$. Для $n = 0$ возьмём $l_0 = 3$. Действительно, $u_0 =
0$, а в последовательности $\thue$ нигде нет трёх 1 подряд (и вообще
нет никаких кубов). Ясно теперь, что для произвольного $n$ подойдёт
$4|u_n| = 2^{n + 2}$, потому что для каждого $n$ последовательность
$\thue$ можно записать как $\thue = u_n\bar u_n\bar u_n u_n \bar u_n
u_n u_n \bar u_n \dots$, причём если заменить всюду $u_n$ на 0 и
$\bar u_n$ на 1, то получится в точности $\thue$.

\subsection{Второй пример: последовательности Штурма}
\label{Sturm}

Другая очень популярная последовательность~---
\emph{последовательность Фибоначчи}\label{def-Fibonacci}~---
определяется как бесконечная вправо неподвижная точка отображения на
бесконечных последовательностях, при котором одновременно заменяется
$0 \to 01$ и $1 \to 0$ (то есть чисто морфическая
последовательность, начинающаяся с 0 и полученная из морфизма $0 \to
01$, $1 \to 0$; см.~разделы~\ref{Automatic}, \ref{Morphic}).
Получается последовательность
  $$
\fib = 010010100100101001010\dots
  $$

Последовательность $\fib$ также служит примером ко многим
утверждениям и обладает самыми разными экстремальными свойствами.
Некоторые из них мы встретим далее в этой статье.

Последовательность Фибоначчи~--- пример последовательности Штурма.
Последовательности Штурма\label{def-Sturm} можно определить
несколькими эквивалентными способами. Мы опишем некоторые из них,
вместе с сопутствующими свойствами. Подробнее о последовательностях
Штурма можно прочитать, например, в~\cite{AlShall03,Lothaire02}.

В этом разделе речь идёт о последовательностях в алфавите $\{0,1\}$.
Напомним, что подсловной сложностью последовательности $x$
называется такая функция $\p_x\colon \N \to \N$, что $\p_x(n)$ равно
количеству слов длины $n$, входящих в последовательность~$x$.
Следующий известный результат в терминах подсловной сложности
характеризует периодические последовательности.

\begin{proposition}[\cite{MorseHedl38}]
\label{periodicity-subword-complexity}
\begin{enumerate}
\item Если последовательность $x$ заключительно периодическая, её
подсловная сложность ограничена: существует такая константа $C$, что
$\p_x(n) \leqslant C$ для всех~$n$
\item Если для некоторого $n$ верно $\p_x(n) \leqslant n$, то
последовательность $x$ заключительно периодическая.
\end{enumerate}
\end{proposition}

Последовательность $x$ называется \emph{последовательностью Штурма},
если $\p_x(n) = n + 1$. Таким образом, последовательности Штурма~---
это последовательности с минимальной возможной для апериодических
последовательностей подсловной сложностью.

Множество конечных слов $X$ называется
\emph{сбалансированным}\label{def-balanced-set}, если для
любых слов $u,v \in X$ одинаковой длины имеем %$\delta(u,v) \leqslant 1$
$||u|_1 - |v|_1| \leqslant 1$. Конечное слово или бесконечная
последовательность называется
\emph{сбалансированным/ой}\label{def-balanced-word}, если множество
его/её подслов сбалансированно.

Для действительных чисел $\alpha$ и $\rho$, таких что $0 \leqslant
\alpha \leqslant 1$ и $0 \leqslant \rho < 1$, определим
последовательности $s_{\alpha,\rho}$ и $s'_{\alpha,\rho}$ как
$s_{\alpha,\rho}(n) = \lfloor \alpha(n + 1) + \rho\rfloor -
\lfloor\alpha n + \rho\rfloor$, $s'_{\alpha,\rho}(n) = \lceil
\alpha(n + 1) + \rho\rceil - \lceil\alpha n + \rho\rceil$. Они
называются, соответственно, \emph{нижняя} и \emph{верхняя
механические}\label{def-mechanical-sequence} с наклоном $\alpha$ и
сдвигом $\rho$. Эти термины легко объяснимы, если посмотреть на
графическую интерпретацию механических слов
(рис.~\ref{mechanical_sequences}). Она получается следующим образом.
Чтобы получить механические последовательности $s_{\alpha,\rho}$ и
$s'_{\alpha,\rho}$, нужно провести прямую $y = \alpha x + \rho$,
после чего на каждой вертикальной прямой $x = n$ отметить самые
близкие к прямой с обеих сторон~--- сверху и снизу~--- целые точки.
Далее точки над прямой нужно соединить ломаной, она и определяет
верхнюю механическую последовательность: горизонтальное звено
ломаной соответствует 0, а наклонённое под углом $45^\circ$
соответствует 1. Нижняя механическая последовательность получается
аналогично по ломаной, соединяющей точки снизу от прямой.

Можно показать, что $\fib = s_{1/\phi^2,1/\phi^2}$, где $\phi =
\frac{1 + \sqrt 5}2$~--- золотое сечение, то есть последовательность
Фибоначчи является последовательностью Штурма. Она изображена на
рис.~\ref{mechanical_sequences}.

\begin{figure}[h]
\centering
\includegraphics{survey_almper_pictures.5}
\caption{Верхняя механическая последовательность
$s'_{1/\phi^2,1/\phi^2} = \fib$.} \label{mechanical_sequences}
\end{figure}

Важно отметить, что верхняя и нижняя механические последовательности
почти совпадают, кроме, возможно, двух символов; это несовпадение
происходит в случае, если найдётся такое $n$, что $\alpha n + \rho
\in \Z$. Механическое слово называется рациональным или
иррациональным в зависимости от рациональности наклона $\alpha$.

Можно рассмотреть другие варианты получения символической
последовательности из прямой, нарисованной на клетчатой бумаге.
Например, можно отметить все пересечения прямой с линиями сетки, и
при пересечении горизонтальной линии писать в последовательность 0,
а при пересечении вертикальной писать 1. Другой вариант: можно
приближать прямую сверху ломаной, составленной не из горизонтальных
и наклонных отрезков, а из горизонтальных и вертикальных: при этом
горизонтальный отрезок будет соответствовать 0, а вертикальный 1.
Оба только что названных варианта (как и некоторые другие не
названные) определяют по-прежнему класс механических
последовательностей, причём от параметров этих вариантов к
параметрам исходного определения можно переходить вычислимо, и
наоборот.

Доказательство следующей теоремы, видимо, уже относящейся к
фольклору, можно найти, например, в~\cite{Lothaire02}.

\begin{theorem}%[\cite{Lothaire02}]
\label{Sturmian-three-definitions}
Для последовательности $x$ следующие условия эквивалентны:\\
(i) $x$ является последовательностью Штурма;\\
(ii) $x$ сбалансированная и апериодическая;\\
(iii) $x$ механическая иррациональная.
\end{theorem}

Таким образом, благодаря теореме~\ref{Sturmian-three-definitions} мы
имеем три эквивалентных определения последовательностей Штурма.

Алгоритмические вопросы в связи с последовательностями Штурма
рассмотрены в разделе~\ref{Decidability-of-theories}.

Заметим, что любая последовательность Штурма является почти
периодической. Действительно, это следует из п.~$(iii)$
теоремы~\ref{Sturmian-three-definitions} и из
теоремы~\ref{dynamics-almost-periodicity}.

Последовательности со сложностью $n + c$, где $c$~--- положительная
константа, описаны в работе~\cite{Cass98}.

%\begin{problem}
%Что можно сказать про последовательности с подсловной сложностью $n
%+ 2$? Верно ли, что их континуум? Могут ли такие последовательности
%не быть почти периодическими?
%\end{problem}

%%%%%%%%%%%%%%%%%%%%%%%%%%%%%%%%%%%%%%%%%%%%%%%%%%%%%%%%%%%%%%%%%%
%%%%%%%%%%%%%%%%%%%%%%%%%%%%%%%%%%%%%%%%%%%%%%%%%%%%%%%%%%%%%%%%%%
%%%%%%%%%%%%%%%%%%%%%%%%%%%%%%%%%%%%%%%%%%%%%%%%%%%%%%%%%%%%%%%%%%

\section{Алгебра почти периодических последовательностей}
\label{Algebra}

%%%%%%%%%%%%%%%%%%%%%%%%%%%%%%%%%%%%%%%%%%%%%%%%%%%%%%%%%%%%%%%%%%

\subsection{Блочное произведение}
\label{BlockProduct}

Последовательность Туэ~--- Морса (см.~раздел~\ref{Thue})~--- это
частный случай так называемого блочного произведения, рассмотренного
в статье~\cite{Keane68}. Пусть $u,v \in \B^*$. Определим
\emph{блочное произведение}\label{def-block-product} $u\otimes v$
индукцией по длине~$v$:
\begin{gather*}
u\otimes\Lambda = \Lambda,\\
u\otimes v0 = (u\otimes v)u,\\
u\otimes v1 = (u\otimes v)\bar u.
\end{gather*}
Легко проверить, что блочное произведение дистрибутивно справа (но
не слева!) и ассоциативно, то есть $u \otimes (vw) = (u \otimes
v)(u\otimes w)$ и $u \otimes (v \otimes w) = (u \otimes v)\otimes w$
для любых слов $u$, $v$, $w$. Пусть теперь $u_k$, $k =
0,1,2,\dots$~--- последовательность непустых слов из $\B^*$, таких
что при $k\geqslant1$ слово $u_k$ начинается с 0. Тогда в
последовательности $u_0, u_0\otimes u_1, u_0\otimes u_1\otimes
u_2,\dots$ каждое из слов является префиксом любого из следующих, и
значит, существует
  $$
\lim_{n\to\infty} \bigotimes_{k=0}^n u_k = \bigotimes_{k=0}^\infty
u_k
  $$
--- бесконечная последовательность в $\B^\N$. Например,
последовательность Туэ~--- Морса~--- это блочное произведение слов
$u_k = 01$.

Блочное произведение подробно изучалось в~\cite{Jac70}. В частности,
там доказано, что $\bigotimes_{k=0}^\infty u_k$ всегда почти
периодична (см.~теорему~\ref{block-product-almost-periodic}).

Кроме последовательности Туэ~--- Морса в качестве примера
последовательности, получающейся при помощи блочного произведения,
в~\cite{Jac70} приводится \emph{тернарная последовательность
Кини}\label{def-Keane-sequence} (``вальс бесконечного порядка'')
$x_K = 0010011100010011101101100\ldots = 001 \otimes 001 \otimes 001
\otimes \dots$ Также вводится частный случай блочного
произведения~--- последовательности вида $0p_1 \otimes 0p_2 \otimes
0p_3 \otimes \dots$, где все $p_i$ равны 0 или 1, причём среди них
бесконечное количество 1; они называются \emph{последовательностями
Какутани}\label{def-Kakutani-sequence}, который их впервые изучал
в~\cite{Kakut67}. Видно, кстати, что множество последовательностей
Какутани~--- это континуальное семейство почти периодических
последовательностей; однако не всякая почти периодическая
последовательность является последовательностью Какутани.

Блочное произведение~--- эффективный способ построения различных
последовательностей, удовлетворяющих нужным свойствам. Примеры таких
конструкций содержатся в~\cite{Jac70}, приведём некоторые результаты
оттуда.

\emph{Относительной частотой
вхождения}\label{def-occurrence-frequence} $T_u(x, i, j)$ блока $u$
в последовательность $x$ на отрезке $[i,j]$ называется количество
его вхождений на этом отрезке (то есть так, чтобы левый конец лежал
на отрезке), делённое на $j - i + 1$. Блок $u$ называется
\emph{чезаровским}\label{def-Cesaro} для $x$, если существует
средняя частота его вхождения $I_u(x) = \lim_{n\to\infty} T_u(x, 0,
n)$; \emph{равномерно чезаровским}, если $I_u(x) =
\lim_{n\to\infty}T_u(x, s, s + n)$ равномерно по $s$.
Последовательность \emph{(равномерно) чезаровская}, если все её
факторы (равномерно) чезаровские.

Если $x$ почти периодична, то для любого её фактора $u$ имеем
строгое неравенство $\liminf_{n\to\infty} \left(\inf_{s\geqslant0}
T_u(x, s, s + n)\right) > 0$. В частности, если $u$ чезаровский, то
$I_u(x) > 0$.

В~\cite{Jac70} для бинарных последовательностей, получаемых с
помощью блочного произведения, строится пример, в котором 0 и 1 не
являются чезаровскими. Приводится критерий того, когда для таких
последовательностей 0 является чезаровским со средней частотой
$\frac12$.

Кроме того, что особенно интересно для наших рассмотрений,
в~\cite{Jac70} обсуждается, насколько последовательности, получаемые
при помощи блочного произведения, можно считать периодическими:
``Хотя почти все наши последовательности не периодичны в строгом
смысле, многие из них обладают свойствами периодичности несколько
иного характера''. Пример: для последовательности Туэ~--- Морса
$\thue$ и множества $F = \{0110, 1001, 0101, 1010\}$ выполнено
$\thue[i,j] \in F \Leftrightarrow 2|i$. Аналогично, для
последовательности Кини и $F' = \{001, 110\}$ выполнено $x_K[i,j]
\in F' \Leftrightarrow 3|i$. Доказана следующая общая теорема.

\begin{theorem}
\begin{enumerate}
\item Пусть $x = 0P_1 \otimes 0P_2 \otimes \dots$ и $C_n = 0P_1
\otimes 0P_2 \otimes \dots \otimes 0P_n$ для некоторого $n$. Если
существует $j$, такое что в $D = 0P_{n + 1} \otimes 0P_{n + 2}
\otimes \dots \otimes 0P_{n + j}$ входит одно из слов 001, 110, 011,
100, то для множества $F = \bigcup_{|u| = 2|D|} C_n \otimes u$
выполняется $x[i,j] \in F \Leftrightarrow |C_n||i$.
\item Если последовательность $x = 0P_1 \otimes 0P_2 \otimes \dots$
не периодична, то для каждого $n$ найдётся множество $F_n$, для
которого $x[i,j] \in F_n \Leftrightarrow r_n|i$, где $C_n = 0P_1
\otimes 0P_2 \otimes \dots \otimes 0P_n$, $r_n = |C_n|$.
\end{enumerate}
\end{theorem}

В конце работы~\cite{Jac70} сформулированы 8 задач, есть как простые
упражнения, так и исследовательские проблемы, предложения обобщить
полученные в статье результаты. Одна из задач заключается в
предложении обобщить блочное произведение на случай произвольного
конечного алфавита. Это сделано явно в~\cite{Hoit04}, однако по
существу результаты в этом направлении можно найти и
в~\cite{Sem83-rus} (см.~также раздел~\ref{Universal}).

К этим открытым вопросам мы можем добавить следующий.

\begin{problem}\label{block-product-investigation-problem}
Изучить более подробно ``свойства периодичности иного характера'', о
которых говорится выше. В частности, возможно, получить критерий
разрешимости монадических теорий таких последовательностей (см.
раздел~\ref{Logic}).
\end{problem}

В заключение приведём следующий любопытный пример. Он нам
понадобится в доказательстве
предложения~\ref{GAP-non-preservation-pushdown}.

\begin{proposition}\label{ap-alternating-0-1-prefixes}
Существует почти периодическая последовательность над~$\B$, у
которой для любого $n$ найдётся префикс, в котором нулей на $n$
больше, чем единиц, и для любого $n$ найдётся префикс, в котором
единиц на $n$ больше, чем нулей.
\end{proposition}
\begin{proof}
Положим $x = 001 \otimes \bigotimes_{i = 1}^\infty 0111$. Пусть $u_m
= 001 \otimes \bigotimes_{i = 1}^m 0111$. Докажем индукцией по $m$,
что $|u_m|_0 - |u_m|_1 = (-1)^m2^m$. Действительно, при $m = 0$
имеем $u_0 = 001$, $|u_0|_0 - |u_1|_1 = 2 - 1 = 1$.

Пусть теперь $|u_m|_0 - |u_m|_1 = (-1)^m2^m$. По определению $u_{m +
1} = u_m \bar u_m \bar u_m \bar u_m$, значит, $|u_{m + 1}|_0 - |u_{m
+ 1}|_1 = |u_m|_0 + 3|u_m|_1 - (|u_m|_1 + 3|u_m|_0) = 2(|u_m|_1 -
|u_m|_0) = (-1)^{m + 1}2^{m + 1}$.
\end{proof}

%%%%%%%%%%%%%%%%%%%%%%%%%%%%%%%%%%%%%%%%%%%%%%%%%%%%%%%%%%%%%%%%%%

\subsection{Универсальный метод построения}
\label{Universal}

Следующий метод построения последовательностей, по существу
предложенный в~\cite{Sem83-rus} и обсуждавшийся также
в~\cite{MuchSemUsh03}, в некотором смысле обобщает блочное
произведение. Он является универсальным, в том смысле, что с его
помощью можно получить любые обобщённо почти периодические
последовательности. Мы изложим немного модифицированную версию этого
метода. Этот метод нам неоднократно понадобится далее в статье.

\label{def-universal}Последовательность $\langle B_n, C_n,
l_n\rangle$, где $B_n$ и $C_n$~--- непустые множества непустых слов
в фиксированном конечном алфавите $A$, $l_n$~--- натуральные числа,
называется \emph{$A$-$\GAP$-схемой}\label{def-GAP-scheme}, если для
неё выполняются следующие
четыре условия для любого $n \in \N$:\\[1mm]
(1) все слова из $B_n$ имеют длину $l_n$;\\
(2) все слова из $C_n$ представимы в виде $v_1v_2$, где $v_1,v_2\in
B_n$, и каждое слово из $B_n$ используется в качестве
$v_i$ в каком-то из слов множества $C_n$;\\
(3) каждое слово из $B_{n+1}$ имеет вид $v_1v_2\dots v_k$, где для
каждого $i < k$ имеем $v_iv_{i+1}\in C_n$ (в частности, все $v_i\in
B_n$), и для каждого $w \in С_n$ найдётся $i < k$, для которого $w =
v_iv_{i+1}$;\\
(4) для каждого слова $u = v_1v_2\dots v_kw_1w_2\dots w_k$ из
$С_{n+1}$ (здесь $v_i,w_i\in B_n$) имеем $v_kw_1\in C_n$.

\smallskip

Мы будем говорить, что последовательность $x \in A^\N$
\emph{$\GAP$-порождена
$A$-$\GAP$-схемой}\label{def-GAP-generated-GAP-scheme} $\langle B_n,
C_n, l_n\rangle$, если для каждого $n \in \N$ найдётся такое $k_n$,
что для всех $i \in \N$ выполнено
  $$
x[k_n + il_n, k_n + (i+2)l_n - 1]\in C_n.
  $$
Будем говорить, что последовательность \emph{$\GAP$-порождена
правильно}\label{def-GAP-generated-perfectly}, если $l_n|k_n$.

Несложно видеть (из компактности), что каждая $\GAP$-схема правильно
$\GAP$-по\-рож\-да\-ет хотя бы одну последовательность. Как доказано
в~\cite{MuchSemUsh03}, каждая последовательность, $\GAP$-порождённая
$\GAP$-схемой, является обобщённо почти периодической. Кроме того,
каждая обобщённо почти периодическая последовательность
$\GAP$-по\-рож\-да\-ет\-ся некоторой $\GAP$-схемой, причём эту
$\GAP$-схему можно выбрать так, чтобы последовательность порождалась
правильно. Отметим также важное свойство: если обобщённо почти
периодическая последовательность $\GAP$-порождена $\GAP$-схемой, то,
зная параметры этой схемы (достаточно последовательностей $k_n$ и
$l_n$), можно вычислять оценку сверху на регулятор почти
периодичности порождаемой последовательности.

Для получения почти периодических последовательностей можно также
пользоваться понятием $\GAP$-схемы. Мы будем говорить, что
последовательность $x \in A^\N$ \emph{$\AP$-порождена
$A$-$\GAP$-схемой}\label{def-AP-generated-GAP-scheme} $\langle B_n,
C_n, l_n\rangle$, если для всех $n\in\N$ и $i\in\N$ выполнено
  $$
x[il_n, (i+2)l_n - 1]\in C_n.
  $$
Другими словами, последовательность $\AP$-порождается, если она
$\GAP$-порождается со всеми $k_n = 0$.

Аналогично предыдущему, каждая $\GAP$-схема $\AP$-порождает хотя бы
одну последовательность. Каждая последовательность,
$\AP$-порождённая $\GAP$-схемой, является почти периодической.
Каждая почти периодическая последовательность может быть
$\AP$-порождена какой-нибудь $\GAP$-схемой. Важное свойство
сохраняется: по $\GAP$-схеме, порождающей последовательность, можно
получить оценку сверху на регулятор почти периодичности этой
последовательности.

Этот способ порождения почти периодических последовательностей можно
упростить, но пожертвовав, по всей видимости, последним свойством.

Последовательность $\langle B_n, l_n\rangle$, где $B_n$~--- непустое
множество непустых слов в фиксированном конечном алфавите $A$,
$l_n$~--- натуральные числа, называется
\emph{$A$-$\AP$-схемой}\label{def-AP-scheme}, если для неё для
любого $n \in \N$ выполнено условие (1), и для любого $n \in \N$
каждое $u \in B_{n+1}$ имеет вид $u = v_1v_2\dots v_k$, где $v_i\in
B_n$, причём для каждого $w\in B_n$ найдётся $i$, такое что~$v_i =
w$. Последовательность $x$ \emph{$\AP$-порождена
$A$-$\AP$-схемой}\label{def-AP-generated-AP-scheme}, если для любых
$i$ и $n$ имеем
  $$
x[il_n, (i+1)l_n - 1] \in B_n.
  $$

Как и раньше, каждая $\AP$-схема $\AP$-порождает хотя бы одну
последовательность. Каждая последовательность, $\AP$-порождённая
$\AP$-схемой, является почти периодической. Каждая почти
периодическая последовательность может быть $\AP$-порождена
какой-нибудь $\AP$-схемой. Однако теперь уже, вообще говоря, может
быть неверно, что по схеме можно оценивать сверху регулятор почти
периодичности порождаемой последовательности. По крайней мере, то же
рассуждение, которое можно было применить раньше, здесь уже не
работает. Например, неясно даже, как, зная схему, дать оценку на
расстояния между соседними вхождениями двухбуквенных слов: такое
слово может оказаться на стыке двух блоков~--- элементов схемы.

\begin{conjecture}\label{scheme-non-computable-regulator}
Существует вычислимая $\AP$-схема и $\AP$-порождённая ею
последовательность $x$, такая что регулятор $\r_x$ не вычислим.
\end{conjecture}

Когда ясно, о каком алфавите, о каком типе схемы или о каком типе
порождения последовательности идёт речь, мы будем опускать
соответствующие приставки.

Применение универсального метода построения проиллюстрируем на
примере следующего результата из \cite{Jac70}, который мы теперь
можем доказать совсем просто.

\begin{theorem}[\cite{Jac70}]\label{block-product-almost-periodic}
Пусть $x = \bigotimes_{k = 0}^\infty u_k$, где слова $u_k \in \B^*$
начинаются с 0 при $k \geqslant 1$. Тогда $x$ почти периодическая.
\end{theorem}
\begin{proof}
Если все $u_k$, начиная с некоторого, состоят из одних 0, то
последовательность $x$ периодическая, а значит, и почти
периодическая. Иначе достаточно рассмотреть случай, при котором
каждое $u_k$ содержит хотя бы один символ 1 (общий случай к нему
легко сводится). Но в этом случае $x$ $\AP$-порождается $\AP$-схемой
$B_n = \{a_n, \bar a_n\}$, где~$a_n = \bigotimes_{k=0}^n u_k$, и
значит, является почти периодической.
\end{proof}

%%%%%%%%%%%%%%%%%%%%%%%%%%%%%%%%%%%%%%%%%%%%%%%%%%%%%%%%%%%%%%%%%%

\subsection{Конечно-автоматные преобразования}
\label{Automata_Mappings}

Представляется интересным рассматривать преобразования
последовательностей и пытаться понять, сохраняют ли эти
преобразования свойства почти периодичности. Простейшими
алгоритмическими преобразованиями можно считать конечно-автоматные.
В то же время конечно-автоматные преобразования можно воспринимать
как наиболее широкий класс преобразований, сохраняющих определённую
алгебраическую структуру последовательностей.

Для различных классов последовательностей известны результаты о
замкнутости относительно конечно-автоматных преобразований~---
например, см.~теоремы~\ref{automatic-preservation}
и~\ref{morphic-preservation}. В настоящем разделе мы обсуждаем
замкнутость относительно таких преобразований последовательностей с
различными свойствами типа почти периодичности.

Другая мотивация заключается в связи конечно-автоматных
преобразований (а точнее, распознающих автоматов на
последовательностях) с монадическими теориями на натуральных числах
с отношением порядка, которым посвящён раздел~\ref{Monadic}
(см.~теорему~\ref{decidability-monadic-GAP}).

\label{def-finite-state-transducer}\emph{Конечно-автоматным
преобразователем} назовём совокупность $M = \langle A, B, Q, \tilde
q, \lambda, \mu \rangle$, где $A$ и $B$~--- конечные множества,
называемые соответственно входной и выходной алфавит, $Q$~---
конечное множество состояний, $\tilde q\in Q$~--- выделенное
состояние, называемое начальным, и
  $$
 \lambda\colon Q\times A\to B^*,\qquad \mu\colon Q\times A\to Q
  $$
--- функции переходов. Пусть $x \in A^\N$. Последовательность
$(p_n)_{n=0}^\infty$ элементов множества $Q$ назовём \emph{ходом
преобразователя $M$ на $x$}\label{def-transducer-run}, если $p_0 =
\tilde q$ и для каждого $n$ выполняется $p_{n + 1} = \mu(p_n,
x(n))$. Последовательность $M(x)$, определяемую как $M(x)(n) =
\lambda(p_n, x(n))$, где $(p_n)_{n=0}^\infty$~--- ход
преобразователя $M$ на $x$, назовём \emph{образом последовательности
$x$ под действием $M$}\label{def-transducer-image}.

Если для каждых $a \in A$, $q \in Q$ выполнено $|\lambda(q,a)| = 1$,
то преобразователь $M$ называется
\emph{равномерным}\label{def-uniform-finite-transducer}. Применение
произвольного конечно-автоматного преобразователя к
последовательности можно представить как последовательное применение
равномерного конечно-автоматного преобразователя и некоторого
морфизма.

\begin{proposition}\label{transducer-uniform-moprhism-decomposition}
Пусть $M = \langle A, B, Q, \tilde q, \lambda, \mu \rangle$~---
конечно-автоматный преобразователь, $x$~--- последовательность.
Тогда существует такой равномерный конечно-автоматный
преобразователь $M'$ и такой морфизм $\phi$, что $M(x) =
\phi(M'(x))$.
\end{proposition}
\begin{proof}
Действительно, положим $M' = \langle A, Q \times A, Q, \tilde q,
\lambda, \mu' \rangle$, так что $\mu'(q,a) = \langle q, a \rangle$
для любых $q \in Q$ и $a \in A$. Определим также морфизм $\phi\colon
Q \times A \to B$, так что $\phi(\langle q, a \rangle) = \mu(q,a)$
для любых $q \in Q$ и $a \in A$. Ясно тогда, что $M(x) =
\phi(M'(x))$.
\end{proof}

Поэтому часто для упрощения ситуации мы ограничиваемся рассмотрением
равномерных конечно-автоматных преобразователей. Если $[i,j]$~---
вхождение слова $u$ в последовательность $x$, причём $p_i = q$, где
$(p_n)_{n=0}^\infty$~--- ход преобразователя $M$ на $x$, то будем
говорить, что преобразователь $M$
\emph{подходит}\label{def-transducer-comes-in-state} к этому
вхождению слова $u$ \emph{в состоянии~$q$}.

По существу, следующая теорема доказана в \cite{Sem83-rus}
(см.~также~\cite{MuchSemUsh03}). Для полноты изложения мы приводим
её с полным доказательством (по~\cite{MuchSemUsh03}).

Для произвольной функции $g$ обозначим $\underbrace{g\circ g\circ
\dots\circ g}_m$ через $g^m$.

\begin{theorem}[\cite{Sem83-rus,MuchSemUsh03}]\label{GAP-preservation}
Пусть $M$~--- конечно-автоматный преобразователь с $m$ состояниями,
и $x \in \GAP$. Тогда верно следующее.
\begin{enumerate}
\item $M(x) \in \GAP$.
\item Пусть $M$~--- равномерный преобразователь. Тогда
$\r_{M(x)}(n) \leqslant h(h(n))$ для всех~$n$, где $h(n) = g^m(n) -
1$ и $g(n) = \r_x(n) + 1$.
\item Если $x$ эффективно обобщённо почти периодическая, то $M(x)$
также эффективно обобщённо почти периодическая.
\end{enumerate}
\end{theorem}
\begin{lemma}\label{GAP-preservation-main-lemma}
Пусть $M$~--- равномерный конечно-автоматный преобразователь с $m$
состояниями, и $x \in \GAP$. Пусть $v = M(x)[i,j]$~--- вхождение
слова длины~$n$ в~$M(x)$, такое что $i \geqslant h(n)$, где $h(t) =
g^m(t) - 1$, $g(t) = \r_x(t) + 1$. Тогда найдётся такое $r$, что $j
- h(n) \leqslant r \leqslant i - 1$ и $M(x)[r, r + n - 1] = v$.
\end{lemma}
\begin{proof}
Будем считать сначала, что $v = M(x)[i,j]$~--- просто достаточно
далёкое от начала вхождение слова $v$ в $M(x)$. Объясним, как найти
искомое $v = M(x)[r,s]$. При этом мы будем делать разные допущения,
которые подытожим и сформулируем точно позднее.

Итак, пусть $v = M(x)[i,j]$ и $u_1 = x[i,j]$~--- прообраз $v$ в $x$.
Пусть преобразователь $M$ подходит к позиции $i$, находясь в
состоянии $q_1$. Если $i$ достаточно велико, найдётся вхождение $u_1
= x[i_2,j_2]$ левее $[i,j]$, но достаточно близко. Если $M$ подходит
к $i_2$ в состоянии $q_1$, то $M(x)[i_2,j_2] = v$. Иначе $M$
подходит к $i_2$ в каком-то состоянии $q_2 \ne q_1$
(рис.~\ref{GAP-preservation-proof-illustration-1}).

\begin{figure}[h]
\centering
\includegraphics{survey_almper_pictures.6}
\caption{Иллюстрация к доказательству
теоремы~\ref{GAP-preservation}.}
\label{GAP-preservation-proof-illustration-1}
\end{figure}

Положим $u_2 = x[i_2,j]$. Если $i_2$ достаточно велико, то найдётся
вхождение $u_2 = x[i_3,j_3]$ левее $[i_2,j]$, но достаточно близко.
Если $M$ подходит к $i_3$ в состоянии $q_1$, то $M(x)[i_3,i_3 + n -
1] = v$, так как $u_2$ начинается с $u_1$. Если $M$ подходит к $i_3$
в состоянии $q_2$, то $M(x)[i_3,j_3] = M(x)[i_2,j]$, и тогда
$M(x)[j_3 - n + 1,j_3] = v$, так как $u_2$ заканчивается
словом~$u_1$. В худшем случае $M$ подходит к $i_3$ в состоянии
$q_3$, таком что $q_3 \ne q_2$ и $q_3 \ne q_1$. При этом положим
$u_3 = x[i_3,j]$ (рис.~\ref{GAP-preservation-proof-illustration-2}).

\begin{figure}[h]
\centering
\includegraphics{survey_almper_pictures.7}
\caption{Иллюстрация к доказательству
теоремы~\ref{GAP-preservation}.}
\label{GAP-preservation-proof-illustration-2}
\end{figure}

Рассуждая так дальше, мы, всё время выбирая худший случай, найдём
такие $i_2, \dots, i_{m + 1}$, что к каким-то двум из $i_1 = i, i_2,
\dots, i_{m + 1}$ преобразователь $M$ подходит в одинаковых
состояниях. Таким образом, слово $v$ обязательно встретится в
$M(x)[i_{m + 1},j - 1]$.

Проанализируем теперь вышеприведённое рассуждение. Вначале мы ищем
вхождение $u_1 = x[i_2,j_2]$. Это можно сделать, если $i \geqslant
\r_x(n)$~--- исходя из определения обобщённой почти периодичности,
это означает, что $u_1$ входит в $x$ бесконечно много раз. При этом
получится $(j - 1) - i_2 + 1 \leqslant \r_x(n)$~--- этого
достаточно, чтобы на отрезке $x[i_2,j - 1]$ нашлось вхождение~$u_1$.
Отсюда $i_2 \geqslant j - \r_x(n)$ и $|u_2| = j - i_2 + 1 \leqslant
\r_x(n) + 1$.

Далее аналогично получаем, что при $i_2 \geqslant \r_x(\r_x(n) + 1)
\geqslant \r_x(|u_2|)$ действительно можно найти вхождение $u_2 =
x[i_3,j_3]$, причём $i_3 \geqslant j - \r_x(\r_x(n) + 1)$ и $|u_3|
\leqslant \r_x(\r_x(n) + 1) + 1$.

Положим $g = \r_x + 1$. Аналогично рассуждая, получим, что при $i_m
\geqslant g^m(n) - 1$ можно выбрать $i_{m + 1} \geqslant j - g^m(n)
+ 1$.
\end{proof}

\begin{proof}[Доказательство теоремы~\ref{GAP-preservation}]
Пусть $x \in \GAP$, и на $x$ действует конечный преобразователь $M$
с $m$ состояниями.

1) Из предложения~\ref{transducer-uniform-moprhism-decomposition}
следует, что достаточно доказать утверждение только для равномерного
$M$, так как морфизмы сохраняют обобщённую почти периодичность. Для
равномерного $M$ из леммы~\ref{GAP-preservation-main-lemma} следует,
что если слово $v$ длины $n$ входит в $M(x)$ бесконечно много раз,
то оно входит на каждом отрезке длины $g^m(n) - 1$ в $M(x)$. Поэтому
$M(x)$ обобщённо почти периодическая.

2) Покажем, как найти оценку сверху на $\r_{M(x)}$. Обозначим $h(n)
= g^m(n) - 1$, где $g(n) = \r_x(n) + 1$.

Пусть слово $v$ длины $n$ входит в $M(x)$ бесконечно много раз.
Тогда по 1) оно встретится на любом отрезке длины $h(n)$.

Пусть теперь слово $v$ длины $n$ входит в $M(x)$ конечное количество
раз. Докажем, что тогда $v$ не входит в $M(x)$ правее
позиции~$h(h(n))$. Действительно, предположим, что это не так. Тогда
на каждом отрезке длины $h(n)$ слова $M(x)[0,h(h(n)) - 1]$ найдётся
вхождение $v$, так как по 1) от каждого такого вхождения можно найти
ещё одно слева не дальше, чем на расстоянии $h(n)$. Но $v$ входит в
$M(x)$ лишь конечное количество раз. Поэтому в $M(x)$ найдётся слово
$w$ длины $h(n)$ (где-то сильно справа), которое не содержит~$v$.
Тогда всюду слева от него на любом отрезке длины $h(h(n))$ найдётся
вхождение $w$, значит, и в $M(x)[0, h(h(n)) - 1]$ тоже.
Противоречие, так как на любом отрезке длины $h(n)$ в $M(x)[0,
h(h(n)) - 1]$ есть вхождение $v$, но $w$ не содержит $v$ и входит в
$M(x)[0, h(h(n)) - 1]$.

Таким образом, объединяя последние два абзаца, получаем оценку
$\r_{M(x)}(n) \leqslant h(h(n))$.

3) Пусть теперь последовательность $x$ эффективно обобщённо почти
периодическая, т.~е. $x$ вычислима и регулятор $\r_x$ вычислим.
Аналогично п.~1), достаточно рассматривать только случай
равномерного $M$, так как морфизмы сохраняют эффективную обобщённую
почти периодичность. Ясно, что зная $M$ и умея вычислять $x$, мы
можем вычислять $M(x)$. П.~2) позволяет также вычислять оценку
сверху на $\r_{M(x)}$. Значит, $M(x)$ эффективно обобщённо почти
периодическая.
\end{proof}

Из теоремы~\ref{GAP-preservation} сразу следует, что образ при
конечно-автоматном преобразовании заключительно почти периодической
последовательности обобщённо почти периодичен. Но оказывается, можно
доказать более сильное утверждение.

\begin{theorem}[\cite{Prit06b-rus,Prit06c-rus}]
\label{EAP-preservation} Пусть $M$~--- конечно-автоматный
преобразователь. Тогда если $x \in \EAP$, то $M(x) \in \EAP$.
\end{theorem}

Доказательство теоремы~\ref{EAP-preservation}, приведённое
в~\cite{Prit06b-rus}, не эффективно в следующем смысле. Допустим, мы
знаем $x\in\AP$ и регулятор $\r_x$. Тогда по
теореме~\ref{EAP-preservation} существует оценка сверху на
$\pr(M(x))$, так как $M(x) \in \EAP$, но доказательство
из~\cite{Prit06b-rus} не позволяет по имеющимся данным найти никакую
такую оценку эффективно (и даже до некоторого момента не позволяло
надеяться, что такой эффективный способ вообще существует).

Следующий эффективный вариант теоремы~\ref{EAP-preservation} был
объявлен в~\cite{Prit06a-rus} и доказан в~\cite{Prit06c-rus}. Для
полноты изложения мы приводим план доказательства.

\begin{theorem}[\cite{Prit06c-rus}]
\label{EAP-preservation-prefix-upper-bound} Пусть $M$~---
равномерный конечно-автоматный преобразователь с $m$ состояниями, и
$x \in \AP$. Тогда $M(x) \in \EAP$ и
  $$
 \pr(M(x)) \leqslant \r\nolimits_x^m(1) +
 \r\nolimits_x^{m - 1}(1) + \dots + \r\nolimits_x(1).
  $$
\end{theorem}

Пусть $M = \langle A, B, Q, \tilde q, \lambda, \mu\rangle$~---
равномерный преобразователь. Доказательство проходит индукцией по
количеству состояний преобразователя. Будем считать без ограничения
общности, что $B = Q \times A$ и для всех $q \in Q$, $a \in A$
выполнено $\lambda(q, a) = \langle q, a \rangle$. Действительно,
общий случай произвольного $B$ и произвольного $\lambda$ получается
из рассматриваемого отождествлением каких-то пар $\langle q, a
\rangle$ между собой.

База индукции~--- обратимый конечно-автоматный преобразователь.

Назовём конечно-автоматный преобразователь $M = \langle A, B, Q,
\tilde q, \lambda, \mu\rangle$
\emph{обратимым}\label{def-automaton-reversible}, если для каждых $q
\in Q$ и $a \in A$ существует ровно одно состояние $q' \in Q$, такое
что $\mu(q', a) = q$. Другими словами, в таком преобразователе
каждая буква входного алфавита осуществляет взаимно однозначное
отображение множества состояний в себя. Находясь в некотором
состоянии и зная последовательность предыдущих входных символов,
можно восстановить и последовательность пройденных состояний (в этом
и заключается свойство обратимости).

\begin{proposition}[\cite{Prit06c-rus}]
\label{AP-preservation-reversible-transducer} Пусть $M$~---
обратимый равномерный конечно-ав\-то\-мат\-ный преобразователь с $m$
состояниями. Тогда если $x \in \AP$, то $M(x) \in \AP$.
\end{proposition}
\begin{proof}
Пусть $v = M(x)[i, j]$~--- вхождение слова $v$ длины $n$ в $M(x)$, и
$u_1 = x[i, j]$~--- прообраз $v$ в $x$, к которому $M$ подходит в
состоянии~$q_1$. Тогда найдётся вхождение $u_1 = x[i_2, j_2]$, такое
что $i_2 > i$, но $j_2 \leqslant i + \r_x(n)$. Если $M$ подходит к
$i_2$ в состоянии $q_1$, то $M(x)[i_2, j_2] = v$. Иначе $M$ подходит
к $i_2$ в каком-то состоянии $q_2 \ne q_1$.

Положим $u_2 = x[i, j_2]$. Имеем $|u_2| \leqslant \r_x(n) + 1$.
Тогда найдётся вхождение $u_2 = x[i_3, j_3]$, такое что $i_3 > i$,
но $j_3 \leqslant i + \r_x(\r_x(n) + 1)$. Если $M$ подходит к $j_3 -
n +1$ в состоянии $q_1$, то $M(x)[j_3 - i + 1, j_3] = v$, так как
$u_2$ заканчивается словом $u_1$. Если $M$ подходит к $j_3 - n + 1$
в состоянии $q_2$, то в силу обратимости $M$ подходит к $i_2$ в
состоянии $q_1$, и тогда $M(x)[i_2, i_2 + n - 1] = v$, так как $u_2$
начинается с $u_1$. В худшем случае $M$ подходит к $j_3 - n + 1$ в
состоянии $q_3$, таком что $q_3 \ne q_1$, $q_3 \ne q_2$
(рис.~\ref{AP-preservation-reversible-transducer-proof-illustration}).

\begin{figure}[h]
\centering
\includegraphics{survey_almper_pictures.8}
\caption{Иллюстрация к доказательству
предложения~\ref{AP-preservation-reversible-transducer}.}
\label{AP-preservation-reversible-transducer-proof-illustration}
\end{figure}

Рассуждая так дальше, построим такие $j_2, \dots, j_{m + 1}$, что
хотя бы к каким-то двум из $i = j - n + 1 = j_1 - n + 1, j_2 - n +
1, \dots, j_{m + 1} - n + 1$ преобразователь подходит в одинаковых
состояниях. При этом $j_{m + 1} \leqslant i + g^m(n) - 1$, где $g =
\r_x + 1$.

Таким образом, каждое слово длины $n$, входящее в $M(x)$ бесконечно
много раз, входит на любом отрезке длины $g^m(n) - 1$.
Следовательно, $M(x) \in \AP$.
\end{proof}

Видно, что доказательство
предложения~\ref{AP-preservation-reversible-transducer} очень похоже
на доказательство теоремы~\ref{GAP-preservation}, приведённое выше,
только рассуждение проходит не ``справа налево'', а ``слева
направо''. Это возможно за счёт обратимости преобразователя.

Для индукционного перехода рассмотрим следующую конструкцию.
\label{def-split}Пусть $x \in A^\N$, и символ $a \in A$ входит в $x$
бесконечное количество раз. Проведём разрез в $x$ после каждого
вхождения символа $a$. Тогда $x$ разрежется на блоки вида $ua$, где
$u\in(A\setminus\{a\})^*$, то есть на слова, содержащие ровно один
символ $a$ на последнем месте. Если символ $a$ встречается в $x$ с
ограниченными интервалами, то количество всевозможных таких блоков
конечно (например, если $x \in \GAP$, то их длины не больше
$\r_x(1)$). Закодируем однозначно эти блоки буквами некоторого
конечного алфавита, обозначим этот алфавит $\b_{a,x}(A)$. Таким
образом, мы из $x$ получили новую последовательность в этом
алфавите. Последовательность, полученную из неё отбрасыванием
первого символа, назовём \emph{$a$-разбиением} последовательности
$x$ и обозначим $\s_a(x)$. Например, 0-разбиение последовательности
3200122403100110\dots~--- это (0)(12240)(310)(0)(110)\dots

Несложно доказать, что если $x \in \AP$ и символ $a$ входит в $x$,
то $\s_a(x) \in \AP$ (см.~\cite{Prit06c-rus}).
%
%\begin{lemma}[\cite{Prit06c-rus}]\label{split}
%Пусть $x\in\AP$, и символ $a$ входит в $x$. Тогда $\s_a(x)\in\AP$.
%\end{lemma}

Индукционный переход в доказательстве
теоремы~\ref{EAP-preservation-prefix-upper-bound} заключается в
следующем. Допустим, исходный преобразователь не обратим. Тогда
образ множества состояний~$Q$ преобразователя под действием какой-то
буквы $a$ входного алфавита строго меньше~$Q$. Рассмотрев теперь
вместо последовательности её $a$-разбиение и, считая преобразователь
определённым на алфавите $\b_{a,x}(A)$, сведём задачу к самой же
себе, но с меньшим количеством состояний в преобразователе. При
этом, переходя от последовательности $x$ к $a$-разбиению $\s_a(x)$,
мы отбрасываем первый символ последовательности блоков, что
соответствует отбрасыванию начального блока длины не более $\r_x(1)$
от последовательности $x$. Сделав так некоторое количество раз, мы
приходим к ситуации, в которой к последовательности применяется
обратимый конечно-автоматный преобразователь (преобразователь с
одним состоянием всегда обратим). Подробнее доказательство
см.~в~\cite{Prit06c-rus}.
%См.~также ниже
%теорему~\ref{GAP-preservation-improved-regulator-upper-bound},
%доказательство которой, следующее аналогичной схеме рассуждения,
%изложено полностью.

Примечательно, что, таким образом, мы имеем два различных
доказательства теоремы~\ref{EAP-preservation}.
%Поскольку приведённые
%рассуждения можно приспособить и для доказательства
%теоремы~\ref{GAP-preservation}
%(см.~лемму~\ref{GAP-preservation-almost-reversible-transducer} и
%замечания после неё%
%%теорему~\ref{GAP-preservation-improved-regulator-upper-bound}
%%ниже
%), то с ней ситуация аналогична.

%%%% recurrent sequences %%%%

Заметим, что, формально говоря, введённые нами классы
последовательностей~--- это классы последовательностей над конечным
алфавитом. Однако все те же самые определения (почти периодической
последовательности, обобщённо почти периодической последовательности
и т.~д.) имеет смысл рассматривать и для бесконечного алфавита.
Например, континуум различных почти периодических
последовательностей над бесконечным алфавитом можно построить с
помощью конструкции на с.~\pageref{Rumyantsev-method} (они будут
точно почти периодическими). Несложно построить и другие
нетривиальные примеры.

На бесконечный алфавит можно обобщить и понятие конечно-автоматного
преобразователя. При этом не требуется никакой эффективности, в
частности, функции переходов могут быть совершенно произвольными.
Все результаты настоящего раздела верны и для конечно-автоматных
преобразователей на последовательностях над бесконечным алфавитом
(но это, вообще говоря, может быть не так для результатов во всей
статье, например, для результатов о разрешимости
раздела~\ref{Logic}).

Идея рассмотрения бесконечного алфавита позволяет получить новый
результат и для класса последовательностей над конечным алфавитом.

\begin{theorem}\label{ER-preservation}
Пусть $M$~--- конечно-автоматный преобразователь. Тогда если $x \in
\ER$, то $M(x) \in \ER$.
\end{theorem}
\begin{proof}[План доказательства]
Мы приводим только план доказательства этого предложения, потому что
доказательство следует схеме доказательства
теоремы~\ref{EAP-preservation-prefix-upper-bound}, изложенной выше,
почти без изменений. Единственное важное замечание~--- утверждение
нужно доказывать над бесконечным алфавитом, по-прежнему индукцией по
количеству состояний преобразователя.

База индукции~--- обратимый преобразователь. Можно доказать,
аналогично предложению~\ref{AP-preservation-reversible-transducer},
что класс рекуррентных последовательностей (даже и над бесконечным
алфавитом) сохраняется под действием обратимых конечно-автоматных
преобразователей.

Для шага индукции нужно опять рассмотреть разбиения, описанные выше
на с.~\pageref{def-split}. Можно доказать, что $a$-разбиение
рекуррентной последовательности, в которую входит символ $a$,
рекуррентно. Однако, как несложно видеть, разбиение рекуррентной
последовательности над конечным алфавитом может стать
последовательностью над бесконечным алфавитом, потому что в
рекуррентной последовательности нет ограничения на расстояние между
вхождениями буквы.
\end{proof}

В виду теоремы~\ref{ER-preservation} естественно спросить следующее.

\begin{problem}
Предложить правильную эффективную версию
теоремы~\ref{ER-preservation}, которая позволит упростить известные
и получить новые результаты о разрешимости монадических теорий
одноместных функций (о монадических теориях
см.~раздел~\ref{Monadic}).
\end{problem}

Представляется интересным разрабатывать все введённые нами понятия и
для бесконечного алфавита.

\begin{problem}
1) Провести более систематическое изучение обобщений понятий и
результатов, введённых в настоящей статье, на бесконечный алфавит.
Здесь может иметься в виду как счётный алфавит, так и алфавит
б\'ольших мощностей.\\
2) На бесконечном алфавите можно вводить дополнительную структуру.
Например, можно отождествить алфавит с множеством натуральных чисел
и ввести порядок и операции. То же можно сделать и для
континуального алфавита. Представляется интересным разработать
соответствующие понятия и изучить их свойства.

В обоих случаях в первую очередь, в контексте настоящего обзора, нас
могут интересовать вопросы разрешимости логических теорий. Однако
можно ожидать и другие интересные и любопытные результаты.
\end{problem}

%%%% end of recurrent sequences %%%%

%%%% linear almost periodic %%%%

Следуя~\cite{Cass01,Pyth02}, будем называть последовательность $x$
\emph{линейно почти
периодической}\footnote{%
В \cite{Cass01,Pyth02} такие последовательности называются линейно
рекуррентными.%
}\label{def-linear-almost-periodic}, если она почти периодическая и
имеет не более чем линейный регулятор почти периодичности, то есть
$\r_x(n) \leqslant Cn$ для некоторого $C \in \R$, $C \geqslant 1$.
(Несложно видеть, что при $C < 1$ последовательностей $x$ c $\r_x(n)
\leqslant Cn$ для всех $n$ не существует.) Заметим, что класс
линейно почти периодических последовательностей также сохраняется
при конечно-автоматных преобразованиях с точностью до префикса.
Естественно при этом определить класс \emph{заключительно линейно
почти периодических
последовательностей}\label{def-linear-eventually-almost-periodic}, у
которых некоторый суффикс линейно почти периодический.

\begin{theorem}\label{linearAP-preservation}
Пусть $M$~--- конечно-автоматный преобразователь с $m$ состояниями.
Тогда если $x$ заключительно линейно почти периодическая, то $M(x)$
также заключительно линейно почти периодическая.
\end{theorem}
\begin{proof}
Имеем $M(x) \in \EAP$ по теореме~\ref{EAP-preservation}. Пусть
$\r_x(n) \leqslant Cn$ для всех $n$ и некоторого $C$. Тогда по
теореме~\ref{GAP-preservation}, п.~2) регулятор $\r_{M(x)}(n)$ можно
оценить сверху величиной $g^{2m}(n)$, где $g(n) = Cn + 1$. Таким
образом, $\r_{M(x)}(n) \leqslant C^{2m}n + C^{2m - 1} + C^{2m - 2} +
\dots + C + 1$ , что означает, что $M(x)$ заключительно линейно
почти периодическая.
\end{proof}

\begin{problem}\label{linearAP-preservation-better-upper-bound-problem}
Можно ли
%с помощью
%предложения~\ref{GAP-preservation-almost-reversible-transducer} и
%замечания после неё
улучшить оценку $\r_{M(x)}(n) \leqslant C^{2m}n + \frac{C^{2m} -
1}{C - 1}$ в теореме~\ref{linearAP-preservation} (см.~также
проблемы~\ref{GAP-preservation-better-upper-bound-problem}
и~\ref{linearAP-preservation-lower-bound-problem})?
\end{problem}

%Заметим, что
%теорема~\ref{GAP-preservation-improved-regulator-upper-bound}
%позволяет улучшить оценку на константу в регуляторе образа $M(x)$ в
%теореме~\ref{linearAP-preservation}.
Минимальный префикс $\pr(M(x))$ в
теореме~\ref{linearAP-preservation} также можно оценить, с помощью
теоремы~\ref{EAP-preservation-prefix-upper-bound}.

%%%% end of linear almost periodic %%%%

%%%% regulator upper bounds %%%%

Теорема~\ref{GAP-preservation}, п.~2) даёт оценку сверху порядка
$\r_{x}^{2m}$ для регулятора выходной последовательности $M(x)$ при
применении равномерного преобразователя $M$ с $m$ состояниями к
последовательности $x$. Если воспользоваться общим планом
доказательства теоремы~\ref{EAP-preservation-prefix-upper-bound}
применительно к случаю обобщённо почти периодических
последовательностей, то, по всей видимости, эту оценку можно
улучшить. Однако, к сожалению, компактной формулы, удобной для
формулировки и изложения доказательства, нам найти не удалось,
поэтому полностью этот план для обобщённо почти периодических
последовательностей мы здесь реализовывать не будем.

%\begin{theorem}\label{GAP-preservation-improved-regulator-upper-bound}
%{\bfseries [будет написано точно вместе с доказательством]} Пусть
%$M$~--- равномерный конечно-автоматный преобразователь с $m$
%состояниями, и $x \in \GAP$. Тогда $M(x) \in \GAP$ и $\r_{M(x)}
%\lessapprox \r_x^m$.
%\end{theorem}
%\begin{proof}
%Пусть $M = \langle A, B, Q, \tilde q, \lambda, \mu\rangle$~---
%равномерный преобразователь. Мы, как договорились и ранее, будем без
%ограничения общности считать, что $B = Q \times A$ и для всех $q$,
%$a$ выполнено $\lambda(q, a) = \langle q, a \rangle$.
%
%Доказательство теоремы следует общей схеме доказательств
%теорем~\ref{EAP-preservation-prefix-upper-bound}
%и~\ref{ER-preservation} и проводится индукцией по~$|Q|$. База
%индукции~--- почти обратимый конечно-автоматный преобразователь.

Тем не менее обобщение ключевой части этого плана~---
теоремы~\ref{AP-preservation-reversible-transducer}~--- представляет
и самостоятельный интерес. Назовём конечно-автоматный
преобразователь $M$ \emph{почти обратимым относительно
последовательности
$x$}\label{def-almost-reversible-finite-transducer}, если каждая
буква $a$, встречающаяся в $x$ бесконечное количество раз,
осуществляет взаимно однозначное отображение на множестве состояний
преобразователя~$M$.

\begin{proposition}\label{GAP-preservation-almost-reversible-transducer}
Пусть $M$~--- почти обратимый относительно последовательности $x$
равномерный конечно-автоматный преобразователь с $m$ состояниями, и
$x \in \GAP$. Тогда $M(x) \in \GAP$, причём $\r_{M(x)}(n) \leqslant
g^m(n) - 1$, где $g(n) = \r_x(n) + 1$.
\end{proposition}
\begin{proof}
Положим $h(n) = g^m(n) - 1$ для всех $n$, где $g(n) = \r_x(n) + 1$
для всех~$n$.

Пусть слово $v$ длины $n$ входит в $M(x)$, и $v = M(x)[i,j]$~---
одно из таких вхождений при $i \geqslant h(n)$.

Заметим, что вхождения символов, встречающихся в $x$ конечное
количество раз, ограничены префиксом длины $\r_x(1)$, поэтому,
начиная с позиции $\r_x(1)$, преобразователь $M$ действует на $x$
как обратимый. При этом $i \geqslant h(n) \geqslant \r_x(1)$.

Докажем, что найдётся вхождение $[r,s]$ слова $v$ в $M(x)$ при $i <
r \leqslant i + h(n)$.

Итак, пусть $v = M(x)[i,j]$, и $u_1 = x[i,j]$~--- прообраз $v$ в
$x$, к которому $M$ подходит, находясь в состоянии~$q_1$. Если бы
$u_1$ входило в $x$ конечное количество раз, то $u_1$ не могло бы
входить в $x[t, \infty)$ для $t \geqslant \r_x(n)$ по определению
регулятора. Но $i \geqslant \r_x(n)$, поэтому $u_1$ входит в $x$
бесконечно много раз. Тогда найдётся вхождение $u_1 = x[i_2, j_2]$,
такое что $i_2 > i$, но $j_2 \leqslant i + \r_x(n)$. Если $M$
подходит к $i_2$ в состоянии $q_1$, то $M(x)[i_2,j_2] = v$. Иначе
$M$ подходит к $i_2$ в некотором состоянии $q_2 \ne q_1$ (в этом
случае $m \geqslant 2$).

Положим $u_2 = x[i,j_2]$. Имеем $|u_2| \leqslant \r_x(n) + 1$.
Тогда, поскольку $i \geqslant h(n) \geqslant \r_x(\r_x(n) + 1)$ (при
$m \geqslant 2$), $u_2$ входит в $x$, бесконечно много раз. Поэтому
найдётся вхождение $u_2 = x[i_3,j_3]$, такое что $i_3 > i$, но $j_3
\leqslant i + \r_x(\r_x(n) + 1)$. Если $M$ подходит к $j_3 - n + 1$
в состоянии $q_1$, то $M(x)[j_3 - j + 1,j_3] = v$, так как $u_2$
заканчивается словом~$u_1$. Если $M$ подходит к $j_3 - n + 1$ в
состоянии~$q_2$, то в силу обратимости на $x[i, \infty)$
преобразователь $M$ подходит к $i_2$ в состоянии $q_1$, и тогда
$M(x)[i_2, i_2 + n - 1] = v$, так как $u_2$ начинается с~$u_1$. В
худшем случае $M$ подходит к $j_3 - n + 1$ в некотором состоянии
$q_3$, таком что $q_3 \ne q_1$ и $q_3 \ne q_2$
(рис.~\ref{AP-preservation-reversible-transducer-proof-illustration}).

Рассуждая так дальше, найдём такие $j_2,\dots,j_{m + 1}$, что хотя
бы к каким-то двум из позиций $i = j - n + 1$, $j_2 - n + 1$, \dots,
$j_{m + 1} - n + 1$ преобразователь подходит в одинаковых
состояниях. При этом $j_{m + 1} \leqslant i + h(n)$.

Таким образом, мы доказали, что если слово длины $n$ входит в
$M(x)[h(n),\infty)$, то оно входит в $M(x)$ бесконечно много раз. По
лемме~\ref{GAP-preservation-main-lemma} мы знаем тогда, что это
слово входит в каждый отрезок длины $h(n)$ в $M(x)$. Отсюда $M(x)
\in \GAP$ и $\r_x(n) \leqslant h(n)$.
\end{proof}

Рассмотрев теперь разбиение $\s_a(x)$ последовательности $x$ для
некоторой буквы $a$, входящей в $x$ бесконечно много раз, можно
провести индукцию по количеству состояний преобразователя и, таким
образом, другим способом доказать теорему~\ref{GAP-preservation},
п.~1). Как уже было сказано, в удобной форме оценку на регулятор
образа типа той, что получена в теореме~\ref{GAP-preservation},
п.~2), нам получить не удалось, хотя с помощью описанного только что
подхода, видимо, такую оценку получить можно, как минимум для
некоторых частных случаев обобщённо почти периодических
последовательностей.

\begin{problem}\label{GAP-preservation-better-upper-bound-problem}
Можно ли улучшить оценку на регулятор образа, полученную в
теореме~\ref{GAP-preservation}, п.~2)? Можно ли это сделать хотя бы
в частных случаях, например, для заключительно почти периодических
последовательностей? Для последовательностей с регулятором
$\r_x(n)$, растущим достаточно быстро с ростом $n$? См.~также
проблему~\ref{linearAP-preservation-better-upper-bound-problem}.
\end{problem}

%Для шага индукции нам необходима следующая лемма о разбиении.
%
%\begin{lemma}\label{GAP-preservation-improved-regulator-upper-bound-lemma}
%Пусть $x \in \GAP$, и $a$ входит в $x$ бесконечно много раз. Тогда
%верно следующее:
%\begin{enumerate}
%\item $\s_a(x) \in \GAP$;
%\item здесь напрямую
%\item а здесь обратно
%\end{enumerate}
%\end{lemma}
%\begin{proof}
%\end{proof}

%[\dots]
%
%\end{proof}

Мы обсуждаем нижние оценки в разделе~\ref{Projection_Product}.

%%%% end of regulator upper bounds %%%%

%%%% generalizations: pushdown automata %%%%

Интересно понять, что произойдёт, если мы несколько расширим класс
рассматриваемых преобразований.

По-видимому, простейшим обобщением можно считать стековый
(по-другому, магазинный) конечно-ав\-то\-мат\-ный преобразователь,
по аналогии со стековыми (или магазинными) автоматами (pushdown
automata), распознающими контекстно-свободные языки (например,
см.~\cite{Sips05}). \emph{Стековый конечно-автоматный
преобразователь}\label{def-pushdown-transducer} неформально можно
описать как конечно-ав\-то\-мат\-ный преобразователь, к которому
добавлен стек. (Стек, или магазин,~--- структура данных, в которой
элементы хранятся в фиксированном порядке, и пользователь имеет
доступ к последнему добавленному элементу, про который говорят, что
он находится в голове стека.) На каждом шаге следующее состояние
преобразователя и выходной символ определяются текущим состоянием,
символом входной последовательности и символом в голове стека. Кроме
того, на каждом шаге преобразователь должен решить, что делать со
стеком: добавить символ, удалить символ или оставить стек без
изменений.

Как и следовало ожидать, стековые конечно-автоматные преобразователи
не сохраняют свойство обобщённой почти периодичности.

\begin{proposition}[\cite{Prit06c-rus}]
\label{GAP-non-preservation-pushdown} Существуют стековый
конечно-автоматный преобразователь и обобщённо почти периодическая
последовательность, которая под действием этого преобразователя
переходит в последовательность, не являющуюся обобщённо почти
периодической.
\end{proposition}
\begin{proof}
По предложению~\ref{ap-alternating-0-1-prefixes} существует почти
периодическая последовательность в алфавите~$\B$, обладающая
следующим свойством: для любого $n \in \N$ найдётся такой её
префикс, в котором нулей на $n$ больше, чем единиц, а также найдётся
такой префикс, в котором единиц на $n$ больше, чем нулей. Применим к
этой последовательности следующий стековый конечно-автоматный
преобразователь. У него два режима, $a$ (начальный) и $b$. В режиме
$a$ он действует так: увидев в последовательности символ~0, он
кладёт его в стек, а увидев~1, убирает 0 из стека. Когда стек
опустошается (что соответствует начальному отрезку
последовательности с равным количеством~0 и~1), преобразователь
переключается в режим $b$. В режиме $b$ преобразователь действует
противоположным образом: при входе 1 кладёт 1 в стек, при входе 0
убирает 1 из стека, при опустошении стека переключается в режим~$a$.
В выходную последовательность преобразователь подаёт всегда номер
своего режима. Видно тогда, что в выходной последовательности
найдутся сколь угодно длинные отрезки подряд идущих $a$, и значит,
она не является обобщённо почти периодической.
\end{proof}

%%%% end of pushdown automata %%%%

%%%%%%%%%%%%%%%%%%%%%%%%%%%%%%%%%%%%%%%%%%%%%%%%%%%%%%%%%%%%%%%%%%

\subsection{О произведении последовательностей и о почти периодичности
при проекции} \label{Projection_Product}

Результаты этого раздела дают любопытные примеры почти периодических
последовательностей, подтверждающие, что свойства почти
периодических последовательностей могут быть очень разнообразными.

На последовательностях можно определить операцию $\times$, которую
мы будем называть \emph{произведением}\label{def-product-sequences}.
Для $x\in A^\N$, $y\in B^\N$ определим $x\times y\in(A\times B)^\N$,
так что $(x\times y)(i) = \langle x(i),y(i)\rangle$. Аналогично
можно определить произведение произвольного количества
последовательностей.

Как несложно видеть, если последовательность $y$ периодическая с
периодом $m$, то $x \times y$ можно получить как результат
равномерного конечно-автоматного преобразования последовательности
$x$ преобразователем с $m$ состояниями. Такой преобразователь мы
будем называть
\emph{циклическим}\label{def-cyclic-finite-transducer}.

Отсюда получаем такое следствие из
предложения~\ref{AP-preservation-reversible-transducer}.

\begin{corollary} \label{APtimesPer}
Если $x\in\AP$ и $y\in\Per$, то $x\times y\in\AP$.
\end{corollary}

Следующий результат
(теорема~\ref{GAP-preservation-regulator-lower-bound}) можно считать
нижней оценкой для п.~2) теоремы~\ref{GAP-preservation}. %, а также для
%теоремы~\ref{GAP-preservation-improved-regulator-upper-bound}.
В~теореме~\ref{GAP-preservation}, п.~2) даётся верхняя оценка на
регулятор почти периодичности образа обобщённо почти периодической
последовательности при конечно-автоматном преобразовании. Если
исходная последовательность $x$ имеет регулятор $\r_x$, и автомат
имеет $m$ состояний, то регулятор образа можно оценить сверху
величиной порядка $\r_x^{2m}$ (где верхний индекс $m$, как и в
разделе~\ref{Automata_Mappings}, понимается в смысле композиции
функций, а не в смысле возведения в степень).
%Теорема~\ref{GAP-preservation-improved-regulator-upper-bound} даёт
%лучшую оценку порядка~$\r_x^m$.
Теорема~\ref{GAP-preservation-regulator-lower-bound} показывает, что
даже для простейшего типа конечно-автоматных преобразований,
осуществляемых циклическими преобразователями, эту оценку на
регулятор образа нельзя существенно улучшить.

Обозначим через $C_m$ последовательность $01\dots(m - 1)01\dots(m -
1)01\dots$ в алфавите из $m$ символов $0, 1, \dots, m - 1$. Мы пишем
$f \geqslant g$ для функций $f$, $g$, если $f(x) \geqslant g(x)$ для
всякого аргумента~$x$.

\begin{theorem}[\cite{Raskin06-rus,PritRas07}]
\label{GAP-preservation-regulator-lower-bound} Для каждого $m
\geqslant 12$ существует бесконечно много последовательностей $x \in
\AP$, таких что $\r_{x \times C_m} \geqslant \r_x^{\lfloor m/4
\rfloor - 3}$.
\end{theorem}

Как сказано выше, по-другому
теорему~\ref{GAP-preservation-regulator-lower-bound} можно
сформулировать как нижнюю оценку на регулятор почти периодичности
образа последовательности $x$ под действием циклического
конечно-автоматного преобразователя. С помощью простой модификации
этого преобразователя можно в качестве следствия из
теоремы~\ref{GAP-preservation-regulator-lower-bound} получить нижнюю
оценку на минимальный префикс для
теоремы~\ref{EAP-preservation-prefix-upper-bound}, в которой даётся
соответствующая верхняя оценка. Подробнее см.~\cite{PritRas07}.

Можно сформулировать следующую гипотезу о нижней оценке в
теореме~\ref{linearAP-preservation}.

\begin{conjecture}\label{linearAP-preservation-lower-bound-problem}
%Если $x$~--- заключительно линейно почти периодическая, $\r_x(n)
%\leqslant Cn$ для всех $n$ и некоторого $C$, и $M$~--- равномерный
%конечно-автоматный преобразователь с $m$ состояниями, то $M(x)$~---
%заключительно почти периодическая, и $\r_{M(x)}(n) \leqslant C^{2m}n
%+ C^{2m - 1} + C^{2m - 2} + \dots + C + 1$
%(см.~теорему~\ref{linearAP-preservation}).
В обозначениях теоремы~\ref{linearAP-preservation}, можно ли
построить пример последовательности и преобразователя (серии
последовательностей и преобразователей), для которых выполнена
нижняя оценка типа~$\r_{M(x)}(n) > C^{\alpha m}n$ для
некоторого~$\alpha$?
\end{conjecture}

В качестве естественного продолжения этой темы, можно задаться
следующим не очень формальным вопросом.

\begin{problem}\label{problem-product}
Пусть $x,y \in \AP$. Что тогда можно сказать про $x\times y$?
\end{problem}

Проблему~\ref{problem-product} можно переформулировать как вопрос о
свойствах декартова произведения двух динамических систем~---
см.~теорему~\ref{dynamics-almost-periodicity} и обсуждение вокруг
неё.

Следующий результат показывает, что ничего простого в качестве
ответа на проблему~\ref{problem-product} сказать не удастся.

\begin{theorem}[\cite{Raskin07}]\label{almost-periodic-product}
Для каждого из множеств $\mathcal M = \AP$, $\EAP\setminus\AP$,
$\GAP\setminus\EAP$, $\B^\N\setminus\GAP$ можно построить
последовательности $x,y \in \AP$, такие что $x \times y \in \mathcal
M$.
\end{theorem}

Таким образом, проблему~\ref{problem-product} можно немного уточнить
и конкретизировать.

\begin{problem}\label{problem-product-concretization}
1) Пусть $x,y \in \AP$. Существует ли какой-то критерий для
определения, верно ли, что $x\times y\in\GAP$? Что $x \times y \in
\AP$?\\
2) Можно ли, как-то ограничив $x,y$ и рассмотрев специальные случаи
почти периодических последовательностей, как-то охарактеризовать $x
\times y$?\\
3) Для результатов типа имеющихся в виду в предыдущем пункте, можно
ли дать оценку сверху на регулятор почти периодичности произведения,
зная оценки сверху на регуляторы почти периодичности компонент
произведения? Оценки такого типа могут позволить получить новые
результаты о разрешимости логических теорий
(см.~раздел~\ref{Logic}).
\end{problem}

Примерами результатов того типа, которые предлагается найти в п.~2
вопроса~\ref{problem-product-concretization}, являются
следствие~\ref{APtimesPer} и теорема~\ref{PAP-times-Per}. Рассмотрим
множество $PP$ последовательностей~$x$, для которых верно следующее:
произведение $x$ с любой почти периодической последовательностью
почти периодично. Например, следствие~\ref{APtimesPer} утверждает,
что множеству $PP$ принадлежат все периодические последовательности.
Ясно, что $PP \subsetneq \AP$. В~\cite{Salim09a,Salim09b} доказана
для разнообразных последовательностей их принадлежность множеству
$PP$, а именно: для достаточно широкого класса чисто морфических
последовательностей, в частности, для последовательности Туэ~---
Морса; для последовательностей Тёплица и их обобщений; для
достаточно широкого класса последовательностей Штурма и их
обобщений. Для всех этих результатов, безусловно, интересно также
ответить и на вопрос типа п.~3
вопроса~~\ref{problem-product-concretization}, то есть получить
оценку на регулятор почти периодичности произведения, зная
регуляторы почти периодичности исходных последовательностей.

В некотором смысле обобщением одного из утверждений
теоремы~\ref{almost-periodic-product} являются следующие три
результата, доказанные явно в~\cite{MuchSemUsh03}, но по существу
полученные в~\cite{Sem83-rus}.

\begin{theorem}[\cite{Sem83-rus,MuchSemUsh03}]
Для любого $m \in \N$ существуют $m + 1$ двоичных
последовательностей, таких что произведение любых $m$ из них
является эффективно почти периодической последовательностью, а
произведение всех $m$ последовательностей не является почти
периодическим.
\end{theorem}

\begin{theorem}[\cite{Sem83-rus,MuchSemUsh03}]
Для любого $m \in \N$ существуют $m + 1$ двоичных
последовательностей, таких что произведение любых $m$ из них
является эффективно почти периодической последовательностью, а
произведение всех $m$ последовательностей является почти
периодической последовательностью, но не является эффективно почти
периодической.
\end{theorem}

Морфизм $h\colon A^* \to B^*$ назовём
\emph{проекцией}\label{def-projection}, если для любой буквы $a \in
A$ выполнено $|h(a)| = 1$, и при этом $|A| > |B|$.

\begin{theorem}[\cite{Sem83-rus,MuchSemUsh03}]
Для любого $m \in \N$ существует вычислимая последовательность
$x\colon \N \to \{1,\dots, m\}$, такая что для любой проекции $h$
последовательность $h(x)$ почти периодична с вычислимой верхней
оценкой на регулятор почти периодичности. Последовательность $x$
может быть построена удовлетворяющей одному из следующих условий:
\begin{enumerate}
\item $x$ не является почти периодической;
\item $x$ является почти
периодической последовательностью, но не является эффективно почти
периодической.
\end{enumerate}
\end{theorem}

В следующей теореме доказывается, что класс точно почти
периодических последовательностей замкнут относительно операции
умножения на периодическую последовательность.

\begin{theorem}\label{PAP-times-Per}
Если $x \in \PAP$ и $y \in \Per$, то $x \times y \in \PAP$.
\end{theorem}
\begin{proof}
Будем считать, что $y = 01\dots(m - 1)01\dots(m - 1)01\dots$, где
$m$~--- период $y$. Случай произвольного $y \in \Per$ получается из
этого кодированием, которое, очевидно, сохраняет $\PAP$. В этом
доказательстве для краткости вместо $u \times [i, i + |u| - 1]$
будем писать $u \times [i,\cdot]$.

Пусть $u$ входит в $x$, и $A = \{i : 0 \leqslant i \leqslant m, u
\times [j, \cdot]$ входит в $z$ для некоторого $j \equiv i \pmod
m\}$. Пусть $v \times [a, \cdot]$~--- такое подслово $z$, которое
для каждого $i \in A$ содержит вхождение $u \times [j,\cdot]$ для
некоторого $j \equiv i \pmod m$. Слово $v$ входит в $x$, поэтому для
некоторых $b,d$ имеем $x[b + dt, b + dt + |v| - 1] = v$ для каждого
$t \in \N$. Ясно при этом, что $B = \{i : 0 \leqslant i \leqslant m,
u \times [j, \cdot]$ входит в $v \times [b, \cdot]$ для некоторого
$j \equiv i \pmod m\} = A$. Действительно, по определению $A$ имеем
$B \subseteq A$, но $B = \{(i + b - a) \mod m : i \in A\}$. Значит,
$\#B = \#A$, откуда $B = A$.

Таким образом, если $u \times [s,\cdot]$ входит в $z$, то для
некоторого $p \equiv s \pmod m$ имеем вхождение $u \times [p,
\cdot]$ в $v \times [b,\cdot]$. Поэтому $u \times [p + tdm,\cdot]$
входит в $z$ для каждого $t \in \N$. Следовательно, $z \in \PAP$.
\end{proof}

В связи с теоремой~\ref{PAP-times-Per} возникает следующий вопрос.

\begin{problem}
1) Верно ли, что класс точно почти периодических последовательностей
замкнут относительно конечно-автоматных преобразований?\\
2) Если ответ на предыдущий вопрос положительный, то можно ли это
доказать в каком-нибудь смысле эффективно и получить более простой
критерий разрешимости монадической теории точно почти периодической
последовательности (см.~раздел~\ref{Logic})?
\end{problem}

%%%%%%%%%%%%%%%%%%%%%%%%%%%%%%%%%%%%%%%%%%%%%%%%%%%%%%%%%%%%%%%%%%
%%%%%%%%%%%%%%%%%%%%%%%%%%%%%%%%%%%%%%%%%%%%%%%%%%%%%%%%%%%%%%%%%%
%%%%%%%%%%%%%%%%%%%%%%%%%%%%%%%%%%%%%%%%%%%%%%%%%%%%%%%%%%%%%%%%%%

\section{Логика почти периодических последовательностей}
\label{Logic}

%%%%%%%%%%%%%%%%%%%%%%%%%%%%%%%%%%%%%%%%%%%%%%%%%%%%%%%%%%%%%%%%%%

\subsection{Логические теории}
\label{Monadic}

Рассмотрим следующие простые вопросы о последовательности $x$:
входит ли в $x$ символ $a$? Входит ли слово $u$? Входит ли слово $u$
бесконечно много раз? Можно сформулировать и более сложные вопросы
про последовательность. Когда на такие вопросы можно отвечать
алгоритмически, получая на вход последовательность?

Различные, в том числе и перечисленные только что, свойства
бесконечных последовательностей могут быть выражены в
\emph{логической теории первого
порядка}\label{def-first-order-theory}. Под такой теорией для
последовательности $x \in A^\N$ мы будем понимать следующее.
Формально, в качестве структуры возьмём $\langle \N, S, <,
X\rangle$, где $\N$~--- множество натуральных чисел, которое
пробегают индивидные переменные, $S$~--- двухместный предикат
следования, $<$~--- двухместный предикат порядка на натуральных
числах, $X$~--- функциональный символ, интерпретируемый как
последовательность $x\colon\N \to A$. В качестве теории берём
обычную теорию первого порядка, истинность формул интерпретируем
естественным образом. Во всех рассматриваемых нами теориях мы
подразумеваем наличие двухместного предиката равенства,
интерпретируемого естественным образом.

Ясно, что у такой реализации есть много вариантов, эквивалентных
между собой по выразительным способностям. Например, можно было
вместо двухместного предиката следования взять в структуру
одноместную функцию следования. Можно было и не брать предикат
следования вообще, потому что он выразим через неравенство. Точно
так же вместо одной функции $X$ можно было взять семейство
предикатов $X_a$ по одному для каждого $a \in A$, истинных ровно
там, где в $x$ стоит буква $a$. Можно обойтись и меньшим~---
логарифмическим по отношению к размеру алфавита
последовательности~--- количеством предикатов. Кроме того, поскольку
константа 0 выразима в определённой выше структуре, ясно, что можно
было, не меняя множество выразимых формул, добавить в структуру все
константы. Часто мы будем переходить от одной конкретной реализации
к другой, если понадобится, при этом чётко фиксируя, какая именно
реализация теории рассматривается, когда это важно, как, например, в
теореме~\ref{GAP-first-order-quantum-elimination}.

Теорию, определённую выше, будем обозначать $\T x$.

Несложно видеть, что простые свойства последовательности $x$, типа
упомянутых в начале раздела, можно выразить в теории $\T x$.
Например, формула $\forall p \exists q \exists r \, q > p \land
S(q,r) \land X(q) = 0 \land X(r) = 1$ означает свойство вхождения в
$x$ бесконечное количество раз слова~01. Тем не менее, некоторые
просто формулируемые свойства, которые, возможно, хотелось бы
выразить, в теории первого порядка выразить нельзя, например,
``слово в алфавите $\{a, b\}$, в котором между любыми
последовательными вхождениями $b$ (такими, между которыми нет других
вхождений $b$) входит нечётное количество букв~$a$''.

Гораздо больше можно выразить в более сильной \emph{монадической
теории второго порядка}\label{def-monadic-theory}. В такой теории,
кроме обычных переменных по натуральным числам $p,q,\dots$,
разрешены также монадические переменные по множествам (или по
одноместным предикатам) $P,Q,\dots$. Вводятся также соответствующие
атомарные формулы вида $P(p), Q(p), \dots$, обозначающие ``$p$
принадлежит $P$'', ``$p$ принадлежит $Q$''\dots Разрешены также
кванторы по монадическим переменным. Такую теорию мы будем
обозначать $\MT x$.

Теории, аналогичные вышеперечисленным, но не расширенные
последовательностью, будем обозначать соответственно
$\mathop{\mathrm{T}\langle \N, <\rangle}$,
$\mathop{\mathrm{MT}\langle \N, <\rangle}$.

Теория $\MT x$ богаче теории $\T x$. Например, упомянутое выше не
выразимое в теории первого порядка свойство в монадической теории
выражается так: $\forall p \forall q (X(p) \land X(q) \land p < q
\land \forall r (p < r \land r < q \to \lnot X(r)) \to \exists Y
(\forall u \forall v (S(u, v) \to (Y(u) \leftrightarrow Y(v))) \land
Y(p) \land Y(q)))$.

Как и в случае с теориями первого порядка, в описанной формализации
монадической теории многое можно реализовать по-другому. Например,
можно отказаться от неравенства, потому что в монадической теории
оно выразимо через следование. Как и в случае теорий первого
порядка, часто мы будем переходить от одной реализации к другой для
удобства.

Для формулы $\phi$ в любом из вышеописанных языков будем обозначать
через $L(\phi)$ множество всех последовательностей $x$, для которых
эта формула верна (то есть верна, если интерпретировать входящую в
неё единственную свободную переменную по функциям $X$ как $x$).

Естественным (и основным для нас) вопросом о любой теории является
вопрос о её разрешимости. Теория
\emph{разрешима}\label{def-decidable-theory}, если существует
алгоритм, который по любой замкнутой формуле определяет её
истинность.

Следующий результат о теориях первого порядка был получен
в~\cite{Sem79-rus}.

Теорию $\Tp x$ (незначительную модификацию теории $\T x$) определим
следующим образом. Пусть $P$~--- система предикатов, задающая
последовательность~$x$. Термами теории будут выражения вида $c$, $x
+ c$, где $c \in \Z$, $x$~--- переменная. Атомарными формулами
теории будут выражения $\tau < \tau'$, $\tau > \tau'$, $p(\tau)$,
где $p$~--- предикат из системы $P$, $\tau$, $\tau'$~--- термы.
Несмотря на то, что значения термов могут не лежать в $\N$,
отношения, задаваемые атомарными формулами, всюду определены
на~$\N$.

Мы говорим, что теория
\emph{бескванторная}\label{def-quantifier-free-theory}, если для
каждой формулы теории найдётся эквивалентная ей бескванторная.

\begin{theorem}[\cite{Sem79-rus}]\label{GAP-first-order-quantum-elimination}
\begin{enumerate}
\item Если теория $\Tp x$ бескванторная, то $x \in \GAP$.
\item Пусть $x \in \GAP$. Тогда теория $\Tp x$ бескванторная,
причём она разрешима, если и только если $x$ эффективно обобщённо
почти периодична.
\end{enumerate}
\end{theorem}

Обратимся теперь к монадическим теориям.
% {\bfseries [сказать про
%результаты о разрешимости теорий помимо почти периодичности; см.
%также абзац во введении на эту тему]}

Оказывается, что благодаря применению теории автоматов для
достаточно широких классов последовательностей можно получить
критерий разрешимости их монадических теорий.

Назовём (недетерминированным) \emph{автоматом
Бюхи}\label{def-Buchi-automaton} совокупность $M = \langle A, Q,
\tilde q, \Delta, F\rangle$, где $A$~--- входной алфавит, $Q$~---
множество состояний, $\tilde q \in Q$~--- начальное состояние,
$\Delta \subseteq Q \times A \times Q$~--- множество переходов, $F
\subseteq Q$~--- множество заключительных состояний. Ходом автомата
$M$ на последовательности $x = x(0)x(1)x(2)\dots$ называется такая
последовательность состояний $\rho = \rho(0)\rho(1)\rho(2)\dots$,
что $\rho(0) = \tilde q$ и $\langle\rho(i), x(i), \rho(i + 1)\rangle
\in \Delta$ для любого~$i$. Мы говорим, что автомат $M$
\emph{принимает $x$}\label{def-Buchi-automaton-accepts}, если
существует хотя бы один ход $\rho$ автомата $M$ на $x$, для которого
хотя бы одно состояние, встречающееся в $\rho$ бесконечное
количество раз, входит в множество заключительных состояний~$F$.
Определяя \emph{детерминированный автомат
Бюхи}\label{def-deterministic-automaton}, множество переходов
$\Delta \subseteq Q \times A \times Q$ можно заменить на функцию
переходов $\delta\colon Q \times A \to Q$ (с естественными
изменениями для определения хода автомата).

Есть немного другой вариант понятия автомата на последовательности.
Назовём (недетерминированным) \emph{автоматом
Мюллера}\label{def-Muller-automaton} совокупность $M = \langle A, Q,
\tilde q, \Delta, \F\rangle$, где $A$~--- входной алфавит, $Q$~---
множество состояний, $\tilde q \in Q$~--- начальное состояние,
$\Delta \subseteq Q \times A \times Q$~--- множество переходов, $\F
\subseteq 2^Q$~--- множество заключительных макросостояний. Здесь
под \emph{макросостоянием}\label{def-macrostate} мы понимаем элемент
множества $2^Q$, то есть произвольное подмножество множества
состояний~$Q$. Ходом автомата $M$ на последовательности $x =
x(0)x(1)x(2)\dots$ называется такая последовательность состояний
$\rho = \rho(0)\rho(1)\rho(2)\dots$, что $\rho(0) = \tilde q$ и
$(\rho(i), x(i), \rho(i + 1)) \in \Delta$ для любого~$i$. Назовём
\emph{пределом (предельным
макросостоянием)}\label{def-limit-macrostate} автомата $M$ на
последовательности $x$ вдоль хода $\rho$ множество всех таких
состояний, которые встречаются в $\rho$ бесконечное количество раз,
обозначим это множество через $\lim_\rho M$. Мы говорим, что автомат
$M$ \emph{принимает $x$}\label{def-Muller-automaton-accepts}, если
существует хотя бы один ход $\rho$ автомата $M$ на $x$, для которого
$\lim_\rho M \in \F$. Другими словами, слово принимается, если хотя
бы на каком-нибудь ходе предельное макросостояние принадлежит
множеству заключительных макросостояний~$\F$. Аналогично
предыдущему, определяя \emph{детерминированный автомат
Мюллера}\label{def-deterministic-Muller-automaton}, множество
переходов $\Delta \subseteq Q \times A \times Q$ можно заменить на
функцию переходов $\delta\colon Q \times A \to Q$ (с естественными
изменениями для определения хода автомата).

Для автомата $M$ Бюхи или Мюллера множество всех
последовательностей, которые принимаются автоматом $M$,
обозначим~$L(M)$.

Для примера рассмотрим множество $L$ последовательностей в алфавите
$\{a, b, c\}$, в которые если $a$ входит бесконечное количество раз,
то и $b$ входит бесконечное количество раз. Автоматы Мюллера и Бюхи,
принимающие в точности слова множества $L$, показаны на
рис.~\ref{infinite-automata-example} (автомат Бюхи понадобился
недетерминированный).

\begin{figure}[h]
\centering
\includegraphics{survey_almper_pictures.1}
\hspace{20mm}
\includegraphics{survey_almper_pictures.2}
\caption{Пример автомата Мюллера (слева) и автомата Бюхи (справа),
принимающих одно и то же множество последовательностей. Множество
принимающих макросостояний автомата Мюллера дано списком.}
\label{infinite-automata-example}
\end{figure}

Это пример общей ситуации.

\begin{theorem}[\cite{McNaught66}]\label{buchi-muller-equivalence}
Недетерминированные автоматы Бюхи, недетерминированные автоматы
Мюллера и детерминированные автоматы Мюллера распознают один и тот
же класс множеств последовательностей. Более того, по автомату
одного типа можно получать эквивалентный автомат другого типа
алгоритмически.
\end{theorem}

Множество последовательностей, распознаваемое автоматом Мюллера или
автоматом Бюхи, будем называть
\emph{регулярным}\label{def-regular-set}. Детерминированные автоматы
Бюхи распознают меньший класс множеств.

Оказывается, что между описанием множеств последовательностей с
помощью конечных автоматов и описанием их формулами монадического
языка есть прямая связь.

\begin{theorem}[\cite{Buchi62}]\label{mso-buchi-automata-correspondence}
Существует алгоритм, который по каждому автомату Бюхи $M$ строит
формулу $\phi$ монадического языка, такую что $L(M) = L(\phi)$, и
наоборот, по любой формуле $\phi$ строит такой автомат Бюхи $M$, что
$L(\phi) = L(M)$.
\end{theorem}

\begin{corollary}[\cite{Buchi62}]
Теория $\mathop{\mathrm{MT}\langle \N, <\rangle}$ разрешима.
\end{corollary}

Нас же интересует ситуация, когда теория $\mathop{\mathrm{MT}\langle
\N, <\rangle}$ расширена последовательностью. Оказывается, и в этом
случае теорема Бюхи
(теорема~\ref{mso-buchi-automata-correspondence}) может помочь.
Несложно видеть, что выполняется следующее следствие.

\begin{corollary}\label{decidability-automata}
Монадическая теория последовательности $x$ разрешима тогда и только
тогда, когда существует алгоритм, который по любому автомату Бюхи
(или любому детерминированному автомату Мюллера) может определить,
принимает ли этот автомат последовательность $x$ или нет.
\end{corollary}

Покажем теперь, как благодаря следствию~\ref{decidability-automata}
получить критерий разрешимости монадических теорий обобщённо почти
периодических последовательностей.

Следующий основной результат в существенной степени мотивирует
определения и результаты раздела~\ref{Automata_Mappings}, а также
теорему~\ref{GAP-preservation}.

\begin{theorem}[\cite{Sem83-rus}]\label{decidability-monadic-GAP}
Теория $\MT x$ обобщённо почти периодической последовательности $x$
разрешима тогда и только тогда, когда $x$ эффективно обобщённо почти
периодична.
\end{theorem}
\begin{proof}
$\Rightarrow$. Пусть для $x \in \GAP$ теория $\MT p$ разрешима.
Тогда для каждого $n$ и для каждого возможного символа $a$
проверяем, не верно ли $x(n) = a$, и так перебором находим значение
$x(n)$. Таким образом, $x$ вычислима. Также перебором можно
вычислять $\r_x$, поскольку для любых $n$ и $l$ можно записать
формулой тот факт, что $\r_x(n) \leqslant l$. Тут нам помогает то,
что слов фиксированной длины конечное число, поскольку
последовательность конечнозначная. И если даже мы не знаем вначале,
какие элементы могут входить в последовательность (потенциальных
претендентов, как мы договорились, может быть бесконечное
количество), мы можем их также найти перебором, поскольку можно
выразить формулой тот факт, что все символы, кроме символов
некоторого фиксированного конечного множества, в последовательности
не встречаются.

\smallskip

$\Leftarrow$. Пусть теперь $x \in \GAP$ вычислима и некоторая $f
\geqslant \r_x$ вычислима.

По следствию~\ref{decidability-automata} для разрешимости $\MT x$
достаточно уметь по любому детерминированному автомату Мюллера,
запущенному на $x$, определять, принимает ли он $x$ или нет.

Пусть $M$~--- какой-нибудь детерминированный автомат Мюллера,
действующий на~$x$. Рассмотрим конечно-автоматный преобразователь
$M'$, полученный из $M$ следующим образом: внутреннее устройство
$M'$ такое же, как и у $M$, при этом про принимающие макросостояния
$M$ мы забываем. На входных последовательностях $M'$ работает точно
так же, как и $M$, а на выход выдаёт на каждом шаге своё текущее
состояние. По теореме~\ref{GAP-preservation}, п.~3) из вычислимости
$x$ и $f$ следует вычислимость $F(x) \in \GAP$ и некоторой $g
\geqslant \r_{M'(x)}$.

Легко вычислить, какие символы входят в $M'(x)$ бесконечно много
раз: для этого достаточно посмотреть, какие символы входят в отрезок
$M'(x)[g(1), 2g(1) - 1]$ (а ``посмотреть'' мы можем, потому что сама
последовательность $M'(x)$ тоже вычислима).

Таким образом, мы можем найти множество всех таких состояний
автомата $M$, которые встречаются бесконечное количество раз в
процессе работы $M$ на $x$. Значит, мы можем проверить, принимает ли
автомат $M$ последовательность~$x$.
\end{proof}

\begin{corollary}
\label{decidability-monadic-AP} Теория $\MT x$ почти периодической
последовательности $x$ разрешима тогда и только тогда, когда $x$
вычислима и множество подслов $\Fac(x)$ разрешимо.
\end{corollary}
\begin{proof}
Следует из теоремы~\ref{decidability-monadic-GAP} и
предложения~\ref{effective-AP-equivalent-definition}.
\end{proof}

\begin{problem}
Предположим, последовательность $x \in \EAP$ вычислима. Верно ли,
что теория $\MT x$ разрешима, если множество подслов $\Fac(x)$
разрешимо? А если дополнительно известно, что можно вычислить про
каждое слово, входит ли оно в $x$ бесконечно много раз? Те же
вопросы про классы $\GAP$, $\Rec$, $\ER$. Для всех этих вопросов
есть и различные равномерные варианты. Например, существует ли
алгоритм, который по последовательности $x \in \EAP$, числу $l
\geqslant \pr(x)$ и множеству подслов последовательности
$x[l,\infty) \in \AP$ разрешает теорию $\MT x$? В этой связи
см.~также результаты раздела~\ref{Computability}.
\end{problem}

По-видимому, ответы на все эти вопросы отрицательны, хотя это можно
утверждать и с различной степенью уверенности для разных из них.

\subsection{Разрешимость теорий конкретных последовательностей}
\label{Decidability-of-theories}

В этом разделе мы показываем, как полученные в разделе~\ref{Monadic}
критерии применить к некоторым ранее рассмотренным примерам
последовательностей.

О последовательности Туэ~--- Морса см.~раздел~\ref{Thue}.

\begin{theorem}\label{Thue-monadic-decidable}
Монадическая теория $\MT \thue$ последовательности Туэ~--- Морса
разрешима.
\end{theorem}
\begin{proof}
Напомним, что последовательность Туэ~--- Морса можно определить как
$\thue = \lim_{n \to \infty} u_n$, где последовательность слов $u_n$
определяется по индукции: $u_0 = 0$ и $u_{n + 1} = u_n\bar u_n$. При
этом $|u_n| = 2^n$. (См.~раздел~\ref{Thue}.)

Ясно, что $\thue$ вычислима. По
следствию~\ref{decidability-monadic-AP} достаточно доказать теперь,
что $\Fac(\thue)$ разрешимо. Пусть $w$~--- произвольное слово.
Найдём какое-нибудь $n$, так чтобы $|w| \leqslant 2^n$. Поскольку
для любого $n$ имеем $\thue = u_n\bar u_n\bar u_n u_n \bar u_n u_n
u_n \bar u_n \dots$, слово $w$ входит в $\thue$ тогда и только
тогда, когда оно входит в одно из слов $u_nu_n$, $u_n\bar u_n$,
$\bar u_n u_n$, $\bar u_n \bar u_n$, что легко проверяется.
\end{proof}

Аналогично явно можно доказать разрешимость монадической теории
последовательности Фибоначчи~$\fib$, но этот результат следует и из
теоремы~\ref{Sturmian-monadic-decidability-criterion} ниже.

О последовательностях Штурма см.~раздел~\ref{Sturm}. Вопрос о
разрешимости монадической теории последовательности $x_n = \lceil
\sin(n)\rceil$ (где $\lceil x \rceil$ для действительного числа $x$
обозначает наименьшее целое число не меньше $x$), поставленный
в~\cite{Sief71}, изначально послужил мотивацией в
работах~\cite{Sem79-rus,Sem83-rus}. Несложно видеть, что
последовательность $x_n$ является последовательностью Штурма. Для
специальных последовательностей Штурма, то есть таких, что сдвиг
$\rho = 0$, критерий эффективной почти периодичности (дающий в
соответствии с теоремой~\ref{decidability-monadic-GAP} критерий
разрешимости монадической теории) был получен в~\cite{Sem79-rus}.
Здесь мы доказываем критерий в более общем случае.
В~\cite{CarThom02} вопрос получения такого критерия сформулирован
как открытый.

Напомним, что действительное число $a$ называется
\emph{вычислимым}\label{def-computable-real-number}, если существует
алгоритм, который по любому положительному рациональному числу
$\epsilon$ выдаёт приближение к $a$ с точностью $\epsilon$, то есть
число в интервале $(a - \epsilon, a + \epsilon)$.

\begin{theorem}\label{Sturmian-monadic-decidability-criterion}
Монадическая теория $\MT{s_{\alpha,\rho}}$ последовательности
Штурма, определённой как нижняя механическая последовательность с
наклоном $\alpha$ и сдвигом~$\rho$, разрешима тогда и только тогда,
когда $\alpha$ и $\rho$ вычислимы.
\end{theorem}
\begin{proof}
$\Leftarrow$. Пусть $\alpha$ и $\rho$ вычислимы. Все, кроме,
возможно, двух символов последовательности, вычислимы: для
вычисления $s_{\alpha,\rho}(n) = \lfloor \alpha(n + 1) + \rho\rfloor
- \lfloor\alpha n + \rho\rfloor$ нужно с достаточной степенью
точности вычислить $\alpha$ и $\rho$, чтобы можно было вычислить
$\lfloor \alpha n + \rho\rfloor$ и $\lfloor \alpha(n + 1) +
\rho\rfloor$, а значит, и $s_{\alpha,\rho}(n)$. Это возможно, если
только не окажется, что $\alpha n + \rho$ или $\alpha(n + 1) + \rho$
целые~--- все такие $n$ (их не больше двух) мы запоминаем особо и в
программу, вычисляющую $s_{\alpha,\rho}$, записываем как отдельный
случай.

Далее, ясно, что если последовательность вычислима, и мы для каждого
$n$ знаем количество подслов длины $n$ в ней~--- в данном случае это
количество равно $n + 1$~--- то множество подслов разрешимо. Отсюда
по следствию~\ref{decidability-monadic-AP} получаем, что
$\MT{s_{\alpha,\rho}}$ разрешима.

$\Rightarrow$. Предположим, что теория $\MT{s_{\alpha,\rho}}$
разрешима. Отсюда следует, что последовательность $s_{\alpha,\rho}$
вычислима. Заметим, что $k_n = \sum_{i = 1}^n s_{\alpha,\rho}(i) =
\lfloor \alpha(n + 1) + \rho\rfloor - \lfloor\rho\rfloor = \lfloor
\alpha(n + 1) + \rho\rfloor$. Отсюда $k_n \leqslant \alpha(n + 1) +
\rho < k_n + 1$, следовательно, $\frac{k_n - 1}{n + 1} < \alpha <
\frac{k_n + 1}{n + 1}$. Это позволяет нам оценить $\alpha$ с
точностью $\frac2{n + 1}$. Таким образом, число $\alpha$ вычислимо,
потому что его можно вычислить с любой наперёд заданной точностью.

Теперь возможны два случая. Первый случай: $\alpha n + \rho$ при
каком-нибудь $n$ является целым числом. Тогда ясно, что число $\rho$
в этом случае вычислимо.

Рассмотрим второй случай, когда $\alpha n + \rho$ не является целым
ни при каком~$n$. Будем все числа рассматривать по модулю 1, то есть
не различать числа, отличающиеся на целое слагаемое. Такие числа
удобно представлять себе отмеченными на окружности единичной длины,
на которой мы в положительную сторону идём против часовой стрелки, а
в отрицательную по часовой. Соответствующим образом понимаются и
двойные неравенства между такими числами: $a < b < c$, если тройка
$(a,b,c)$ ориентирована против часовой стрелки. Заметим, что
$s_{\alpha,\rho} = 0$, если $0 < \alpha n + \rho < - \alpha$, и
$s_{\alpha,\rho} = 1$, если $- \alpha < \alpha n + \rho < 0$.
Следовательно, $s_{\alpha,\rho} = 0$ при $-\alpha n < \rho < -
\alpha (n + 1)$ и $s_{\alpha,\rho} = 1$ при $-\alpha(n + 1) < \rho <
\alpha n$.

Чтобы посчитать $\rho$ с точностью $2\epsilon$, нужно сначала для
достаточно большого $N$ отметить на окружности точки $\alpha k$ для
всех $k = 0, 1, 2, \dots, N$, так чтобы расстояния между соседними
отмеченными точками не превышали $2\epsilon$. Для этого $\alpha$
должно быть вычислено с точностью $\epsilon/N$, и расстояния между
вычисленными значениями не должны превышать $\epsilon$. Такое
достаточно большое $N$ найдётся, так как $\alpha$ иррационально, а
при иррациональных $\alpha$ последовательность $\alpha n$ всюду
плотна на окружности. После этого, вычислив значения
$s_{\alpha,\rho}(k)$ для $k = 0, 1, 2, \dots, N$, мы сможем
ограничить $\rho$ одним из получившихся интервалов длины не более
$2\epsilon$ между точками.
\end{proof}

Рассмотрим равномерные аналоги утверждений, доказанных в
теореме~\ref{Sturmian-monadic-decidability-criterion}. Предположим,
мы хотим построить алгоритм, который, получив на вход числа $\alpha$
и $\rho$, разрешает теорию $\MT{s_{\alpha,\rho}}$. Если бы
существовала процедура, распознающая факт существования такого $n$,
что $\alpha n + \rho$ целое, то её можно было бы использовать так,
как описано в доказательстве
теоремы~\ref{Sturmian-monadic-decidability-criterion}. Но известно,
что такой процедуры не существует. Однако это ещё не доказывает, что
желаемого алгоритма не существует. Обратно аналогично: при наличии
вышеупомянутой процедуры можно построить алгоритм, который по
алгоритму разрешения теории $\MT{s_{\alpha,\rho}}$ мог бы
вычислять~$\alpha$ и~$\rho$. Но известное нам отсутствие такой
процедуры ещё не влечёт отсутствия желаемого алгоритма. Это
порождает следующий вопрос.

\begin{problem}\label{Sturmian-monadic-decidability-criterion-effective-problem}
1) Существует ли алгоритм, который, имея алгоритм разрешения теории
$\MT{s_{\alpha,\rho}}$, может вычислять $\alpha$ и $\rho$?\\
2) Существует ли алгоритм, который по алгоритму вычисления
параметров $\alpha$ и $\rho$ может разрешать теорию
$\MT{s_{\alpha,\rho}}$?
\end{problem}

%%%%%%%%%%%%%%%%%%%%%%%%%%%%%%%%%%%%%%%%%%%%%%%%%%%%%%%%%%%%%%%%%%
%%%%%%%%%%%%%%%%%%%%%%%%%%%%%%%%%%%%%%%%%%%%%%%%%%%%%%%%%%%%%%%%%%
%%%%%%%%%%%%%%%%%%%%%%%%%%%%%%%%%%%%%%%%%%%%%%%%%%%%%%%%%%%%%%%%%%

\section{Вычислимость для почти периодических последовательностей}
\label{Computability}

%%%%%%%%%%%%%%%%%%%%%%%%%%%%%%%%%%%%%%%%%%%%%%%%%%%%%%%%%%%%%%%%%%%

\subsection{Неразрешимость некоторых свойств почти периодических
последовательностей} \label{Computability_Sequences}

Во многих вопросах, связанных с почти периодическими и с обобщённо
почти периодическими последовательностями, естественным образом
возникает алгоритмическая составляющая: можно ли ту или иную
характеристику или то или иное свойство проверять алгоритмически,
получая на вход последовательность. Иногда эти вопросы являются
эффективными аналогами уже известных результатов, например,
теорема~\ref{EAP-preservation-prefix-upper-bound}~--- эффективный
аналог теоремы~\ref{EAP-preservation}. В настоящем разделе в
основном будут рассмотрены результаты, когда ответ на эти вопросы
отрицательный. Мы доказываем, что некоторые свойства почти
периодичности не имеют эффективных аналогов.

Формально, мы рассматриваем \emph{вычислимые операторы на
последовательностях}\label{def-computable-operator}. Отметим, что
тут возможны два подхода. В первом мы рассматриваем операторы,
определённые на всевозможных последовательностях. Такие операторы в
процессе вычисления могут обращаться к любым элементам
последовательности. Основное свойство вычислимых операторов~---
непрерывность относительно топологии, индуцированной метрикой~$d_C$.
Другими словами, перед тем, как выдать ответ, оператор может
прочитать лишь конечное количество элементов последовательности.
Таким образом, для доказательства неразрешимости некоторого свойства
таким оператором достаточно показать, что это свойство не является
непрерывным.

При втором подходе мы рассматриваем операторы, определённые только
на вычислимых последовательностях. На вход оператору подаётся
процедура вычисления последовательности. Однако можно доказать, что
при таком определении оператор также будет обладать свойством
непрерывности (см.~\cite{Rogers67-rus}). Поэтому достаточно иметь в
виду только первый подход, а все результаты настоящего раздела верны
и при втором подходе.

Почти никакие осмысленные свойства нельзя распознать, имея на входе
только саму последовательность. Например, про $x \in \B^\N$ нельзя
даже сказать, входит ли в неё символ 1: если алгоритм проверил
некоторое конечное количество символов, и все они оказались 0, он не
может гарантировать, что далее не встретится 1. Вопрос об
эффективности станет более интересным, если мы на вход будем
подавать какую-то дополнительную информацию. В случае с обобщённо
почти периодическими последовательностями в качестве этой информации
естественно взять регулятор почти периодичности. (Ясно, что точно
так же можно рассматривать операторы, которым на вход подаются и
функции, а не только последовательности.)

Теперь мы видим, что при таком подходе сформулированная выше задача
может быть решена эффективно: прочитав первые $f(1)$ символов
последовательности, мы можем понять, входит ли в неё 1, а прочитав
следующие $f(1)$ символов, сможем сказать даже, входит ли в неё 1
бесконечно много раз. Более того, как следует из
теоремы~\ref{decidability-monadic-GAP}, при таком подходе любое
свойство последовательности $x$, выразимое формулой $\phi(x)$
монадического языка, оказывается разрешимым по~$x$.

Следующие несколько теорем~--- примеры результатов о
неэффективности. Особенно интересно сопоставить утверждения
теоремы~\ref{EAPprefixNumber} и
теоремы~\ref{EAP-preservation-prefix-upper-bound}.
Теорема~\ref{find_pr_by_EAP} также связана с
теоремой~\ref{EAP-preservation-prefix-upper-bound}. Теоремы
\ref{EAPprefixNumber}, \ref{GAP-separate-EAP}, \ref{find_pr_by_EAP},
п.~1) предложения~\ref{GAP-separate-Per-automatic-morphic} (для
класса $\AP$, что в данном случае несущественно) были впервые
объявлены в~\cite{Prit06a-rus} и доказаны в~\cite{Prit06c-rus}.
Схема доказательства теоремы~\ref{GAP-separate-countable},
сформулированная для п.~2)
предложения~\ref{GAP-separate-Per-automatic-morphic}, изложена
в~\cite{Prit07c-rus} (здесь мы приводим полное доказательство).

Под сходимостью $f_n \to f$, где $f_n,f\colon \N \to \N$, будем
понимать условие $\forall i\,\exists n\,\forall{m > n}\,f_m(i) =
f(i)$.

\begin{theorem}[\cite{Prit06a-rus,Prit06c-rus}]\label{EAPprefixNumber}
По $x \in \EAP$ и $f \geqslant \r_x$ невозможно алгоритмически
определять какое-либо $l\geqslant \pr(x)$.
\end{theorem}

Приведём доказательство этой теоремы, чтобы продемонстрировать
используемое во всех теоремах такого типа рассуждение.

Напомним, что последовательность Туэ~--- Морса можно определить как
$\thue = \lim_{n \to \infty} u_n$, где последовательность слов $u_n$
определяется по индукции: $u_0 = 0$ и $u_{n + 1} = u_n\bar u_n$. При
этом $|u_n| = 2^n$. Последовательность $\thue$ не содержит кубов
(подробнее о последовательности $\thue$ см.~раздел~\ref{Thue}). (На
самом деле, как будет видно из доказательства, нам достаточно даже
более слабого утверждения~--- отсутствия кубов специального вида.)

\begin{proof}[Доказательство теоремы~\ref{EAPprefixNumber}]
Для доказательства достаточно построить $x_n\in\EAP$, $x\in\AP$ с
оценками на регуляторы $f_n \geqslant \r_{x_n}$ и $f \geqslant
\r_x$, такие что $x_n\to x$, $f_n\to f$, но $\pr(x_n)\to\infty$.

Действительно, предположим, указанный в условии теоремы алгоритм
существует. Пусть, получив на вход $\langle x, f\rangle$, он выдаёт
число $l\geqslant0$ (оно может быть произвольным, так как
$x\in\AP$). Во время вычисления $l$ алгоритм прочитал лишь конечное
количество символов $x$ и значений $f$, поэтому существует такое $N
> l$, что алгоритм не знает $x(k)$ и $f(k)$ для $k > N$ (формально это
означает, что на всех входах, различие с которыми только в позициях
$k > N$, алгоритм работает так же, как и на $\langle x, f\rangle$).
Поскольку $\pr(x_n)\to\infty$, найдётся $n$, для которого $\pr(x_n)
> N$. Ясно тогда, что на входе $\langle x_n, f_n\rangle$
алгоритм будет работать так же, как и на $\langle x, f\rangle$, и
значит, выдаст $l$, но $\pr(x_n)
> N > l$.

Положим $x = \thue$, $x_n = u_nu_nu_n x$. Тогда
$\pr(x_n)\geqslant2^n$. Действительно, если $\pr(x_n) < 2^n$, то
$u_nu_n x = u_nu_nu_n\bar u_n\bar u_nu_n\ldots\in\AP$, и значит,
$u_nu_nu_n$ входит в $x = \thue$~--- противоречие с
бескубностью~$\thue$.

Осталось показать, что можно выбрать оценку на регуляторы $f_n$,
$f$, так что $f_n\to f$. Для этого достаточно найти общий регулятор
$g$ для всех $x_n$~--- потом мы можем увеличить $g$ и добиться того,
чтобы $g$ была общей оценкой на регулятор для всех $x_n$ и для $x$.
Положим $g = 4\cdot\r_x$. Пусть слово $v$, $|v| = k$ входит в $x_n =
u_nu_nu_nx$ бесконечно много раз. Возьмём отрезок $[i,j]$
последовательности $x$ длины $4\cdot\r_x(k)$ и покажем, что $v$ в
него входит. Если $j\geqslant 3\cdot2^n + \r_x(k)$, то $v$
встречается на отрезке $x[3\cdot2^n, 3\cdot2^n + \r_x(k)]$ (по
определению регулятора). Иначе $j < 3\cdot2^n + \r_x(k)$, откуда $i
\leqslant 3\cdot2^n - 3\r_x(k)$. Но $i\geqslant0$, значит, в этом
случае $\r_x(k) \leqslant 2^n = |u_n|$. Таким образом, $x_n[i, i +
\r_x(k)]$ целиком содержится в $u_nu_n$. Но $u_nu_n$ входит в $x$, и
значит, $x_n[i, i + \r_x(k)]$ входит в $x$, поэтому $v$ входит в
$x[i, i + \r_x(k)]$, так как $v$ входит в $x$ бесконечно много раз,
а значит, входит в любое подслово $x$ длины $\r_x(k)$.

Однако $g$~--- ещё не искомая. Необходимо проследить за словами,
которые встречаются в $x_n$ конечное количество раз. Но ясно, что
если какое-то $v$ входит в $x_n$ конечное количество раз, то $|v| =
k > 2^n$ (иначе $v$ входит в блок из двух последовательных слов
$u_n$ или $\bar u_n$, а значит, и в $x$), то есть такое может быть
лишь для конечного множества различных $n$. Поэтому, рассмотрев все
ситуации, когда слова фиксированной длины $k$ входят в какие-то
$x_n$ конечное количество раз, нам, возможно, придётся увеличить
значение $g(k)$, но лишь конечное количество раз. Значит, искомая
оценка регуляторов найдётся.
\end{proof}

Как мы уже отмечали в разделе~\ref{Classes}, имеем $\EAP \subsetneq
\GAP$ \cite{Prit06b-rus}. Можно показать, что эти классы невозможно
разделять эффективно.

\begin{theorem}[\cite{Prit06a-rus,Prit06c-rus}]\label{GAP-separate-EAP}
По $x\in\GAP$ и $f\geqslant \r_x$ невозможно алгоритмически
определять, верно ли, что $x\in\EAP$.
\end{theorem}

Следующее утверждение говорит о том, что даже приписав к почти
периодической последовательности один символ, мы, вообще говоря, уже
не можем проверять, является ли она по-прежнему почти периодической.

\begin{theorem}[\cite{Prit06a-rus,Prit06c-rus}]\label{find_pr_by_EAP}
По $x \in \EAP$, $f\geqslant \r_x$ и некоторому $l \geqslant \pr(x)$
невозможно алгоритмически находить $\pr(x)$.
\end{theorem}

Следующий довольно общий результат был изначально мотивирован
вопросом о регулярности множества автоматных последовательностей~---
см.~предложение~\ref{GAP-separate-Per-automatic-morphic}, п.~2) и
следствие~\ref{nonregular-Per-automatic-morphic}, п.~2), доказанные
в~\cite{Prit07c-rus}.

\begin{theorem}\label{GAP-separate-countable}
Пусть $\M$~--- счётное множество последовательностей, такое что
$\Per \subseteq \M$. Тогда по $x\in\GAP$ и $f\geqslant \r_x$
невозможно алгоритмически определять, верно ли, что $x \in \M$.
\end{theorem}
\begin{proof}
Достаточно построить последовательность $x_n$ бесконечных слов и
бесконечное слово $x$, так что $x_n$ стремится к $x$, все слова
$x_n$, $x$ обобщённо почти периодичны и имеют общую оценку на
регулятор почти периодичности $g \geqslant \r_{x_n}$, $g \geqslant
\r_x$, причём все $x_n \in \M$, а $x \notin \M$. Действительно
(повторим основную схему рассуждения ещё раз), предположим, алгоритм
из условия теоремы (распознающий принадлежность $\M$ обобщённо почти
периодических последовательностей) существует. Подадим ему на вход
$x$ и $g$. Он выдаст отрицательный ответ, так как $x \notin \M$. Во
время своей работы, перед тем как выдать ответ, алгоритм прочитал
лишь какое-то конечное количество символов из $x$.
Последовательность $x_n$ стремится к $x$, значит, те же самые
символы, которые алгоритм успел прочитать в $x$, стоят на тех же
местах в каком-то $x_m$ для некоторого $m$. Значит, тот же самый
отрицательный ответ алгоритм должен выдать и на $x_m$ и $g$~---
противоречие, так как $x_m \in \M$.

Построим теперь нужные $x_n$, $x$.

В разделе~\ref{Universal} мы описали универсальный метод построения
почти периодических последовательностей. Напомним один из его
вариантов.

Последовательность $\langle B_n, l_n\rangle$, где $B_n$~--- непустое
множество непустых слов в фиксированном конечном алфавите $A$,
$l_n$~--- натуральные числа, является $A$-$\AP$-схемой, если для неё
выполнено для любого $n \in \N$:\\
(1) все слова из $B_n$ имеют длину $l_n$;\\
(2) любое слово $u\in B_{n+1}$ представимо в виде $u = v_1v_2\dots
v_k$, где $v_i\in B_n$, и для любого $w\in B_n$ существует $i$,
такое что $v_i = w$. Последовательность $x$ $\AP$-порождена
$A$-$\AP$-схемой, если для любых $i$ и $n$ имеем $x[il_n, (i+1)l_n -
1] \in B_n$ (далее приставки $\AP$- и $A$-$\AP$- мы опускаем).

Как мы уже говорили, каждая схема порождает какую-то
последовательность, и каждая последовательность, порождённая схемой,
является почти периодической. Более того, каждая почти периодическая
последовательность порождается некоторой схемой.

Теперь мы усилим главное условие на $B_n$, а именно, будем
рассматривать схемы, для которых
верно:\\[1mm]
$(*)$ для любого $n > 1$ каждое $u \in B_n$ имеет вид $u =
v_1v_2\dots v_k$, где $v_i \in B_{n - 1}$, причём для каждых $w_1,
w_2\in B_n$ найдётся $i < k$, такое что~$v_iv_{i+1} = w_1w_2$.

Отметим, что так мы теряем свойство универсальности: очевидно, не
каждая почти периодическая последовательность может быть порождена
схемой, удовлетворяющей условию~$(*)$.

\begin{lemma}
Существует схема, для которой выполнено условие $(*)$, и которая
порождает континуум различных последовательностей.
\end{lemma}
\begin{proof}
Утверждение леммы почти очевидно, но тем не менее докажем его и
построим необходимую $\B$-$\AP$-схему.

Определим рекурсивно схему $\langle l_n, B_n\rangle$: положим $l_0 =
1$ и $B_0 = \{0, 1\}$, и дальше для каждого $n \geqslant 1$ положим
$l_n = (\#B_{n - 1})^2 l_{n - 1}$. Также положим $B_n$ состоящим из
всех слов длины $l_n$, которые удовлетворяют условию~$(*)$.

Теперь объясним, почему эта схема порождает континуум различных
последовательностей. Начнём строить какую-нибудь последовательность,
порождаемую этой схемой. Для этого выберем последовательность слов
$w_i$ рекурсивно следующим образом: слово $w_0$ из множества $B_0$
выберем произвольно, а далее слово $w_i$ для $i \geqslant 1$ будем
выбирать произвольно из множества $B_i$, так чтобы оно являлось
продолжением слова $w_{i - 1}$. Заметим, что длина $l_i$ выбрана
достаточной для того, чтобы для любого $v \in B_{i - 1}$ можно было
найти более одного слова из $B_i$, являющегося продолжением~$v$. Это
значит, что каждый раз, когда мы выбираем $w_i$ для какого-нибудь
$i$, у нас есть хотя бы два различных варианта, независимо от всех
предыдущих выборов. Значит, мы можем выбрать последовательность слов
$w_i$ континуумом различных способов. Осталось только заметить, что
при любом таком выборе предельное бесконечное слово $\lim_{i \to
\infty} w_i$ порождено схемой $\langle l_n, B_n\rangle$ по
построению.
\end{proof}

Как следует из этой леммы, среди порождённых такими усиленными
схемами найдутся последовательности, не принадлежащие $\M$, так как
$\M$ счётно (это единственное место, где используется
счётность~$\M$). Возьмём одну из них~--- $x$, порождённую некоторой
схемой $\langle B_n, l_n\rangle$.

Определим $p_n = x[0,l_n]$. Таким образом, $p_n \in B_n$ и
$\lim_{n\to\infty}p_n = x$. Положим $x_n = p_np_np_n\ldots \in
\Per$. Ясно, что $x_n \to x$, кроме того, все $x_n \in \M$. Осталось
выбрать общий регулятор $f$ для всех~$x_n$. Увеличив его потом, если
нужно, сделаем так, чтобы он подходил и для $x$.

Сейчас мы получим конечное (для каждого $k$) количество требований
вида $f(k) \geqslant \alpha$, после чего в качестве $f(k)$ можем
взять максимум по всем таким $\alpha$.

Пусть $v = x_n[i,j]$, $|v| = k$ (это сразу означает, что $v$ входит
в $x_n$ бесконечно много раз, так как $x_n\in\Per$). Неравенство $k
\geqslant l_{n - 1}$ может выполняться лишь для конечного количества
различных $n$ (при фиксированном $k$), что даёт лишь конечное
количество условий на $f(k)$. Теперь можно считать, что $k < l_{n -
1}$. Возьмём такое $t$, что $l_{t-1} < k \leqslant l_t$ (важно, что
$t$ не зависит от $n$ и однозначно определяется по $k$). Тогда $t
\leqslant n - 1$. Существует $m$, такое что $ml_t \leqslant i$ и
$j\leqslant (m+2)l_t - 1$, то есть $v$ содержится в некотором $ab$,
где $a,b\in B_t$, и значит, благодаря свойству $(*)$, $v$ входит в
любое $c\in B_{t+1}$. Но на каждом отрезке длины $2l_{t+1}$
последовательности $x_n$ найдётся вхождение какого-нибудь $c\in
B_{t+1}$ (целиком входящего в какое-то $p_n$), а значит, и вхождение
$v$. Таким образом, достаточно $f \geqslant 2l_{t+1}$.
\end{proof}

\begin{proposition}\label{GAP-separate-Per-automatic-morphic}
\begin{enumerate}
\item \cite{Prit06a-rus,Prit06c-rus} По
$x \in \GAP$ и $f\geqslant \r_x$ невозможно алгоритмически
определять, верно ли, что $x \in \Per$.

\item \cite{Prit07c-rus} По $x \in
\GAP$ и $f\geqslant \r_x$ невозможно алгоритмически определять,
верно ли, что $x$ автоматна.

\item По $x \in \GAP$ и $f\geqslant \r_x$ невозможно алгоритмически
определять, верно ли, что $x$ морфическая.
\end{enumerate}
\end{proposition}

Из конструкции в доказательстве теоремы~\ref{GAP-separate-countable}
следует существование бесконечного количества периодических
последовательностей $x_n$ с общей верхней оценкой на регулятор почти
периодичности. Эта серия примеров любопытна тем, что одна
естественная мера повторяемости в этих последовательностях~---
оценка на регулятор почти периодичности~--- постоянна, тогда как
другая такая мера~--- период последовательности~--- может быть сколь
угодно большой. По-другому, это замечание можно сформулировать так:
если естественным образом обобщить понятие регулятора почти
периодичности на конечные слова и взять последовательность слов
$p_n$~--- периодов $x_n$, то это будет пример сколь угодно длинных
конечных слов с общей оценкой сверху на регулятор почти
периодичности.

%%%%%%%%%%%%%%%%%%%%%%%%%%%%%%%%%%%%%%%%
%%%%%%%%%%%%%%%%%%%%%%%%%%%%%%%%%%%%%%%%

\subsection{Связь с монадическими теориями}
\label{Monadic-connection}

Любопытным образом оказывается теперь, что из вышеперечисленных
результатов и из теоремы~\ref{decidability-monadic-GAP} можно
получить следствия в терминах монадических теорий следующим образом.

\begin{corollary}
В монадической теории при $x \in \EAP$ невыразимо никакое число $l
\geqslant \pr(x)$ равномерно по~$x$. Другими словами, не существует
такой формулы $\phi(x, l)$ в монадическом языке, которая при $x \in
\EAP$ истинна, только если~$l \geqslant \pr(x)$, причём при каждом
$x \in \EAP$ хотя бы одно такое $l$, что $\phi(x, l)$ истинна,
найдётся.
\end{corollary}
\begin{proof}
Пусть $x \in \EAP$ и пусть существует формула $\phi(x, l)$ из
условия теоремы. Тогда по $x$ и какому-то $f \geqslant \r_x$, можно
перебором найти $l \geqslant \pr(x)$. Для этого нужно перебирать
подряд все натуральные числа $n$ и проверять, верно ли $\phi(x,
n)$~--- это можно проверить по
теореме~\ref{decidability-monadic-GAP}. Таким образом, это дало бы
алгоритмический способ находить $l \geqslant \pr(x)$ по $x$ и $f
\geqslant \r_x$~--- противоречие с теоремой~\ref{EAPprefixNumber}.
\end{proof}

Аналогичным образом получаются и все остальные следствия. Скажем,
что \emph{свойство $P$ выразимо}\label{def-expressible-property} при
$x \in C$ в теории $\MT x$, если существует такая формула $\phi(x)$
в монадическом языке, которая в ограничении на множество $C$
выражает свойство $P$

\begin{corollary}
Если $x \in \GAP$, то в теории $\MT x$ невыразимо свойство $x \in
\EAP$.
\end{corollary}
\begin{proof}
Следует из теоремы~\ref{GAP-separate-EAP}.
\end{proof}

\begin{corollary}
Если $x \in \EAP$, то в теории $\MT x$ с дополнительной константой,
равной какому-то $l \geqslant \pr(x)$, невыразимо число $\pr(x)$.
\end{corollary}
\begin{proof}
Следует из теоремы~\ref{find_pr_by_EAP}.
\end{proof}

\begin{corollary}
\begin{enumerate}
\item Если $x \in \GAP$, то в теории $\MT x$ невыразимо свойство $x \in
\Per$.

\item Если $x \in \GAP$, то в теории $\MT x$
невыразимо свойство ``$x$~--- автоматная последовательность''.

\item Если $x \in \GAP$, то в теории $\MT x$ невыразимо свойство ``$x$~---
морфическая последовательность''.
\end{enumerate}
\end{corollary}
\begin{proof}
Следует из предложения~\ref{GAP-separate-Per-automatic-morphic}.
\end{proof}

\subsection{Нерегулярность некоторых множеств последовательностей}
\label{Nonregularity}

Теперь мы покажем, как, используя теоремы этого раздела, можно
доказать про некоторые множества последовательностей, что они не
распознаются конечным автоматом (имеется в виду автомат Бюхи или
автомат Мюллера, см.~раздел~\ref{Monadic}), то есть не являются
регулярными. Возможно, у этих фактов есть и более прямое
доказательство. В изложении этого раздела мы в существенной степени
следуем работе~\cite{Prit07c-rus}.

Как утверждает теорема~\ref{decidability-monadic-GAP}, по обобщённо
почти периодической последовательности, какой-нибудь оценке сверху
на её регулятор и автомату Бюхи можно алгоритмически определять,
принимает ли данный автомат Бюхи данную последовательность. Говоря
более точно, существует алгоритм, который, получив на вход оракул
для последовательности $x \in \GAP$, оракул для некоторой функции $g
\geqslant \r_x$ и автомат Бюхи $F$, выдаёт ответ, верно ли, что $F$
принимает~$x$.

Предположим, что некоторое множество $\M$ последовательностей
является регулярным, то есть существует принимающий его автомат
Бюхи. Тогда по теореме~\ref{decidability-monadic-GAP} существует
алгоритм, который по обобщённо почти периодической
последовательности и оценке сверху на её регулятор определяет,
принадлежит ли эта последовательность множеству~$\M$.

Таким образом, для того чтобы показать, что множество $\M$
нерегулярно, достаточно показать, что не существует алгоритма,
который по оракулу для обобщённо почти периодической
последовательности и для какой-то оценки сверху на её регулятор
определяет принадлежность этой последовательности к множеству $\M$.
(Обобщённо почти периодические последовательности в этом утверждении
можно заменить на почти периодические.)

Так, из теоремы~\ref{GAP-separate-EAP} и
предложения~\ref{GAP-separate-Per-automatic-morphic} можно получить
следующие результаты.

\begin{corollary}
Множество $\EAP$ не регулярно.\label{nonregular-EAP}
\end{corollary}

\begin{corollary}\label{nonregular-Per-automatic-morphic}
\begin{enumerate}
\item Множество $\Per$ не регулярно.
\item Множество автоматных последовательностей не регулярно.
\item Множество морфических последовательностей не регулярно.
\end{enumerate}
\end{corollary}

В качестве естественного продолжения можно сформулировать следующую
гипотезу.

\begin{conjecture}
Множества $\AP$, $\GAP$, $\Rec$, $\ER$ не регулярны.
\end{conjecture}

Скорее всего, во всех этих случаях можно применить технику,
продемонстрированную в этом разделе.

%%%%%%%%%%%%%%%%%%%%%%%%%%%%%%%%%%%%%%%%%%%%%%%%%%%%%%%%%%%%%%%%%%
%%%%%%%%%%%%%%%%%%%%%%%%%%%%%%%%%%%%%%%%%%%%%%%%%%%%%%%%%%%%%%%%%%
%%%%%%%%%%%%%%%%%%%%%%%%%%%%%%%%%%%%%%%%%%%%%%%%%%%%%%%%%%%%%%%%%%

\section{Сложность почти периодических последовательностей}
\label{Complexity}

В разделе~\ref{Regulator} мы кратко говорим о сложных результате и
гипотезе, связанных с предельным поведением регулятора почти
периодичности последовательности, а в разделе~\ref{Complex_subwords}
обсуждаем сложность последовательностей в терминах колмогоровской
сложности и подсловной сложности.

%%%%%%%%%%%%%%%%%%%%%%%%%%%%%%%%%%%%%%%%%%%%%%%%%%%%%%%%%%%%%%%%%%

\subsection{Предельное поведение регулятора почти периодичности}
\label{Regulator}

Сложность почти периодической последовательности можно понимать как
сложность устройства её регулятора почти периодичности. Следующий
результат стал положительными ответом на долгое время остававшийся
открытым вопрос Морса и Хедлунда, сформулированный
в~\cite{MorseHedl38}.

\begin{theorem}[\cite{CassChekh06}]\label{aperiodic-non-limit}
Если последовательность $x$ апериодическая, то не существует
конечного предела последовательности $\frac{\r_x(n)}n$ при $n \to
\infty$.
\end{theorem}

Удобно кроме регулятора почти периодичности последовательности
определить также функцию $\rd_x(n) = \r_x(n) - n + 1$, которая равна
максимальному расстоянию между соседними вхождениями слова $u$ длины
$n$ (расстояние берётся между левыми концами вхождений, а максимум
берётся по всем словам длины $n$ и по всем расстояниям).
Следуя~\cite{Cass01}, определим \emph{коэффициент почти
периодичности}\label{def-almost-periodicity-quotient} $\rho_x$
последовательности $x$ как $\rho_x = \limsup_{n \to \infty}
\frac{\rd_x(n)}{n}$. Несложно видеть, что для периодической
последовательности $x$ коэффициент почти периодичности равен 1.
Кроме того, последовательность линейно почти периодическая
(см.~с.~\pageref{def-linear-almost-periodic}) тогда и только тогда,
когда её коэффициент почти периодичности конечен.

В~\cite{MorseHedl38} поставлен вопрос нахождения наилучшей нижней
оценки на $\rd_x$ для непериодических $x$. Из
предложения~\ref{periodicity-subword-complexity} несложно видеть,
что $\rho_x \geqslant 2$ для непериодических~$x$. В~\cite{Cass01}
доказано, что $\rho_x \geqslant 3$ для непериодических~$x$. Рози и
Кассэнь выдвигают следующую гипотезу.

\begin{conjecture}[\cite{Cass01,Rauzy83}]\label{aperiodic-lower-bound}
Для любой непериодической последовательности $x$ выполнено $\rho_x
\geqslant \frac{5 + \sqrt 5}2 = 3{,}618\dots$
\end{conjecture}

Поскольку для последовательности Фибоначчи $\rho_\fib = \frac{5 +
\sqrt 5}2$ (см.~\cite{Cass01}), доказательство этой гипотезы стало
бы полным ответом на упомянутый вопрос из~\cite{MorseHedl38}.

Основным средством в доказательстве
теоремы~\ref{aperiodic-non-limit}, а также в предполагаемом уже
существующем, но пока не опубликованном доказательстве
гипотезы~\ref{aperiodic-lower-bound} (см. комментарии
в~\cite{Cass01}) являются графы Рози последовательностей, широко
используемые в комбинаторике слов (определение и некоторые свойства
можно найти, например, в~\cite{CassChekh06}).

%%%%%%%%%%%%%%%%%%%%%%%%%%%%%%%%%%%%%%%%%%%%%%%%%%%%%%%%%%%%%%%%%%

\subsection{Последовательности со сложными подсловами}
\label{Complex_subwords}

\emph{Колмогоровская сложность}\label{def-Kolmogorov-complexity}
конечного слова, неформально,~--- это количество информации в этом
слове. Более формально, это длина кратчайшего бинарного описания
этого слова, при таком способе описания, при котором слова
получаются из своих коротких описаний применением универсального
алгоритма. По-другому, колмогоровская сложность слова~--- это длина
кратчайшей закодированной в бинарном алфавите программы без входа в
каком-нибудь естественном языке программирования, печатающей это
слово. Колмогоровская сложность определена с точностью до аддитивной
константы (зависящей от языка программирования или конкретного
универсального алгоритма, который берётся в определении). Более
детально с фундаментальным понятием колмогоровской сложности можно
ознакомиться по~\cite{LiVit97} или~\cite{UspShenVer-rus} (см.
также~\cite{Prit09a}).

Мы обозначаем через $K(u)$ сложность слова~$u$. В этом разделе все
конечные слова и бесконечные последовательности предполагаются
бинарными. Одно из базовых свойств колмогоровской сложности, которое
нам понадобится, следующее: для всех слов $u \in \B^*$ выполнено
$K(u) < |u| + C$, поскольку каждое слово можно описать как минимум
им самим.

Хотя колмогоровская сложность измеряет количество информации в
словах (и в других конечных объектах), и следовательно, её
применение и связи с комбинаторикой слов представляются любопытными
и естественными, очень мало известно таких связей. Здесь можно
отметить отметить работу~\cite{RumUsh06} (см. обсуждение дальше),
\cite{Rum07}, где найдено новое доказательство существования
последовательностей с произвольным значением критической экспоненты
(изначально результат получен в \cite{KrieShal07}) и, наконец,
раздел 5 работы~\cite{Prit07b}, расширенной версией которого и
является настоящий раздел.

Основной целью данного раздела является обсуждение следующей
гипотезы.

\begin{conjecture}[Ан.~Мучник, 2005]
\label{Muchnik-complex_almost_periodic-conjecture} Для любого $0 <
\alpha < 1$ существует почти периодическая последовательность $x$,
для которой одновременно выполнено следующее:
\begin{enumerate}
\item некоторая верхняя оценка на $\r_x$ вычислима;
\item неравенство $K(u) > \alpha|u|$ выполнено для любого
подслова~$u$ последовательности~$x$.
\end{enumerate}
\end{conjecture}

Гипотеза не доказана даже хотя бы для какого-нибудь
фиксированного~$\alpha$.

Свойства последовательности из этой гипотезы объединяют в себе до
некоторой степени вычислимость (вычислимый регулятор почти
периодичности) и до некоторой степени случайность (почти
максимальная возможная колмогоровская сложность подслов). Однако эта
последовательность не может быть ни по-настоящему вычислимой (это
означало бы логарифмическую сложность всех начал), ни по-настоящему
случайной в смысле Мартин--Лёфа (это означало бы, что в неё входят
все возможные слова, в том числе простые).

Было получено много результатов, близких к
гипотезе~\ref{Muchnik-complex_almost_periodic-conjecture}. Следующее
утверждение является полезным средством.

\begin{theorem}[\cite{DurLevShen01}]\label{Levin-lemma}
Для любого $0 < \alpha < 1$ существует последовательность $x \in
\B^\N$, такая что неравенство $K(u) > \alpha|u|$ выполнено для всех
подслов $u$ последовательности~$x$.
\end{theorem}

Так как по предложению~\ref{minimality-property} для каждой
последовательности существует почти периодическая
последовательность, все подслова которой являются подсловами
исходной, в качестве следствия из теоремы~\ref{Levin-lemma} выводим
существование почти периодической последовательности со сложными
подсловами. Теперь задача в том, чтобы найти такую
последовательность с вычислимой верхней оценкой на регулятор почти
периодичности.

На самом деле, несколько другая, более слабая, характеристика почти
периодической последовательности может быть здесь сделана вычислимо
оцениваемой сверху. Пусть $x \in \AP$. Определим $\r'_x(n)$ как
минимальное такое $l$, что слово $x[0, n - 1]$ входит в любое
подслово длины $l$ в~$x$ (см. также~\cite{Cass01}). Очевидно, $\r_x
\geqslant \r'_x$, поэтому для любого $x \in \AP$ если какая-то
верхняя оценка на $\r_x$ вычислима, то то же верно и про $\r'_x$. По
существу, следующее утверждение доказано в \cite{RumUsh06}.

\begin{theorem}[\cite{RumUsh06}]
\label{Rumyantsev-Ushakov-complex_almost_periodic} Для любого $0 <
\alpha < 1$ существует почти периодическая последовательность $x$,
для которой одновременно выполнено следующее:
\begin{enumerate}
\item какая-то верхняя оценка на $\r'_x$ вычислима;
\item неравенство $K(u) > \alpha|u|$ выполнено для любого
подслова~$u$ последовательности~$x$.
\end{enumerate}
\end{theorem}

При доказательстве
теоремы~\ref{Rumyantsev-Ushakov-complex_almost_periodic}
используется следующий простой метод\label{Rumyantsev-method}
построения почти периодических последовательностей. Пусть $n_i$~---
возрастающая последовательность натуральных чисел, такая что $n_i |
n_{i + 1}$ для любого $i$. Возьмём теперь произвольную
последовательность $x$ и изменим в ней некоторые элементы, так чтобы
она стала почти периодической. Первым шагом заменим все отрезки вида
$x[in_1,in_1 + n_0 - 1]$ для $i \geqslant 1$ на отрезок $x[0,n_0 -
1]$. Далее заменим все отрезки вида $x[in_2,in_2 + n_1 - 1]$ для $i
\geqslant 1$ на отрезок $x[0,n_1 - 1]$. И так далее: на $k$-м шаге
процедуры мы заменяем все отрезки вида $x[in_{k + 1},in_{k + 1} +
n_k - 1]$ для $i \geqslant 1$ на отрезок $x[0,n_k - 1]$. За счёт
условия $n_i | n_{i + 1}$ для любого $i$ все переписывания друг с
другом согласованны: если какой-то символ на некотором шаге был
переписан, то дальше он уже переписан не будет. Отсюда следует, что
последовательность, полученная такой процедурой, будет почти
периодической: мы обеспечили достаточную встречаемость для каждого
префикса. Более того, мы обеспечили каждому префиксу встречаемость
по арифметической прогрессии, из чего следует, что каждая
последовательность, полученная по вышеописанной процедуре, будет
точно почти периодической. Однако так получаются не все точно почти
периодические последовательности.

%Как показывает следующий (довольно естественно ожидаемый) результат,
%теорема~\ref{Rumyantsev-Ushakov-complex_almost_periodic} не
%позволяет вывести
%гипотезу~\ref{Muchnik-complex_almost_periodic-conjecture}.
%
%\begin{theorem}
%Существует почти периодическая последовательность $x$, для которой
%какая-то верхняя оценка на $\r'_x$ вычислима, но никакая верхняя
%оценка на $\r_x$ не вычислима.
%\end{theorem}
%\begin{proof}
%\end{proof}

Другое ослабление гипотезы Мучника может быть получено, если мы
потребуем близости к случайности не от всех подслов
последовательности, а только от всех префиксов. По существу,
следующий результат получен в \cite{MuchSemUsh03}.

\begin{theorem}[\cite{MuchSemUsh03}]
\label{Muchnik-Semenov-Ushakov-complex_almost_periodic} Для любого
$0 < \alpha < 1$ существует почти периодическая последовательность
$x$, для которой одновременно выполнено следующее:
\begin{enumerate}
\item какая-то верхняя оценка на $\r_x$ вычислима;
\item неравенство $K(u) > \alpha|u|$ выполнено для любого
префикса~$u$ последовательности~$x$.
\end{enumerate}
\end{theorem}

Техника, с помощью которой этот результат доказывается в
работе~\cite{MuchSemUsh03}, использует универсальный метод
построения почти периодических последовательностей (см.
раздел~\ref{Universal}). Это один из возможных способов пытаться
доказать гипотезу~\ref{Muchnik-complex_almost_periodic-conjecture}.
(Строго говоря, в работе~\cite{MuchSemUsh03} дана конструкция,
использующая $\AP$-схемы, которая, по-видимому, не даёт вообще
говоря оценки на регулятор почти периодичности~---
см.~гипотезу~\ref{scheme-non-computable-regulator}. Но эту
конструкцию можно модифицировать, заменив использование $\AP$-схем
$\GAP$-схемами и тем самым обеспечив вычислимость оценки на
регулятор.)

Естественное понятие подсловной сложности часто используется для
характеризации конечных слов и бесконечных последовательностей.
Следующие леммы описывают связи между колмогоровской и подсловной
сложностью.

\begin{lemma}\label{subword-greater-than-Kolm}
Для произвольной $x \in \B^\N$ для любого $n$ существует такое $C$,
что для любого $m \geqslant n$ для каждого $u \in \Fac_m(x)$
выполнено $K(u) \leqslant \lceil \frac mn\rceil \log(p_x(n)) + C$.
\end{lemma}
\begin{proof}
Возьмём слово длины $m$, входящее в последовательность. Разделим его
на блоки длины $n$ (возможно, последний блок будет неполной
длины)~--- получится $\lceil \frac mn\rceil$ блоков. Каждый из
получившихся блоков~--- слово длины $n$ (последнее, возможно, короче
$n$), входящее в последовательность $x$. Всего в последовательность
$x$ входит $p_x(n)$ слов длины $n$, поэтому на задание каждого из
блоков достаточно потратить $\log(p_x(n))$ битов. Кроме того, для
задания длины последнего блока достаточно $\log\log(p_x(n))$ битов,
что есть константа от~$m$.
\end{proof}

\begin{lemma}\label{Kolm-greater-than-subword}
Для произвольной $x \in \B^\N$ для любого $n$ существует $u \in
\Fac_n(x)$, такое что $K(u) \geqslant \log(p_x(n))$.
\end{lemma}
\begin{proof}
Всего слов, имеющих сложность меньше $\log(p_x(n))$, не больше, чем
$2^{\log(p_x(n))} - 1 = p_x(n) - 1$, поэтому хотя бы одно из слов
множества $\Fac_n(x)$ имеет сложность не меньше, чем $\log(p_x(n))$.
\end{proof}

Предел $\lim_{n \to \infty} \frac1n \log(p_x(n))$ существует для
любой последовательности $x$ и называется её \emph{топологической
энтропией}\label{def-topological-entropy}
(см.~\cite{BealPerr97,Fer99}); мы обозначаем эту величину через
$E_t(x)$. Это действительное число между 0 и 1, некоторым образом
описывающее, насколько детерминирована последовательность: чем ближе
$E_t(x)$ к 0, тем более детерминирована последовательность~$x$ и тем
менее хаотична. Если пытаться определять аналогичную числовую
характеристику в терминах колмогоровской сложности, то естественным
выбором будет $E_k(x) = \lim_{n \to \infty} \frac1n \max\{K(u) : u
\in \Fac_n(x)\}$. Из этого определения неясно даже, когда существует
$E_k(x)$, но из лемм~\ref{subword-greater-than-Kolm}
и~\ref{Kolm-greater-than-subword} можно вывести следующее.

\begin{theorem}\label{topological-Kolmogorov}
Для любой последовательности $x$ число $E_k(x)$ существует и
равно~$E_t(x)$.
\end{theorem}
\begin{proof}
Из леммы~\ref{Kolm-greater-than-subword} сразу следует, что
$\frac1n\max\{K(u) : u \in \Fac_n(x)\} \geqslant \frac1n
\log(p_x(n))$, откуда $\liminf_{n \to \infty}\frac1n\max\{K(u) : u
\in \Fac_n(x)\} \geqslant E_t(x)$.

Теперь зафиксируем $n$. Тогда для некоторой константы $C$ для любого
$m \geqslant n$ для любого $u \in \Fac_m(x)$ имеем $K(u) \leqslant
\lceil \frac mn\rceil \log(p_x(n)) + C$. Отсюда $\frac1m\max\{K(u) :
u \in \Fac_m(x)\} \leqslant \frac1m\lceil \frac mn\rceil
\log(p_x(n)) + \frac Cm$. Следовательно, $\limsup_{m \to \infty}
\frac1m\max\{K(u) : u \in \Fac_m(x)\} \leqslant \frac1n\log(p_x(n))$
для любого $n$, откуда следует, что $\limsup_{m \to \infty}
\frac1m\max\{K(u) : u \in \Fac_m(x)\} \leqslant E_t(x)$.

Отсюда получаем, что $E_t(x) \leqslant \liminf_{n \to
\infty}\frac1n\max\{K(u) : u \in \Fac_n(x)\} $\\ $\leqslant
\limsup_{n \to \infty} \frac1n\max\{K(u) : u \in \Fac_n(x)\}
\leqslant E_t(x)$. Следовательно, $E_k(x) = E_t(x)$.
\end{proof}

\begin{corollary}
Для любого $0 < \alpha < 1$ существует почти периодическая
последовательность $x$ с вычислимой оценкой сверху на регулятор
почти периодичности, такая что $E_t(x) > \alpha$.
\end{corollary}
\begin{proof}
Действительно, последовательность из
теоремы~\ref{Muchnik-Semenov-Ushakov-complex_almost_periodic}
подходит, что следует из теоремы~\ref{topological-Kolmogorov}.
\end{proof}

%Другим важным следствием леммы Левина и
%леммы~\ref{Kolm-greater-than-subword} является тот факт, что для
%любого $0 < \alpha < 1$ существует почти периодическая $x$, для
%которой $E_t(x) > \alpha$. Более того, применяя свойство
%минимальности к
%теореме~\ref{Muchnik-Semenov-Ushakov-complex_almost_periodic} и
%пользуясь теоремой~\ref{topological-Kolmogorov}, получаем, что такая
%последовательность может быть выбрана с вычислимо оцениваемым сверху
%регулятором почти периодичности.

Однако, гипотеза Мучника до сих пор остаётся открытой.

%%%%%%%%%%%%%%%%%%%%%%%%%%%%%%%%%%%%%%%%%%%%%%%%%%%%%%%%%%%%%%%%%%
%%%%%%%%%%%%%%%%%%%%%%%%%%%%%%%%%%%%%%%%%%%%%%%%%%%%%%%%%%%%%%%%%%
%%%%%%%%%%%%%%%%%%%%%%%%%%%%%%%%%%%%%%%%%%%%%%%%%%%%%%%%%%%%%%%%%%

\section{Автоматные и морфические последовательности.
Соотношения с почти периодичностью} \label{Automatic-and-Morphic}

%%%%%%%%%%%%%%%%%%%%%%%%%%%%%%%%%%%%%%%%%%%%%%%%%%%%%%%%%%%%%%%%%%

\subsection{Автоматные последовательности}
\label{Automatic}

Любая заключительно периодическая последовательность, например, $ 3
1 4 2 8 5 7 1 4 2 8 5 7 1 4 2  \dots$ (цифры десятичной записи числа
22/7) может быть порождена машиной с конечной памятью. Достаточно
иметь информацию о предпериоде и периоде последовательности: в нашем
случае 3 и 142857. И наоборот, любая машина с конечной памятью,
печатающая символы конечного алфавита, порождает заключительно
периодическую последовательность. Действительно, во время её работы
в какой-то момент конфигурация машины полностью совпадёт с какой-то
из уже встречавшихся, и так начнётся период в выходной
последовательности.

Если разрешить машине обращаться к символам, написанным ранее, класс
порождаемых последовательностей существенно возрастёт. При некоторых
ограничениях на порядок, в котором ранее написанные символы читаются
снова, получаем класс автоматных последовательностей, введённый
Кобхэмом в~\cite{Cobh72}.

Например, последовательность Туэ~--- Морса (см.~раздел~\ref{Thue})
$\thue = 0110100110010110\dots$ удовлетворяет следующему условию:
если на $n$-м месте стоит 0, то на $2n$-м и $(2n + 1)$-м будет 0 и 1
соответственно, а если на $n$-м месте 1, то, наоборот, на $2n$-м и
$(2n + 1)$-м будет 1 и 0 соответственно.

Кобхэм вводит иерархию на автоматных последовательностях, в
зависимости от того, как далеко надо заглядывать назад, чтобы узнать
текущий символ. В $k$-автоматных последовательностях $n$-й символ
определяет, что стоит на местах $kn, kn + 1, kn + 2, \dots, (k + 1)n
- 1$.

Более формально, автоматные последовательности можно определить
следующим образом. На самом деле, мы дадим даже два (эквивалентных)
определения.

Рассмотрим конечный автомат, действующий на словах в алфавите
$\Sigma_k = \{0, 1, \dots, k-1\}$. Каждому состоянию автомата
соответствует буква в некотором другом алфавите~$A$ (разным
состояниям могут соответствовать одинаковые буквы). Автомат
действует так: получает на вход слово в алфавите $\Sigma_k$,
производит вычисления и выдаёт ту букву алфавита $A$, которая
соответствует последнему состоянию в вычислении. Последовательность
$x$ букв алфавита $A$ называется
\emph{$k$-автоматной}\label{def-automatic}, если существует конечный
автомат вышеуказанного вида, который, будучи запущенным на числе
$n$, записанном в $k$-ичной системе счисления, выдаёт букву~$x(n)$.
Когда ясно или неважно, о каком $k$ идёт речь, приставку ``$k$-'' мы
будем опускать.

Теперь опишем другой подход.

Напомним, что отображение $\phi\colon A^*\to B^*$ называется
морфизмом ($A$, $B$~--- конечные алфавиты), если для любых $u,v\in
A^*$ выполнено $\phi(uv) = \phi(u)\phi(v)$. Ясно, что морфизм
полностью определяется своими значениями на однобуквенных словах.
Морфизм называется $k$-равномерным, если $|\phi(a)| = k$ для всех $a
\in A$. 1-равномерный морфизм называется кодированием.

В этом разделе мы рассматриваем морфизмы вида $A^* \to A^*$. Пусть
$\phi(s) = su$ для некоторых $s\in A$, $u\in A^*$. Тогда для всех
натуральных $m < n$ слово $\phi^n(s)$ начинается со слова
$\phi^m(s)$, так что можно корректно определить $\phi^\infty(s) =
\lim_{n \to \infty} \phi^n(s) = su\phi(u)\phi^2(u)\phi^3(u)\dots$
Если $\forall n\ \phi^n(u) \ne \Lambda$, то $\phi^\infty(s)$
бесконечна. В этом случае говорят, что морфизм $\phi$
\emph{продолжаем на~$s$}\label{def-prolongable}. Последовательности
вида $h(\phi^\infty(s))$, где $h\colon A \to B$~--- кодирование,
называются \emph{морфическими}\label{def-morphic}, вида
$\phi^\infty(s)$~--- \emph{чисто
морфическими}\label{def-pure-morphic}.

\begin{theorem}[\cite{Cobh72,AlShall03}]\label{automatic-equivalence}
Последовательность $k$-автоматна тогда и только тогда, когда она
является морфической, полученной из $k$-равномерного морфизма.
\end{theorem}

Благодаря теореме~\ref{automatic-equivalence} мы имеем два
эквивалентных определения автоматных последовательностей.

Из определения видно, что автоматная последовательность может быть
конечно описана~--- достаточно либо задать конечный автомат, который
её порождает, либо задать порождающий равномерный морфизм, букву, с
которой надо начинать итерировать, и кодирование. Это значит, что
можно корректно формулировать вопросы о разрешимости различных
свойств автоматных последовательностей.

В контексте этой статьи для нас наиболее интересным представляется
сравнивать различные определения последовательностей, близких к
периодическим. Ясно, например, что существуют почти периодические
последовательности, не являющиеся автоматными~--- почти
периодических последовательностей континуум, в то время как
автоматных, очевидно, счётное количество. И наоборот, существуют
автоматные последовательности, не являющиеся почти периодическими.
Действительно, последовательность
011010001000000010000000000000001\dots, у которой на каждом $2^n$-м
месте стоит 1, а на остальных 0, 2-автоматна (легко построить
автомат, отделяющий числа вида 100\dots0 в двоичной системе
счисления от всех остальных), но не является почти периодической.

Естественно задать вопрос о разрешимости свойства почти
периодичности для автоматных последовательностей. Оказывается, на
него можно ответить положительно.

\begin{theorem}[\cite{Prit07a,NicPrit07}]\label{automatic-almper-decidable}
Существует полиномиальный по времени алгоритм, который по автоматной
последовательности определяет, является ли она почти периодической.
\end{theorem}

Близкими к периодическим автоматные последовательности можно назвать
и ещё по одной причине. Напомним, что подсловной сложностью
последовательности $x$ называется такая функция $\p\colon \N \to
\N$, что $\p(n)$ равно количеству слов длины $n$, входящих в
последовательность~$x$. Для последовательностей в алфавите из $m$
символов подсловная сложность может варьироваться от 1 до $m^n$. Как
доказано в~\cite{MorseHedl38} (см.
предложение~\ref{periodicity-subword-complexity}), подсловная
сложность последовательности ограничена тогда и только тогда, когда
последовательность является заключительно периодической.
Оказывается, что подсловная сложность автоматных последовательностей
тоже не слишком велика.

\begin{theorem}[\cite{Cobh72}]\label{automatic-subwcompl-linear}
Подсловная сложность автоматной последовательности не более чем
линейна.
\end{theorem}

Тем не менее, теорема~\ref{automatic-highAM} показывает, что
автоматные последовательности могут быть достаточно далеки от
периодических.

Интересно также следующее свойство замкнутости для автоматных
последовательностей.

\begin{theorem}[\cite{Cobh72}]\label{automatic-preservation}
Множество $k$-автоматных последовательностей замкнуто относительно
равномерных конечно-автоматных преобразований.
\end{theorem}

О конечно-автоматных преобразованиях
см.~раздел~\ref{Automata_Mappings}. (Тут важно отметить, что
``автоматы'' в ``конечно-автоматных преобразованиях'' и ``автоматных
последовательностях'' имеют разный смысл.) Доказательство
теоремы~\ref{automatic-preservation} можно найти также
в~\cite{AlShall03}.

%%%%%%%%%%%%%%%%%%%%%%%%%%%%%%%%%%%%%%%%%%%%%%%%%%%%%%%%%%%%%%%%%%

\subsection{Мера апериодичности}
\label{Aperiodicity-measure}

Периодические последовательности имеют самую простую структуру среди
всех последовательностей над конечным алфавитом, поэтому естественно
было бы научиться измерять, насколько данная последовательность
далека от любой периодической. В~\cite{PritUl09} вводится понятие
меры апериодичности последовательности. Неформально, мера
апериодичности последовательности~--- это максимальное такое число
$\alpha$, что последовательность с любым своим нетривиальным сдвигом
имеет хотя бы долю $\alpha$ различий.

Формальное определение этого понятия основано на дискретном аналоге
расстояния Безиковича $d_B$, использованном в~\cite{MorseHedl38} для
определения почти периодических по Безиковичу последовательностей,
см.~с.~\pageref{def-Besicovitch-distance}. Тот же подход был
использован в~\cite{DurRomShen08} для определения
$\alpha$-апериодических последовательностей. По существу,
в~\cite{DurRomShen08} было замечено, в нашей терминологии, что если
$\am(x)> \alpha$ для последовательности $x$, то $x$ находится на
расстоянии хотя бы $\alpha/2$ от любой заключительно периодической
последовательности, в смысле расстояния $d_B$.

Напомним (см.~раздел~\ref{Introduction}), что расстояние Безиковича
определяется как $d_B(x, y) = \liminf \frac1n \#\{i: 0 \leqslant i
\leqslant n - 1, x(i) \ne y(i)\}$. Определим \emph{меру
апериодичности}\label{def-aperiodicity-measure} $\am(x) =
\inf\{d_B(x,L^nx) : n \geqslant 1\}$, где $L$ обозначает операцию
левого сдвига. Другими словами, $\am(x)$~--- это максимальное такое
число от 0 до 1, что можно утверждать, что у последовательности $x$
с любым её собственным сдвигом хотя бы доля $\am(x)$ символов
различны.

Одним из мотивов для изучения понятия меры апериодичности стала
следующая гипотеза Б.~Дюрана и А.~Шеня: \emph{Для любого $\alpha <
1$ существует автоматная последовательность $x$, такая что $\am(x)
\geqslant \alpha$}. Эта гипотеза доказана в~\cite{PritUl09},
см.~теорему~\ref{automatic-highAM}. Теорема~\ref{automatic-highAM}
позволяет упростить конструкцию сильно апериодического замощения
из~\cite{DurRomShen08}.

%\begin{conjecture}
%\label{automatic-highAM-conjecture}
%Для любого $\alpha < 1$
%существует автоматная последовательность $x$, такая что $\am(x)
%\geqslant \alpha$.
%\end{conjecture}

Кроме того, понятие меры апериодичности достаточно естественно и
просто, чтобы представлять самостоятельный интерес.

Ясно, что если последовательность $x$ заключительно периодическая с
периодом $p$, то $d_B(x,L^px) = 0$ и, следовательно, $\am(x) = 0$.
Обратное не всегда верно~--- несложно привести пример апериодической
последовательности, мера апериодичности которой равна~0. Тем не
менее естественно считать, что чем меньше число $\am(x)$, тем
ближе~$x$ к периодической последовательности. В~\cite{PritUl09}
доказано, что $\am(x) = 0$, если $x$~--- последовательность Штурма
(теорема~\ref{AM-Sturmian}).

Если в вышеупомянутой гипотезе не требовать автоматности
последовательности, то утверждение становится почти очевидным.

\begin{theorem}[\cite{PritUl09}]\label{highAM}
Пусть $x$~--- случайная последовательность в алфавите из $k$
символов. Тогда $\am(x) = \frac1k$.
\end{theorem}

%В теореме~\ref{highAM} случайность можно понимать разными способами,
%например, как случайную по Мартин--Лёфу последовательность
%(см.~\cite{UspShenVer-rus}). (Обычно случайные по Мартин--Лёфу
%последовательности определяют для алфавита из двух символов, но это
%можно сделать и для любого конечного алфавита, с сохранением многих
%свойств.)

Несложно показать, что для фиксированного алфавита можно дать и
верхнюю оценку на меру апериодичности.

\begin{theorem}[\cite{PritUl09}]\label{AM-upper-bound}
Если последовательность $x$ состоит из не более чем $k$ символов, то
$\am(x) \leqslant 1 - \frac1{2k}$.
\end{theorem}

Интересно также посчитать меру апериодичности некоторых хорошо
известных последовательностей.

\begin{theorem}[\cite{PritUl09}]\label{AM-Thue}
$\am(\thue) = \frac13$.
\end{theorem}

Заметим, что по существу в~\cite{MorseHedl38} было доказано
$\am(\thue) \geqslant 1/4$.

\begin{theorem}[\cite{PritUl09}]\label{AM-Sturmian}
Если $x$~--- последовательность Штурма, то $\am(x) = 0$.
\end{theorem}

Теперь приведём конструкцию автоматной последовательности с мерой
апериодичности, сколь угодно близкой к~1.

Пусть $k \geqslant 3$ и $\phi\colon \{0,\dots,k - 1\}^* \to
\{0,\dots,k - 1\}^*$~--- морфизм, такой что $(\phi(i))(j) = i + 1 +
2 + \dots + (j - 1) + j$ для $0 \leqslant i,j \leqslant k - 1$, где
$+$ всегда берётся по модулю $k$, а $u(i)$ обозначает $i$-й символ
конечного слова~$u$. Положим $x_k = \phi^\infty(0)$. Например, если
$k = 5$, то $\phi$ получается таким:
\begin{equation*}
\begin{split}
&\phi(0) = 01310 \\
&\phi(1) = 12421 \\
&\phi(2) = 23032 \\
&\phi(3) = 34143 \\
&\phi(4) = 40204,
\end{split}
\end{equation*}
и $x_5 = 01310 12421 34143 12421 01310 12421\dots$

Можно доказать (см. \cite{PritUl09}), что если $k$ простое, то
$\am(x_k) = 1 - \frac2k$. Таким образом, получаем следующий
результат.

\begin{theorem}[\cite{PritUl09}]\label{automatic-highAM}
Для любого $\alpha < 1$ существует автоматная
последовательность~$x$, такая что $\am(x) \geqslant \alpha$.
\end{theorem}

В заключение приведём интересные открытые вопросы о мере
апериодичности бесконечных последовательностей.

\begin{problem}
1) \label{AM-fixed-alphabet-problem} Для каждого $k \geqslant 2$
каково максимальное значение $\am(x)$ для $x$ в алфавите из $k$
символов (интересно сравнить теорему~\ref{highAM} с
теоремой~\ref{AM-upper-bound})? Мы предполагаем, что это значение
равно $1 - \frac1k$.\\
2) \label{AM-fixed-alphabet-automatic-problem} Для каждого $k
\geqslant 2$ каково максимальное значение $\am(x)$ для автоматных
последовательностей $x$ в алфавите из $k$ символов?\\
3) \label{AM-k-automatic-problem} Для каждого $k \geqslant 2$ каково
максимальное значение $\am(x)$ для $k$-автоматных
последовательностей~$x$?
\end{problem}

\begin{problem}
Для каких последовательностей $x$ можно взять предел вместо верхнего
предела в выражении $d_B(x,L^nx) = \limsup \frac1m \#\{i \in [m -
1]: x(i) \ne x(i + n)\}$, то есть когда этот предел существует? В
частности, верно ли это для автоматных последовательностей? Верно
ли, что если этот предел существует для $n = 1$, то он существует
для всех~$n$?
\end{problem}

\begin{problem}
Посчитайте меру апериодичности других известных последовательностей
и классов последовательностей, например, последовательностей Тёплица
(см.~раздел~\ref{Toeplitz}), каких-нибудь автоматных и морфических
последовательностей (см.~разделы~\ref{Automatic} и \ref{Morphic}),
каких-нибудь обобщений последовательностей Штурма
(см.~раздел~\ref{Sturm}), и т.~д.
\end{problem}

%%%%%%%%%%%%%%%%%%%%%%%%%%%%%%%%%%%%%%%%%%%%%%%%%%%%%%%%%%%%%%%%%%

\subsection{Морфические последовательности}
\label{Morphic}

В разделе~\ref{Automatic} мы ввели понятие автоматных
последовательностей. Один из способов их определить использует
морфические последовательности, порождённые равномерными морфизмами.
В качестве естественного обобщения можно рассматривать морфические
последовательности, порождённые морфизмами произвольного вида.

Класс морфических последовательностей, очевидно, шире класса
автоматных последовательностей. Например, последовательность
Фибоначчи $\fib = 01 0 01 010 01001\dots$ (см. раздел~\ref{Sturm})
является чисто морфической, порождённой морфизмом $0 \to 01$, $1 \to
0$, но не является автоматной.

Ясно также, что класс морфических последовательностей находится в
общем положении с классом почти периодических последовательностей.
Как и в разделе~\ref{Automatic}
(см.~теорему~\ref{automatic-almper-decidable}), возникает
естественный вопрос о разрешимости почти периодичности для
морфических последовательностей. Однако в этом случае проблема до
сих пор не решена.

\begin{conjecture}
\label{morphic-almost-periodic-decision-algorithm-conjecture}
Существует алгоритм, который по морфической последовательности
определяет, является ли она почти периодической.
\end{conjecture}

В одном из частных случаев~--- случае автоматных
последовательностей~--- ответ на вопрос о разрешимости положительный
(теорема~\ref{automatic-almper-decidable}). Ответ положительный и в
другом частном случае чисто морфических последовательностей.

\begin{theorem}[\cite{Prit07a,NicPrit07}]\label{puremorphic-almper-decidable}
Существует полиномиальный по времени алгоритм, который по чисто
морфической последовательности определяет, является ли она почти
периодической.
\end{theorem}

Во многом свойства морфических последовательностей проще свойств
почти периодических. Например, монадическая теория морфической
последовательности всегда разрешима (о монадических теориях
см.~раздел~\ref{Monadic}).

\begin{theorem}[\cite{CarThom02}]\label{morphic-monadic-decidable}
Монадическая теория морфической последовательности разрешима.
\end{theorem}

Теорема~\ref{morphic-monadic-decidable} сформулирована и доказана
в~\cite{CarThom02}, однако несложно следует и из более ранних
результатов. Действительно, в~\cite{Dekk94} доказано следующее
утверждение.

\begin{theorem}[\cite{Dekk94}]\label{morphic-preservation}
Класс морфических последовательностей замкнут относительно
конечно-автоматных преобразований.
\end{theorem}

О конечно-автоматных преобразованиях
см.~раздел~\ref{Automata_Mappings}. Доказательство
теоремы~\ref{morphic-preservation} можно найти также
в~\cite{AlShall03}. Из доказательства несложно видеть, что
утверждение этой теоремы эффективно, то есть по конечно-автоматному
преобразователю $M$ и описанию морфической последовательности $x$
можно получить описание морфической последовательности $M(x)$. Ясно
также, что существует (полиномиальный) алгоритм, который по любому
символу и морфической последовательности определяет, входит ли этот
символ в последовательность бесконечное количество раз
(доказательство можно найти в~\cite{NicPrit07}). Из этого следует
(аналогично тому, как это получается в~\ref{Monadic} для обобщённо
почти периодических последовательностей), что монадическая теория
морфической последовательности разрешима.

Также интересно понять, что можно сказать про подсловную сложность
морфических последовательностей. В серии работ, завершающейся
работой~\cite{Pans84}, доказано, что подсловная сложность чисто
морфической последовательности может удовлетворять одной из пяти
следующих асимптотик: $O(1)$, $\Theta(n)$, $\Theta(n\log\log n)$,
$\Theta(n\log n)$, $\Theta(n^2)$. Про подсловную сложность
морфических последовательностей произвольного вида долгое время было
ничего неизвестно. В~\cite{Pans85} показано, что существуют примеры
морфических последовательностей с подсловной сложностью вида
$\Theta(n^{1 + \frac1k})$ для каждого $k \in \N$. Наконец, в
\cite{Dev08a} сформулировано и в~\cite{Dev08b} будет доказано
следующее.

\begin{theorem}[\cite{Dev08a,Dev08b}]\label{morphic-subwcompl-theorem}
Подсловная сложность морфической последовательности имеет один из
следующих видов: $O(n\log n)$, $\Theta(n^{1 + \frac1k})$ для
некоторого $k \in \N$. Каждый из указанных классов непуст.
\end{theorem}

Таким образом, для полного описания возможных асимптотик подсловной
сложности морфической последовательности осталось разобраться
подробнее со случаем $O(n \log n)$. Можно сформулировать следующую
гипотезу.

\begin{conjecture}[\cite{Dev08a,Dev08b}]
\label{morphic-subwcompl-conjecture} Подсловная сложность
морфической последовательности имеет один из следующих видов:
$O(1)$, $\Theta(n)$, $\Theta(n\log\log n)$, $\Theta(n\log n)$,
$\Theta(n^{1 + \frac1k})$ для некоторого $k \in \N$. Каждый из
указанных классов непуст.
\end{conjecture}

Оказывается, про подсловную сложность морфической последовательности
можно сказать гораздо больше, если наложить дополнительные условия.

\begin{theorem}[\cite{NicPrit07}]
\label{morphic-almost-periodic-subword-complexity} Подсловная
сложность морфической почти периодической последовательности не
более чем линейна.
\end{theorem}

См.~также следствие~\ref{transcedence-morphic-almost-periodic}.

План доказательства
теоремы~\ref{morphic-almost-periodic-subword-complexity} содержится
в нескольких следующих леммах. Видимо, ключевой нужно считать
лемму~\ref{reduction-primitive}. Другие важные вспомогательные
леммы~--- это леммы~\ref{primitive-complexity} и
\ref{reduction-growing}.
% и~\ref{minimality-property}.
Леммы~\ref{nonerase-morphism-complexity} и \ref{growing-prolongable}
технические.

Для понимания плана доказательства понадобится несколько
дополнительных определений. Пусть $\phi\colon A^* \to A^*$~---
некоторый морфизм. Назовём его
\emph{примитивным}\label{def-morphism-primitive}, если существует
такое $n$, что для всех $a,b \in A$ слово $\phi^n(a)$ содержит~$b$.
Слово $w \in A^*$ называется
\emph{$\phi$-ограниченным}\label{def-bounded-word}, если
последовательность конечных слов $(w, \phi(w), \phi^2(w), \phi^3(w),
\dots)$ периодична с некоторого места. Слово $w \in A^*$ называется
\emph{$\phi$-возрастающим}\label{def-growing-word}, если
$|\phi^n(w)| \to \infty$ при $n \to \infty$. Ясно, что каждое слово
из $A^*$ либо $\phi$-ограниченное, либо $\phi$-возрастающее. Морфизм
$\phi$ называется \emph{возрастающим}\label{def-growing-morphism},
если $|\phi(a)| \ge 2$ для любого $a \in A$.

\begin{lemma}
\label{primitive-complexity} Подсловная сложность чисто морфической
последовательности, порождённой примитивным морфизмом, не более чем
линейна.
\end{lemma}

Лемма~\ref{primitive-complexity} в явном виде доказана в
\cite{Queff89} или в \cite{AlShall03} (теорема~10.4.12), но также
следует из результатов работы~\cite{Pans84}.

\begin{lemma}
\label{nonerase-morphism-complexity} Пусть $A$, $B$~--- конечные
алфавиты, $f \colon A^* \to B^*$~--- нестирающий морфизм, и пусть
максимальная длина $f(a)$ по всем $a \in A$ равна $M$. Тогда
$p_{f(x)}(n) \leqslant M p_x(n)$ для каждой последовательности $x
\in A^\N$ и $n \in \N$.
\end{lemma}

Лемму~\ref{nonerase-morphism-complexity} можно найти в
\cite{AlShall03} (теорема~10.2.4).

\begin{lemma}[\cite{Pans84}]\label{reduction-growing}
Пусть $A$~--- конечный алфавит, $s \in A$, и $\phi\colon A^* \to
A^*$~--- морфизм, продолжаемый на $s$. Предположим, что множество
всевозможных $\phi$-ограниченных подслов последовательности
$\phi^\infty(s)$ конечно. Тогда $\phi^\infty(s)$ может быть
представлена как образ под действием нестирающего морфизма чисто
морфической последовательности, порождённой возрастающим морфизмом.
\end{lemma}

\begin{lemma} \label{growing-prolongable}
Пусть $B$~--- конечный алфавит, $\phi\colon B^* \to B^*$~---
возрастающий морфизм. Тогда существует натуральное $n$ и такая буква
$t \in B$, что морфизм $\phi^n$ продолжаем на~$t$.
\end{lemma}

\begin{lemma} \label{reduction-primitive}
Для каждой чисто морфической последовательности $x$, порождённой
возрастающим морфизмом, существует чисто морфическая
последовательность $y$, порождённая примитивным морфизмом, такая что
$\Fac(y) \subseteq \Fac(x)$.
\end{lemma}

При помощи указанных лемм можно доказать
теорему~\ref{morphic-almost-periodic-subword-complexity}.
Доказательство теоремы и доказательства лемм можно найти
в~\cite{NicPrit07}.

В дальнейшем представляется интересным продолжить изучение
подсловной сложности морфических последовательностей, в том числе
морфических последовательностей с разными дополнительными
ограничениями.

%%%%%%%%%%%%%%%%%%%%%%%%%%%%%%%%%%%%%%%%%%%%%%%%%%%%%%%%%%%%%%%%%%

\subsection{Нормальные числа и гипотеза Бореля}
\label{Normal-Borel}

\begin{problem}\label{sqrt-2-problem}
Рассмотрим десятичную запись числа $\sqrt{2}$ как последовательность
букв конечного алфавита. Является ли эта последовательность почти
периодической? Является ли она морфической? Автоматной?
\end{problem}

По всей видимости, проблема~\ref{sqrt-2-problem} очень сложная. Ниже
даны некоторые причины в подтверждение такой точки зрения. В
изложении этого раздела мы в основном следуем обзорной
статье~\cite{Wald08}. Для некоторых теорем мы приводим упрощённые
ослабленные формулировки. Подробнее см.~\cite{Wald08}.

Для простоты все действительные числа в этом разделе будем считать
лежащими на интервале $[0,1)$. Пусть $x$~--- действительное число,
$b \in \N$. Запишем $x$ в системе счисления с основанием $b$: $x =
0{,}x(0)x(1)x(2)\dots$. Последовательность $x(0)x(1)x(2)\dots$ в
алфавите $\Sigma_b = \{0, 1, \dots, b - 1\}$ обозначим тогда $x_b$.

Назовём действительное число $x$
\emph{нормальным}\label{def-normal-number}, если оно обладает
следующим свойством: для любого $b \in \N$ все слова длины $n$ в
алфавите $\Sigma_b$ входят в последовательность $x_b$ бесконечно
много раз с одинаковой частотой~$1/b^n$. Борель доказал, что почти
все (по мере Лебега) действительные числа являются
нормальными~\cite{Borel09}. Таким образом, нормальность можно
воспринимать как свойство, которому удовлетворяет любое случайно
взятое число.

Напомним, что действительное число называется
\emph{алгебраическим}\label{def-algebraic-number}, если оно является
корнем некоторого многочлена от одной переменной с целыми
коэффициентами. Верно ли, что алгебраические числа не случайны?
Борель предположил, что нет, в следующем смысле.

\begin{conjecture}[\cite{Borel50}]
Любое алгебраическое иррациональное число является нормальным.
\end{conjecture}

Эта гипотеза по-прежнему остаётся недоказанной. В частности, из
гипотезы Бореля следует, что десятичная запись числа $\sqrt{2}$ (как
и запись по любому основанию, большему 1, любого алгебраического
иррационального числа) не является почти периодической. Но доказать
это, по всей видимости, довольно сложно.

Не доказан и следующий, более слабый вариант гипотезы Бореля.

\begin{conjecture}
Для любого $b \geqslant 3$ и любого алгебраического иррационального
числа $x$ каждый символ алфавита $\Sigma_b$ входит в $x_b$ хотя бы
один раз.
\end{conjecture}

Также не известно ни одного явного примера тройки $x,a,b$, где
$x$~--- алгебраическое иррациональное число, $b$~--- натуральное
число, $a \in \Sigma_k$, для которых можно доказать, что $a$ входит
в $x_b$ бесконечно много раз.

По всей видимости, гипотеза Бореля очень сложна. Однако можно
пытаться для различных классов последовательностей доказывать, что
ни одна последовательность из этого класса не может быть записью
алгебраического иррационального числа. Следующая ослабленная
формулировка гипотезы Бореля также не доказана. Напомним, что $\p_y$
обозначает подсловную сложность последовательности~$y$.

\begin{conjecture}
Если $x$~--- алгебраическое иррациональное число, и $b \in \N$, то
$\p_{x_b}(n) = b^n$.
\end{conjecture}

Рассмотрим следующую переформулировку.

\medskip

\noindent {\bfseries Гипотеза \theconjecture$'$.} {\itshape Пусть
$x$~--- действительное число, $b \in \N$ и $\p_{x_b}(n) < b^n$.
Тогда $x$ рационально или трансцендентно. }

\medskip

Даже и эта гипотеза, по всей видимости, является очень сложной.
Известны следующие гораздо более слабые результаты.

\begin{theorem}[\cite{FerMaud97}]
Пусть $x$~--- действительное число. Если $x_2$ является
последовательностью Штурма, то $x$ трансцендентно.
\end{theorem}

В частности, например, двоичная запись $\sqrt 2$ не является
последовательностью Штурма. Впоследствии было получено более сильное
утверждение.

\begin{theorem}[\cite{AdamBug07}]
\label{transcedence-linear-subword-complexity} Пусть $x$~---
действительное число, $b \in \N$ и $\p_{x_b}(n) = O(n)$. Тогда $x$
рационально или трансцендентно.
\end{theorem}

Отсюда авторы сразу получают такое следствие.

\begin{corollary}\label{transcedence-automatic}
Пусть $x$~--- действительное число, $b \in \N$, и $x_b$~---
автоматная последовательность. Тогда $x$ рационально или
трансцендентно.
\end{corollary}
\begin{proof}
Следует из теоремы~\ref{automatic-subwcompl-linear} и
теоремы~\ref{transcedence-linear-subword-complexity}.
\end{proof}

Мы же теперь можем получить и ещё одно любопытное следствие.

\begin{corollary}\label{transcedence-morphic-almost-periodic}
Пусть $x$~--- действительное число, $b \in \N$, и $x_b$~---
морфическая почти периодическая последовательность. Тогда $x$
рационально или трансцендентно.
\end{corollary}
\begin{proof}
Следует из теоремы~\ref{morphic-almost-periodic-subword-complexity}
и теоремы~\ref{transcedence-linear-subword-complexity}.
\end{proof}

В~\cite{AdamBug07} также формулируется следующая гипотеза.

\begin{conjecture}
Пусть $x$~--- действительное число, $b \in \N$ и $x_b$~---
морфическая последовательность. Тогда $x$ рационально или
трансцендентно.
\end{conjecture}

Частные случаи этой гипотезы доказаны в~\cite{AdamBug07}, один из
таких случаев~--- следствие~\ref{transcedence-automatic} (которое
также является ответом на вопрос Кобхэма из~\cite{Cobh68}).
Существенное продвижение в этой гипотезе (в частности, покрывающее
результат следствия~\ref{transcedence-morphic-almost-periodic})
получено в~\cite{Albert06}.
%Это является
%дополнительной мотивацией к изучению
%гипотезы~\ref{morphic-almost-periodic-decision-algorithm-conjecture}.

И наконец, самым сильным результатом в этой серии на настоящий
момент является следующий.

\begin{theorem}[\cite{BugEver08}]
Пусть $x$~--- действительное число, $b \in \N$ и $\p_{x_b}(n) = O(n
(\log n)^\theta)$ для некоторого положительного действительного
числа $\theta < 1/11$. Тогда $x$ рационально или трансцендентно.
\end{theorem}

Видно, что он ещё по-прежнему очень далёк от гипотезы Бореля.

%%%%%%%%%%%%%%%%%%%%%%%%%%%%%%%%%%%%%%%%%%%%%%%%%%%%%%%%%%%%%%%%%%
%%%%%%%%%%%%%%%%%%%%%%%%%%%%%%%%%%%%%%%%%%%%%%%%%%%%%%%%%%%%%%%%%%
%%%%%%%%%%%%%%%%%%%%%%%%%%%%%%%%%%%%%%%%%%%%%%%%%%%%%%%%%%%%%%%%%%

\section{Близость к периодичности: альтернативные подходы}
\label{Alternative}

В этом разделе мы говорим о несколько менее известных и менее
изученных, но также интересных вариантах понятий, обобщающих понятие
периодической последовательности.

%%%%%%%%%%%%%%%%%%%%%%%%%%%%%%%%%%%%%%%%%%%%%%%%%%%%%%%%%%%%%%%%%%

\subsection{Квазипериодические последовательности}
\label{Quasiperiodic}

В~\cite{ApostErenf93,Marcus04} была предложена следующая идея
обобщения понятия периодичности. Будем говорить, что конечное слово
$w$ является квазипериодом $x$ (конечного слова или бесконечной
последовательности), если каждую позицию в $x$ можно покрыть
вхождением $w$ в $x$. Более формально, для каждого $i: 1 \leqslant i
\leqslant |x|$ найдутся $k,l: k \leqslant i \leqslant l$, такие что
$x[k,l] = w$ (для бесконечного $x$ можно положить $|x| = \infty$).

Конечное слово \emph{квазипериодично}, если у него есть хотя бы один
квазипериод, отличный от него самого; в противном случае слово
называют минимальным (или в~\cite{LeveRich04} суперпримитивным). У
каждого конечного слова есть ровно один минимальный квазипериод
(см.~\cite{LeveRich04}).

Последовательность называется
\emph{квазипериодической}\label{def-quasiperiodic}, если у неё
существует хотя бы один квазипериод. Класс всех квазипериодических
последовательностей обозначим~$\QP$. Существуют последовательности с
бесконечным количеством минимальных квазипериодов (например,
последовательность Фибоначчи, см.~\cite{LeveRich04}).
Последовательности с бесконечным количеством различных квазипериодов
назовём
\emph{муль\-ти-ква\-зи\-пе\-ри\-о\-ди\-чес\-ки\-ми}\label{def-multi-quasiperiodic}
(multi-scale quasiperiodic, см.~\cite{MontMar06}). Класс
мульти-квазипериодических последовательностей обозначим~$\MQP$. Ясно
также, что существуют последовательности вообще без квазипериодов.
Как показано в~\cite{LeveRich04}, любая промежуточная ситуация
реализуется: для каждого $n$ можно построить последовательность, у
которой ровно $n$ минимальных квазипериодов.

Определение квазипериодичности, несмотря на свою естественность,
было впервые рассмотрено относительно недавно. В~\cite{ApostErenf93}
появилось определение для конечных слов, которое затем было
расширено на последовательности в~\cite{Marcus04}. Там же поставлены
различные открытые вопросы. В~\cite{LeveRich04} авторы получают
ответы на некоторые из них. В~\cite{LeveRich06} они получают полное
описание квазипериодических последовательностей Штурма. В работе
\cite{MontMar06} рассмотрено понятие мульти-квазипериодичности. Вот,
пожалуй, практически полный список работ о квазипериодических
последовательностях.

Оказывается, класс $\QP$ находится в общем положении с классом
$\AP$. Неравенство $\QP \setminus \AP \ne \varnothing$ очевидно.
Действительно, например, в $\QP$ лежит произвольная
последовательность из 0 и 1, начинающаяся с 1 и не содержащая 00 и
111 (с квазипериодом 101), и, очевидно, среди таких
последовательностей есть не почти периодические. Неравенство $\AP
\setminus \QP \ne \varnothing$ следует из
результатов~\cite{LeveRich04}, где авторы приводят примеры
последовательностей Штурма, не являющихся квазипериодическими (в
ответ на вопрос из~\cite{Marcus04}).

Таким образом, квазипериодичность и почти периодичность~---
принципиально разные подходы к обобщению периодичности, что является
дополнительным стимулом к изучению нового класса квазипериодических
последовательностей. Ясно однако, что имеет место включение $\MQP
\subseteq \AP$, а также невырожденность $\Per \subsetneq \MQP$
(см.~\cite{MontMar06}; из результатов оттуда следует также, что
класс $\MQP$ континуален). Таким образом, некоторая связь между
почти периодичностью и квазипериодичностью (правда, уже уточнённой)
всё же есть. Неравенство $\MQP \subsetneq \AP$ сразу следует из
характеризации подсловной сложности произвольного слова $x \in
\MQP$: $\liminf_{n \to \infty} \frac{\p_x(n)}{n^2} \leqslant 1$
(см.~\cite{MontMar06}).

Как показано в~\cite{MontMar06}, подсловная сложность
квазипериодических последовательностей асимптотически может быть
почти произвольной, с точностью до эквивалентности специального
вида. В~\cite{LeveRich04} приведён несложный пример
последовательности, у которой подсловная сложность экспоненциальна,
с основанием экспоненты примерно 1{,}32. Остаётся открытым следующий
вопрос.

\begin{problem}
Какая максимальная возможная подсловная сложность может быть у
квазипериодической последовательности?
\end{problem}

Про два вопроса из~\cite{Marcus04} авторы~\cite{LeveRich04} пишут,
что они нуждаются в более точной формулировке, поскольку в них идёт
речь о последовательностях, в которых все факторы квазипериодичны,
но ясно, что слово длины 1 никогда не квазипериодично. Вопрос~6
(Существует ли неквазипериодическая последовательность, у которой
все факторы квазипериодичны?) переформулируется так: можно ли найти
не квазипериодическую последовательность, у которой найдётся
бесконечное количество квазипериодических префиксов?
В~\cite{LeveRich04} на него даётся положительный ответ.

В контексте настоящей статьи естественно задать следующий вопрос.

\begin{problem}
1) Верно ли, что класс квазипериодических последовательностей
замкнут относительно конечно-автоматных преобразований?\\
2) Если ответ на предыдущий вопрос положительный, то можно ли это
доказать в каком-нибудь смысле эффективно и получить критерий
разрешимости монадической теории квазипериодической
последовательности (см.~раздел~\ref{Logic})?
\end{problem}

%%%%%%%%%%%%%%%%%%%%%%%%%%%%%%%%%%%%%%%%%%%%%%%%%%%%%%%%%%%%%%%%%%

\subsection{Разреженные периодичности}
\label{TilingPeriodicity}

В~\cite{KarhLifRyt06} предложен любопытный подход к обобщению
понятия периодичности, сходный с тем, который изложен в
разделе~\ref{Quasiperiodic}. Более общо, вопросы замощения
бесконечной клетчатой полоски или многомерных клетчатых пространств
клетчатыми фигурками много и активно изучались, но в таком
контексте, видимо, не рассматривались.

Назовём шаблоном слово в алфавите $A \cup \{\square\}$, где $\square
\notin A$ означает для нас пробел, пропуск. Конечное слово $u \in
A^*$ назовём \emph{разреженно
периодическим}\label{def-tiling-periodicity}, если найдётся такой
шаблон $v$, что его копиями можно покрыть всё слово $u$, причём
каждый символ покрыт ровно один раз; при этом мы считаем, что при
наложении шаблона на слово символ $\square$ не покрывает находящийся
под ним символ. Например, слово $0011$ можно покрыть шаблоном
$0\square1$.

В отличие от обычной периодичности, бывают слова с несколькими
несравнимыми разреженными периодами (если сравнивать по
вложенности)~--- в~\cite{KarhLifRyt06} приводится кратчайший
известный пример такого слова из 24 символов. Тем не менее, можно
найти связь между обычным периодом и разреженным: минимальный
разреженный период слова является также минимальным разреженным
периодом его минимального настоящего периода. Кроме того,
представлена классификация всех разреженных периодов слова и
получена рекурсивная формула для максимального количества
разреженных периодов для слов длины~$n$. Наконец, получен алгоритм
нахождения всех минимальных периодов слова длины $n$, время работы
$O(n \log(n) \log\log(n))$.

В конце работы \cite{KarhLifRyt06} приведено несколько открытых
вопросов. По мнению авторов, наиболее важным является следующий.

\begin{problem}[\cite{KarhLifRyt06}]
Предложить естественный способ обобщения понятия ``чистой'' (pure)
разреженной периодичности (по аналогии с pure и ultimate
(заключительной) периодичностью). Это может помочь описать
регулярности в большем количество слов и получить новые алгоритмы
компрессии.
\end{problem}

Для нас наиболее существенным является следующее замечание
негативного характера: при прямом обобщении понятия разреженной
периодичности на бесконечные последовательности получаются только
периодические последовательности, как следует, например, из теоремы
2 работы~\cite{Newman77}.

Таким образом, остаётся следующая

\begin{problem}
Предложить нетривиальный способ обобщения понятия разреженной
периодичности на бесконечные последовательности, который позволит
получить для них критерий разрешимости монадической теории или
теории первого порядка.
\end{problem}

%%%%%%%%%%%%%%%%%%%%%%%%%%%%%%%%%%%%%%%%%%%%%%%%%%%%%%%%%%%%%%%%%%

\subsection{Последовательность Колакоски и
периодически альтернируемые морфизмы} \label{AlternatingMorphisms}

Последовательность Колакоски~--- одна из самых загадочных
последовательностей в комбинаторике слов, про неё есть огромное
количество открытых вопросов.

Определяется последовательность Колакоски так\label{def-Kolakoski}.
Это бесконечная вправо последовательность, состоящая из 1 и 2,
причём если записать длины блоков из подряд идущих одинаковых
символов, то получится сама эта последовательность.
Последовательность Колакоски начинается как
  $$
\kolak = 2, 2, 1, 1, 2, 1, 2, 2, 1, 2, 2, 1, 1, 2, 1, 1, 2, 2, 1, 2,
1, 1, 2, \dots
  $$
Сначала идёт блок из 2 двоек, потом 2 единицы, 1 двойка, 1 единица,
2 двойки, и т.~д. Можно начать записывать последовательность
Колакоски и с 1, руководствуясь тем же самым определением, но тогда
получится в точности последовательность~$1\kolak$.

Последовательность $\kolak$ впервые появилась в заметке
Колакоски~\cite{Kolak65}. В~\cite{Kolak66} он доказывает, что эта
последовательность не периодическая.
См.~также~\cite{MathWorld-Kolakoski}.

Главная гипотеза про последовательность $\kolak$ заключается в
следующем.

\begin{conjecture}
Частота символа 1 (а значит, и символа 2) в последовательности
$\kolak$ равна 1/2.
\end{conjecture}

До сих пор неизвестно, верно ли это. Более того, неизвестно,
существует ли вообще частота 1, причём известно, что если она
существует, то обязательно равна 1/2. Компьютерные эксперименты
подтверждают гипотезу о частоте 1/2. Более того, если $o_n$~---
количество единиц в начальном отрезке $\kolak$ длины $n$, то
компьютерные эксперименты показывают, что, скорее всего, $o_n =
0{,}5n + O(\log n)$ (см.~\cite{PlanetMath-Kolakoski}), в то время
как для случайной последовательности можно было бы утверждать лишь
$o_n = 0{,}5n + O(\sqrt n)$.

\begin{problem}
1) Верно ли, что если какое-то слово входит в $\kolak$ (например,
22112), то оно входит ещё хотя бы раз?\\
2) Верно ли, что если какое-то слово входит в $\kolak$, то входит и
слово, полученное из этого заменой 1 на 2 и 2 на 1?\\
3) Верно ли, что если какое-то слово входит в $\kolak$, то входит и
слово, полученное из этого, если его записать задом наперёд?
\end{problem}

Ответы на все эти вопросы неизвестны
(см.~\cite{Kimber-Kolakoski-1,Kimber-Kolakoski-2}). В частности,
неизвестно, является ли $\kolak$ рекуррентной или почти
периодической. Однако известно, что из рекуррентности $\kolak$
следовало бы существование частоты 1 в ней. Неизвестна и подсловная
сложность последовательности Колакоски.

К этой серии открытых и, по всей видимости, довольно сложных проблем
мы добавим такую.

\begin{problem}\label{Kolakoski-theory-problem}
Разрешима ли (монадическая или первого порядка) теория
последовательности $\kolak$?
\end{problem}

Формально из разрешимости монадической теории последовательности
Колакоски не следует ответ на сложные открытые вопросы, приведённые
выше. Это может являться аргументом за то, что
проблема~\ref{Kolakoski-theory-problem} проще. Однако если доказать,
что монадическая теория последовательности $\kolak$ разрешима, это
дало бы возможность для дополнительных экспериментов. В частности,
из этого следовало бы существование алгоритма, который по каждому
слову проверяет, входит ли оно в $\kolak$, входит ли оно в $\kolak$
бесконечно много раз, и т.~п.

Тем не менее кое-что про последовательность Колакоски известно. Как
уже было отмечено выше, она не периодическая. Кроме того, она не
содержит кубов, а квадраты, которые в неё входят, имеют длину 2, 4,
6, 18 или 54 (см.~\cite{Sloane-Kolakoski}). Кроме того,
последовательность Колакоски не является чисто морфической
\cite{CulKarhuLep92}. Является ли она морфической, неизвестно (про
морфические последовательности см. разделы~\ref{Automatic},
\ref{Morphic}).

Одним из возможных подходов к изучению последовательности Колакоски
может стать детальное изучение одного из методов, которым можно
построить эту последовательность, являющегося обобщением морфических
последовательностей~--- периодически альтернируемые итерируемые
морфизмы (введено в работе~\cite{CulKarhu92}).

Пусть $H = \{h_0, h_1, \dots, h_{p - 1}\}$~--- произвольное
множество нестирающих морфизмов $A^* \to A^*$. Определим отображение
$H$ на слове $a_0a_1\dots a_n \in A^*$ как
  $$
H(a_0a_1\dots a_n) = \alpha_0\alpha_1\dots \alpha_n, \quad \text{
где } \alpha_i = h_k(a_i), \quad k \equiv i \pmod p.
  $$
Если для какой-то буквы $s$ слово $H(s)$ начинается с $s$, то можно
корректно определить последовательность $H^\infty(s) = \lim_{n \to
\infty}H^n(s)$. Будем говорить, что эта последовательность
\emph{порождается периодически альтернируемыми морфизмами $h_0, h_1,
\dots, h_{p - 1}$}\label{def-periodicaly-alternating-morphisms}.

Ясно, что все чисто морфические последовательности порождаются
периодически альтернируемыми морфизмами, точнее, одним морфизмом.
Рассмотрим менее тривиальный пример~--- последовательность
Колакоски. Пусть $h_0(1) = 2$, $h_0(2) = 22$, $h_1(1) = 1$, $h_1(2)
= 11$. Будем итерировать соответствующее отображение $H$ на символе
2:
  $$
\begin{tabular}{l}
$H(2) = h_0(2) = 22$ \\
$H(22) = h_0(2)h_1(2) = 2211$ \\
$H(2211) = h_0(2)h_1(2)h_0(1)h_1(1) = 221121$ \\
$H(221121) = h_0(2)h_1(2)h_0(1)h_1(1)h_0(2)h_1(1) = 221121221$ \\
$H(221121221) = h_0(2)h_1(2)h_0(1)h_1(1)h_0(2)h_1(1)h_0(2)h_1(2)h_0(1) = 22112122122112$ \\
\dots
\end{tabular}
  $$
Можно доказать, что $H^\infty(2) = \kolak$ \cite{CulKarhuLep92}. И
значит, для понимания устройства последовательности Колакоски очень
важно получше понять устройство последовательностей, порождаемых
периодически альтернируемыми морфизмами.

В~\cite{Lepisto93} доказано, что для любого полинома существуют
последовательности, порождаемые периодически альтернируемыми
морфизмами, с подсловной сложностью не менее этого полинома. В
частности, такие последовательности не все морфические, поскольку,
как мы помним из раздела~\ref{Morphic}, подсловная сложность
морфической последовательности не более чем квадратичная.
Неизвестно, может ли подсловная сложность последовательностей,
порождённых периодически альтернируемыми морфизмами, быть
экспоненциальной.

Сформулируем обобщение проблемы~\ref{Kolakoski-theory-problem}.

\begin{problem}\label{alternating-morphisms-theory-problem}
Разрешима ли (монадическая или первого порядка) теория
последовательностей, порождаемых периодически альтернируемыми
морфизмами?
\end{problem}

Напомним, что монадическая теория морфической последовательности
всегда разрешима (теорема~\ref{morphic-monadic-decidable}). Можно
надеяться, что и в случае периодически альтернируемых морфизмов это
так. По аналогии с планом доказательства разрешимости монадических
теорий, описанным в разделе~\ref{Logic}, можно детализировать
проблему~\ref{alternating-morphisms-theory-problem}.

\begin{problem}\label{alternating-morphisms-theory-problem-detailed}
1) Верно ли, что класс последовательностей, порождённых периодически
альтернируемыми морфизмами, сохраняется при конечно-автоматных
преобразованиях?\\
2) Если ответ на предыдущий вопрос положительный, можно ли
доказать эффективный вариант этого утверждения?\\
3) Существует ли алгоритм проверки по описанию последовательности,
порождённой периодически альтернируемыми морфизмами, того, входит ли
в неё данный символ бесконечное количество раз?
\end{problem}

%%%%%%%%%%%%%%%%%%%%%%%%%%%%%%%%%%%%%%%%%%%%%%%%%%%%%%%%%%%%%%%%%%

\subsection{Последовательности Тёплица}
\label{Toeplitz}

В работе~\cite{Toepl28} Тёплиц ввёл конструкцию для определения
почти периодических функций на действительной прямой. В
работе~\cite{JacKea69} авторы модифицировали эту конструкцию для
определения бесконечных слов и назвали эти слова словами Тёплица (мы
для единообразия будем говорить о последовательностях). Краткий
нижеследующий обзор о последовательностях Тёплица сделан в основном
по работе~\cite{CassKarhu97}.

Пусть шаблон (как и в разделе~\ref{TilingPeriodicity})~--- слово в
алфавите $A \cup \{\square\}$, где $\square \notin A$ означает для
нас пробел или пропуск. Из шаблона $w$ можно получить
последовательность следующим образом. Пусть $T_w^0 = w^\infty$~---
периодическая последовательность в алфавите $A \cup \{\square\}$ с
периодом $w$. Для каждого $i$ последовательность $T_w^{i + 1}$
получается из $T_w^i$, если все пробелы $\square$ последовательно
заполнять символами последовательности $T_w^0 = w^\infty$.
Последовательность $T_w = \lim_{i \to \infty} T_w^i$ в алфавите $A$
называется \emph{последовательностью Тёплица}\label{def-Toeplitz},
полученной из шаблона $w$. Если $|w| = p$ и $|w|_\square = q$, то
последовательность $T_w$ называется последовательностью Тёплица типа
$(p,q)$.

Например, пусть $w = 1\square0\square$. Тогда
$$
\begin{tabular}{l}
$T_w^0 = 1\square0\square1\square0\square1\square0\square1\square0
\square1\square0\square1\square0\square1\square0\square1\square0\square\dots$,\\
$T_w^1 = 110\square100\square110\square100
\square110\square100\square110\square100\square\dots$,\\
$T_w^2 = 1101100\square1100100\square1101100\square1100100\square\dots$,\\
$T_w^3 = 110110011100100\square110110001100100\square\dots$,\\
$T_w^4 = 1101100111001001110110001100100\square\dots$,\\
$T_w^4 = 11011001110010011101100011001001\dots$,
\end{tabular}
$$
то есть $T_w = 11011001110010011101100011001001\dots$ Это известная
\emph{``последовательность сгибания
бумаги''}\label{def-paperfolding-sequence} (paperfolding sequence),
например, см.~\cite{AlShall03}. (Эта последовательность получается
следующим образом: возьмём полоску бумаги и согнём её посередине;
потом получившееся опять согнём посередине, и т.~д. Каждый раз
необходимо сгибать в одном и том же направлении. После сгибания
некоторое количество раз, разогнув полоску, мы получим начальный
отрезок последовательности сгибания бумаги: 1 соответствует
``горке'' на полоске, а 0 соответствует ``ямке''.)

Как несложно видеть, любая последовательность Тёплица почти
периодична. В~\cite{CassKarhu97} получен критерий периодичности для
последовательностей Тёплица. Кроме того, там дана следующая
классификация.

\begin{theorem}[\cite{CassKarhu97}]\label{Toeplitz-classification}
1) Последовательности Тёплица типа $(p,1)$ являются чисто морфическими.\\
2) Последовательности Тёплица типа $(tq,q)$ являются морфическими.\\
3) Произвольные последовательности Тёплица порождаются периодически
альтернируемыми морфизмами.
\end{theorem}

Поэтому естественным представляется, особенно в виду
теоремы~\ref{morphic-monadic-decidable} и
проблемы~\ref{alternating-morphisms-theory-problem}, сформулировать
следующую гипотезу.

\begin{conjecture}\label{Toeplitz-theory-conjecture}
Монадическая теория любой последовательности Тёплица разрешима.
\end{conjecture}

Скорее всего, эту гипотезу можно доказать, доказав, что множество
подслов любой последовательности Тёплица разрешимо
(последовательность Тёплица, очевидно, вычислима, и тогда можно
воспользоваться следствием~\ref{decidability-monadic-AP}). Более
того, достаточно доказать, что функция подсловной сложности
вычислима. По всей видимости, это можно сделать, проанализировав
доказательство результата из~\cite{CassKarhu97}, в котором найдёна
асимптотика подсловной сложности последовательностей Тёплица. При
этом отметим, что из описания асимптотики функции подсловной
сложности не следует автоматически вычислимость функции подсловной
сложности.

%%%%%%%%%%%%%%%%%%%%%%%%%%%%%%%%%%%%%%%%%%%%%%%%%%%%%%%%%%%%%%%%%%
%%%%%%%%%%%%%%%%%%%%%%%%%%%%%%%%%%%%%%%%%%%%%%%%%%%%%%%%%%%%%%%%%%

\subsection*{Указатель терминов}
\tocontents{Указатель терминов}

В этом разделе собраны в виде таблицы ссылки на основные
используемые в статье термины и понятия.
%Понятия разделены на смысловые блоки.
Для каждого понятия даётся соответствующее обозначение (если есть) и
номер страницы, на которой появляется определение понятия.

\begin{longtable}{p{115mm}ll}
Понятие & Обозначение & Страница \\[3mm]
топологическая динамическая система & & \pageref{def-topological-dynamical-system} \\
операция левого сдвига & $L$ & \pageref{def-left-shift} \\
почти периодическая относительно меры близости функция & &
\pageref{def-almost-periodic-function} \\
почти периодическая по Бору функция & &
\pageref{def-almost-periodic-function-Bohr} \\
почти периодическая по Безиковичу функция & & \pageref{def-almost-periodic-function-Besicovitch} \\
относительно плотное множество & & \pageref{def-relatively-dense-set} \\
близкая к периодической относительно меры близости
последовательность & & \pageref{def-close-to-periodic} \\
канторовское расстояние & $d_C(x, y)$ & \pageref{def-Cantor-distance} \\
расстояние Безиковича & $d_B(x, y)$ & \pageref{def-Besicovitch-distance} \\
почти периодическая по Безиковичу последовательность & & \pageref{def-almost-periodic-Besicovich} \\
алгебраическое множество & & \pageref{def-algebraic-set} \\
полуалгебраическое множество & & \pageref{def-semialgebraic-set} \\
минимальная динамическая система & & \pageref{def-minimal-dynamical-system} \\
алфавит & & \pageref{def-alphabet} \\
буква & & \pageref{def-letter} \\
символ & & \pageref{def-symbol} \\
последовательность & & \pageref{def-sequence} \\
слово & & \pageref{def-word} \\
бесконечное слово & & \pageref{def-infinite-word} \\
вхождение слова в последовательность & & \pageref{def-occurrence} \\
фактор & & \pageref{def-factor} \\
подслово & & \pageref{def-subword} \\
префикс & & \pageref{def-prefix} \\
суффикс & & \pageref{def-suffix} \\
морфизм & & \pageref{def-morphism} \\
нестирающий морфизм & & \pageref{def-non-erasing-morphism} \\
$k$-равномерный морфизм & & \pageref{def-uniform-morphism} \\
примитивный морфизм & & \pageref{def-morphism-primitive} \\
возрастающее относительно морфизма слово & & \pageref{def-growing-word} \\
ограниченное относительно морфизма слово & & \pageref{def-bounded-word} \\
возрастающий морфизм & & \pageref{def-growing-morphism} \\
кодирование & & \pageref{def-coding} \\
проекция & & \pageref{def-projection} \\
периодическая последовательность & $\Per$ & \pageref{def-periodic} \\
период & & \pageref{def-period} \\
предпериод & & \pageref{def-preperiod} \\
заключительно периодическая последовательность & $\EP$ & \pageref{def-eventually-periodic} \\
апериодическая последовательность & & \pageref{def-aperiodic} \\
почти периодическая последовательность & $\AP$ & \pageref{def-almost-periodic} \\
минимальная последовательность & & \pageref{def-minimal-sequence} \\
равномерно рекуррентная последовательность & & \pageref{def-uniformly-recurrent-sequence} \\
эффективно почти периодическая последовательность & & \pageref{def-effective-almost-periodic} \\
заключительно почти периодическая последовательность & $\EAP$ & \pageref{def-eventually-almost-periodic} \\
минимальный префикс & $\pr(x)$ & \pageref{def-minimal-prefix} \\
обобщённо почти периодическая последовательность & $\GAP$ & \pageref{def-generalized-almost-periodic} \\
эффективно обобщённо почти периодическая последовательность & & \pageref{def-effective-generalized-almost-periodic} \\
точно почти периодическая последовательность & $\PAP$ & \pageref{def-precisely-almost-periodic} \\
рекуррентная последовательность & $\Rec$ & \pageref{def-recurrent} \\
заключительно рекуррентная последовательность & $\ER$ & \pageref{def-eventually-recurrent} \\
регулятор почти периодичности & $\r_x$ & \pageref{def-almost-periodicity-regulator} \\
линейно почти периодическая последовательность & & \pageref{def-linear-almost-periodic} \\
заключительно линейно почти периодическая последовательность & & \pageref{def-linear-eventually-almost-periodic} \\
коэффициент почти периодичности & & \pageref{def-almost-periodicity-quotient} \\
автоматная последовательность & & \pageref{def-automatic} \\
продолжаемый на букве морфизм & & \pageref{def-prolongable} \\
морфическая последовательность & & \pageref{def-morphic} \\
чисто морфическая последовательность & & \pageref{def-pure-morphic} \\
периодически альтернируемые морфизмы & & \pageref{def-periodicaly-alternating-morphisms} \\
квазипериодические последовательности & $\QP$ & \pageref{def-quasiperiodic} \\
мульти-квазипериодические последовательности & $\MQP$ & \pageref{def-multi-quasiperiodic} \\
разреженная периодичность & & \pageref{def-tiling-periodicity} \\
квадрат слова & & \pageref{def-square} \\
куб слова & & \pageref{def-cube} \\
бесквадратная последовательность & & \pageref{def-square-free} \\
бескубная последовательность & & \pageref{def-cube-free} \\
последовательность Туэ~--- Морса & $\thue$ & \pageref{def-Thue-Morse} \\
последовательность Фибоначчи & $\fib$ & \pageref{def-Fibonacci} \\
последовательности Штурма & & \pageref{def-Sturm} \\
тернарная последовательность Кини & & \pageref{def-Keane-sequence} \\
последовательности Какутани & & \pageref{def-Kakutani-sequence} \\
последовательность сгибания бумаги & &
\pageref{def-paperfolding-sequence} \\
сбалансированное множество слов & & \pageref{def-balanced-set} \\
сбалансированное слово или последовательность & & \pageref{def-balanced-word} \\
нижняя механическая последовательность & $s_{\alpha,\rho}$ & \pageref{def-mechanical-sequence} \\
верхняя механическая последовательность & $s'_{\alpha,\rho}$ & \pageref{def-mechanical-sequence} \\
последовательности Тёплица & $T_w$ & \pageref{def-Toeplitz} \\
последовательность Колакоски & $\kolak$ & \pageref{def-Kolakoski} \\
блочное произведение & $u \otimes v$ & \pageref{def-block-product} \\
чезаровский блок & & \pageref{def-Cesaro} \\
равномерно чезаровский блок & & \pageref{def-Cesaro} \\
чезаровская последовательность & & \pageref{def-Cesaro} \\
равномерно чезаровская последовательность & & \pageref{def-Cesaro} \\
$A$-$\GAP$-схема & & \pageref{def-GAP-scheme} \\
$\GAP$-порождённая $A$-$\GAP$-схемой последовательность & & \pageref{def-GAP-generated-GAP-scheme} \\
правильно $\GAP$-порождённая последовательность & & \pageref{def-GAP-generated-perfectly} \\
$\AP$-порождённая $A$-$\GAP$-схемой последовательность & & \pageref{def-AP-generated-GAP-scheme} \\
$A$-$\AP$-схема & & \pageref{def-AP-scheme} \\
$\AP$-порождённая $A$-$\AP$-схемой последовательность & & \pageref{def-AP-generated-AP-scheme} \\
конечно-автоматный преобразователь & & \pageref{def-finite-state-transducer} \\
ход преобразователя & & \pageref{def-transducer-run} \\
образ последовательности под действием преобразователя & & \pageref{def-transducer-image} \\
преобразователь подходит ко вхождению слова в состоянии~$q$ & & \pageref{def-transducer-comes-in-state} \\
равномерный конечно-автоматный преобразователь & & \pageref{def-uniform-finite-transducer} \\
обратимый конечно-автоматный преобразователь & & \pageref{def-automaton-reversible} \\
циклический конечно-автоматный преобразователь & &
\pageref{def-cyclic-finite-transducer} \\
разбиение & $\s_a(x)$ & \pageref{def-split} \\
почти обратимый конечно-автоматный преобразователь & &
\pageref{def-almost-reversible-finite-transducer} \\
стековый (магазинный) конечно-автоматный преобразователь & & \pageref{def-pushdown-transducer} \\
произведение последовательностей & $x \times y$ & \pageref{def-product-sequences} \\
теория первого порядка последовательности & $\T x$ & \pageref{def-first-order-theory} \\
монадическая теория последовательности & $\MT x$ & \pageref{def-monadic-theory} \\
разрешимая теория & & \pageref{def-decidable-theory} \\
бескванторная теория & & \pageref{def-quantifier-free-theory} \\
автомат Бюхи & & \pageref{def-Buchi-automaton} \\
автомат Бюхи принимает последовательность & & \pageref{def-Buchi-automaton-accepts} \\
детерминированный автомат Бюхи & & \pageref{def-deterministic-automaton} \\
автомат Мюллера & & \pageref{def-Muller-automaton} \\
макросостояние & & \pageref{def-macrostate} \\
предельное макросостояние & & \pageref{def-limit-macrostate} \\
автомат Мюллера принимает последовательность & & \pageref{def-Muller-automaton-accepts} \\
детерминированный автомат Мюллера & & \pageref{def-deterministic-Muller-automaton} \\
регулярное множество последовательностей & & \pageref{def-regular-set} \\
вычислимое действительное число & & \pageref{def-computable-real-number} \\
вычислимый оператор на последовательностях & & \pageref{def-computable-operator} \\
выразимое свойство & & \pageref{def-expressible-property} \\
мера апериодичности & $\am(x)$ & \pageref{def-aperiodicity-measure} \\
нормальное число & & \pageref{def-normal-number} \\
алгебраическое число & & \pageref{def-algebraic-number} \\
подсловная сложность & $\p_x(n)$ & \pageref{def-subword-complexity} \\
колмогоровская сложность & $K(u)$ & \pageref{def-Kolmogorov-complexity} \\
топологическая энтропия & $E_t(x)$ &
\pageref{def-topological-entropy} \\
\end{longtable}

%%%%%%%%%%%%%%%%%%%%%%%%%%%%%%%%%%%%%%%%%%%%%%%%%%%%%%%%%%%%%%%%%%
%%%%%%%%%%%%%%%%%%%%%%%%%%%%%%%%%%%%%%%%%%%%%%%%%%%%%%%%%%%%%%%%%%
%%%%%%%%%%%%%%%%%%%%%%%%%%%%%%%%%%%%%%%%%%%%%%%%%%%%%%%%%%%%%%%%%%

\subsection*{Благодарности}
\tocontents{Благодарности}

Авторы благодарны С.~Августиновичу, И.~Богданову, Н.~Верещагину,
Р.~Девятову, М.~Раскину, А.~Румянцеву, Ю.~Уляшкиной, А.~Фрид,
А.~Шеню, а также B.~Adamczewski, J.~Cassaigne, V.~Diekert,
B.~Durand, J.~Karhum\"aki, O.~Kupferman, F.~Nicolas, N.~Rampersad,
K.~Saari за многочисленные полезные обсуждения на темы, затронутые в
настоящей обзорной статье. Темы и результаты этой статьи
неоднократно обсуждались на Колмогоровском семинаре МГУ
\cite{KolmSem-rus}, а также на семинаре ``Алгоритмические вопросы
алгебры и логики'' под руководством академика С.~И.~Адяна, авторы
благодарны всем участникам за внимание.

\bigskip

Работа над текстом продолжалась и когда один из его авторов~---
Андрей Альбертович Мучник~--- скончался. Его вклад в деятельность
двух других соавторов далеко не ограничивается формальными
публикациями, перечисленными в настоящем обзоре и
некрологе~\cite{MuchnikObituary07}. Завершение нашей работы над
текстом обзора является для нас посильным вкладом в память об Андрее
и продолжение его работ и идей в мировой математике.~---
А.~Л.~Семёнов, Ю.~Л.~Притыкин.

%\subsection*{Непроцитированная литература}
%\cite{KolmFom04-rus,Muchnik01-rus,Sem84-diss}

%\bibliographystyle{unsrt}
\tocontents{Литература}

%\bibliography{survey_almper}

\end{document}